\documentclass[fleqn,10pt,dvipsnames]{wlscirep}
\usepackage[utf8]{inputenc}
\usepackage[T1]{fontenc}

\title{Review: Deep Learning in Electron Microscopy}

\author[1,*]{Jeffrey M. Ede}
\affil[1]{University of Warwick, Department of Physics, Coventry, CV4 7AL, UK}
\affil[*]{j.m.ede@warwick.ac.uk}

\newcommand*\linkcolours{CornflowerBlue}

\usepackage{times}
\usepackage{graphicx}
\usepackage{amssymb}
\usepackage{gensymb}
\usepackage{amsmath}
\usepackage{breakurl}

\usepackage{url,hyperref}
\hypersetup{
colorlinks,
linkcolor=\linkcolours,
citecolor=\linkcolours,
filecolor=\linkcolours,
urlcolor=\linkcolours
}

\usepackage{algorithm}
\usepackage{algorithmic}

\usepackage{mathtools, cuted}

\usepackage{tabularx}

\usepackage{multirow}

\usepackage{float}

\newcolumntype{Y}{>{\centering\arraybackslash}X}

\usepackage[]{placeins}

\usepackage{pgf}

\usepackage{csquotes}

\newcommand*\diff{\mathop{}\!\mathrm{d}}

\usepackage{array}
\newcolumntype{C}[0]{>{\centering\arraybackslash}c}

\usepackage{tikz}

\begin{abstract}
Deep learning is transforming most areas of science and technology, including electron microscopy. This review paper offers a practical perspective aimed at developers with limited familiarity. For context, we review popular applications of deep learning in electron microscopy. Afterwards, we discuss hardware and software needed to get started with deep learning and interface with electron microscopes. We then review neural network components, popular architectures, and their optimization. Finally, we discuss future directions of deep learning in electron microscopy.

\end{abstract}

\begin{document}

\flushbottom
\maketitle
\thispagestyle{empty}

\section{Introduction}\label{sec:introduction}

Following decades of exponential increases in computational capability\cite{leiserson2020there} and widespread data availability\cite{sun2017revisiting, hey2020machine}, scientists can routinely develop artificial neural networks\cite{sengupta2020review, shrestha2019review, dargan2019survey, alom2019state, zhang2018survey, hatcher2018survey, lecun2015deep, schmidhuber2015deep} (ANNs) to enable new science and technology\cite{ge2020deep, carleo2019machine, wei2019machine, barbastathis2019use, schleder2019dft, von2020introducing}. The resulting deep learning revolution\cite{sejnowski2018deep, alom2018history} has enabled superhuman performance in image classification\cite{wang2018deepsexual, kheradpisheh2016deep, he2015delving, lu2015surpassing}, games\cite{alphastarblog, firoiu2017beating, lample2017playing, silver2016mastering, mnih2013playing, tesauro2002programming}, medical analysis\cite{han2018deep, wang2016deepbreast}, relational reasoning\cite{santoro2017simple}, speech recognition\cite{xiong2016achieving, weng2014single} and many other applications\cite{lee2017superhuman, weyand2016planet}. This introduction focuses on deep learning in electron microscopy and is aimed at developers with limited familiarity. For context, we therefore review popular applications of deep learning in electron microscopy. We then review resources available to support researchers and outline electron microscopy. Finally, we review popular ANN architectures and their optimization, or  \enquote{training}, and discuss future trends in artificial intelligence (AI) for electron microscopy. 

Deep learning is motivated by universal approximator theorems\cite{kidger2019universal, lin2018resnet, hanin2017approximating, lu2017expressive, pinkus1999approximation, leshno1993multilayer, hornik1991approximation, hornik1989multilayer, cybenko1989approximation}, which state that sufficiently deep and wide\cite{kidger2019universal, johnson2018deep, lu2017expressive} ANNs can approximate functions to arbitrary accuracy. It follows that ANNs can always match or surpass the performance of methods crafted by humans. In practice, deep neural networks (DNNs) reliably\cite{lin2017does} learn to express\cite{guhring2020expressivity, raghu2017expressive, poole2016exponential, hanin2019deep} generalizable\cite{cao2020generalization, geiger2020scaling, dziugaite2020revisiting, cao2019generalization, xu2018understanding, neyshabur2017exploring, wu2017towards, kawaguchi2017generalization} models without a prior understanding of physics. As a result, deep learning is freeing physicists from a need to devise equations to model complicated phenomena\cite{iten2020discovering, wu2019toward, carleo2019machine, wei2019machine, schleder2019dft}. Many modern ANNs have millions of parameters, so inference often takes tens of milliseconds on graphical processing units (GPUs) or other hardware accelerators\cite{chen2020survey}. It is therefore unusual to develop ANNs to approximate computationally efficient methods with exact solutions, such as the fast Fourier transform\cite{garrido2019hardware, velik2008discrete, moreland2003fft} (FFT). However, ANNs are able to leverage an understanding of physics to accelerate time-consuming or iterative calculations\cite{breen2020newton, ryczko2019deep, sinitskiy2018deep, zhang2018fast}, improve accuracy of methods\cite{ede2019improving, han2018deep, wang2016deepbreast}, and find solutions that are otherwise intractable\cite{alphastarblog, krizhevsky2012imagenet}.

\begin{figure*}[tbh!]
\centering
\includegraphics[width=0.9\textwidth]{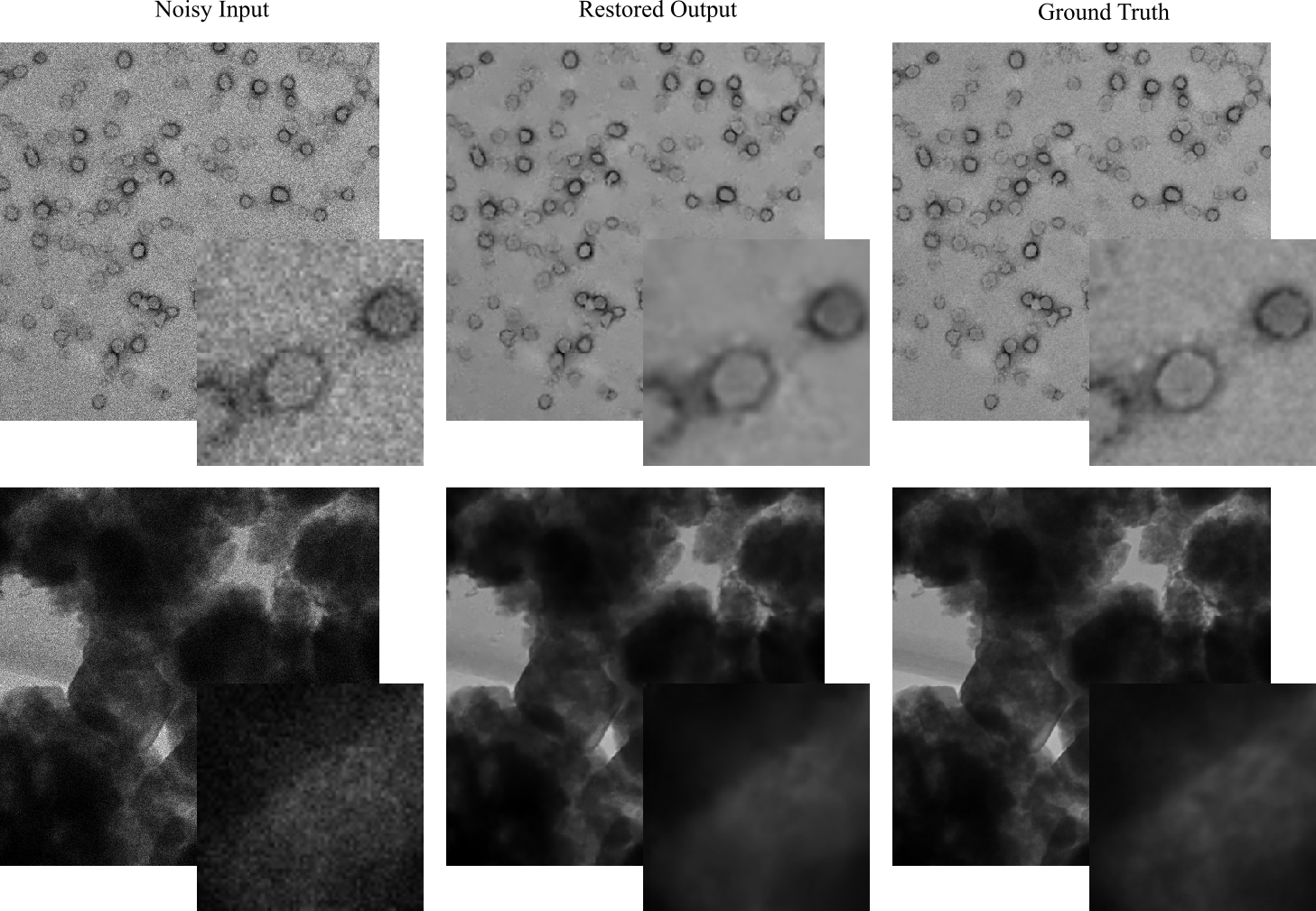}
\caption{ Example applications of a noise-removal DNN to instances of Poisson noise applied to 512$\times$512 crops from TEM images. Enlarged 64$\times$64 regions from the top left of each crop are shown to ease comparison. This figure is adapted from our earlier work\cite{ede2018improvingarxiv} under a Creative Commons Attribution 4.0\cite{cc2020by} license. }
\label{fig:denoiser_example}
\end{figure*}

\subsection{Improving Signal-to-Noise}

A popular application of deep learning is to improve signal-to-noise\cite{liu2019overview, tian2019deep}. For example, of medical electrical\cite{yoon2019deep, antczak2018deep}, medical image\cite{bai2020probabilistic, jifara2019medical, feng2019speckle}, optical microscopy\cite{de2019deepoptical, manifold2019denoising, devalla2019deep, choi2019cycle}, and speech\cite{azarang2020review, choi2020phase, alamdari2019self, han2015learning} signals. There are many traditional denoising algorithms that are not based on deep learning\cite{goyal2020image, girdher2019image, fan2019brief}, including linear\cite{gedraite2011investigation, deng1993adaptive} and non-linear\cite{chang2019automatic, chaudhury2015image, anantrasirichai2014adaptive, tomasi1998bilateral, budhiraja2018efficient, nair2019robust, perona1990scale, wang1999progressive, yang1995optimal} spatial domain filters, Wiener\cite{kodi2017wiener, elsner2013efficient, robinson1967principles} filters, non-linear\cite{bayer2019iterative, mohideen2008image, luisier2007new, jansen2001empirical, chang2000adaptive, donoho1994ideal} wavelet domain filters, curvelet transforms\cite{ma2010curvelet, starck2002curvelet}, contourlet transforms\cite{ahmed2015nonparametric, do2005contourlet}, hybrid algorithms\cite{diwakar2019wavelet, thakur2015hybrid, nagu12014image, bae2014spatial, knaus2013dual, danielyan2011bm3d, dabov2007image} that operate in both spatial and transformed domains, and dictionary-based learning\cite{jia2017image, shao2013heuristic, chatterjee2009clustering, aharon2006k, elad2006image}. However, traditional denoising algorithms are limited by features (often laboriously) crafted by humans and cannot exploit domain-specific context. In perspective, they leverage an ever-increasingly accurate representation of physics to denoise signals. However, traditional algorithms are limited by the difficulty of programmatically describing a complicated reality. As a case in point, an ANN was able to outperform decades of advances in traditional denoising algorithms after training on two GPUs for a week\cite{ede2019improving}.

Definitions of electron microscope noise can include statistical noise\cite{pairis2019shot, seki2018theoretical, lee2014electron, timischl2012statistical, sim2004effect, boyat2015review, meyer2000characterisation, kujawa1992performance}, aberrations\cite{rose2008optics}, scan distortions\cite{fujinaka2020understanding, sang2016dynamic, ning2018scanning, ophus2016correcting}, specimen drift\cite{jones2013identifying}, and electron beam damage\cite{karthik2011situ}. Statistical noise is often minimized by either increasing electron dose or applying traditional denoising algorithms\cite{roels2020interactive, narasimha2008evaluation}. There are a variety of denoising algorithms developed for electron microscopy, including algorithms based on block matching\cite{mevenkamp2015poisson}, contourlet transforms\cite{ahmed2015nonparametric, do2005contourlet}, energy minimization\cite{bajic2016blind}, fast patch reorderings\cite{bodduna2020image}, Gaussian kernel density estimation\cite{jonic2016denoising}, Kronecker envelope principal component analysis\cite{chung29202sdr} (PCA), non-local means and Zernike moments\cite{wang2013zernike}, singular value thresholding\cite{furnival2017denoising}, wavelets\cite{sorzano2006improved}, and other approaches\cite{ouyang2018cryo, du2015nonlinear, jones2013identifying, kushwaha2012noising, hanai1997maximum}. Noise that is not statistical is often minimized by hardware. For example, by using aberration correctors\cite{rose2008optics, pennycook2017impact, ramasse2017twenty, hawkes2009aberration}, choosing scanning transmission electron microscopy (STEM) scan shapes and speeds that minimize distortions\cite{sang2016dynamic}, and using stable sample holders to reduce drift\cite{goodge2020atomic}. Beam damage can also be reduced by using minimal electron voltage and electron dose\cite{egerton2019radiation, egerton2013control, egerton2012mechanisms}, or dose-fractionation across multiple frames in multi-pass transmission electron microscopy\cite{mankos2019electron, koppell2019design, juffmann2017multi} (TEM) or STEM\cite{jones2018managing}.

Deep learning is being applied to improve signal-to-noise for a variety of applications\cite{krull2019noise2void, guo2019toward, lefkimmiatis2018universal, weigert2018content, zhang2018ffdnet, weigert2017isotropic, zhang2017beyond, tai2017memnet, mao2016image}. Most approaches in electron microscopy involve training ANNs to either map low-quality experimental\cite{buchholz2019cryo}, artificially deteriorated\cite{ede2019improving, fang2019deep} or synthetic\cite{mohan2020deep, giannatou2019deep, chaudhary2019line, vasudevan2019deep} inputs to paired high-quality experimental measurements. For example, applications of a DNN trained with artificially deteriorated TEM images are shown in figure~\ref{fig:denoiser_example}. However, ANNs have also been trained with unpaired datasets of low-quality and high-quality electron micrographs\cite{wang2020noise2atom}, or pairs of low-quality electron micrographs\cite{bepler2019topaz, lehtinen2018noise2noise}. Another approach is Noise2Void\cite{krull2019noise2void}, ANNs are trained from single noisy images. However, Noise2Void removes information by masking noisy input pixels corresponding to target output pixels. So far, most ANNs that improve electron microscope signal-to-noise have been trained to decrease statistical noise\cite{wang2020noise2atom, buchholz2019cryo, ede2019improving, tegunov2019real, chaudhary2019line, vasudevan2019deep, bepler2019topaz, mohan2020deep, giannatou2019deep, chaudhary2019line} as other approaches have been developed to correct electron microscope scan distortions\cite{zhang2017joint, jin2015correction} and specimen drift\cite{tong2019image, jin2015correction, jones2013identifying}. However, we anticipate that ANNs will be developed to correct a variety of electron microscopy noise as ANNs have been developed for aberration correction of optical microscopy\cite{krishnan2020optical, cumming2020direct, wang2020correction, tian2019dnn, rivenson2018deep, nguyen2017automatic} and photoacoustic\cite{jeon2020deep} signals.

\begin{figure*}[tbh!]
\centering
\includegraphics[width=\textwidth]{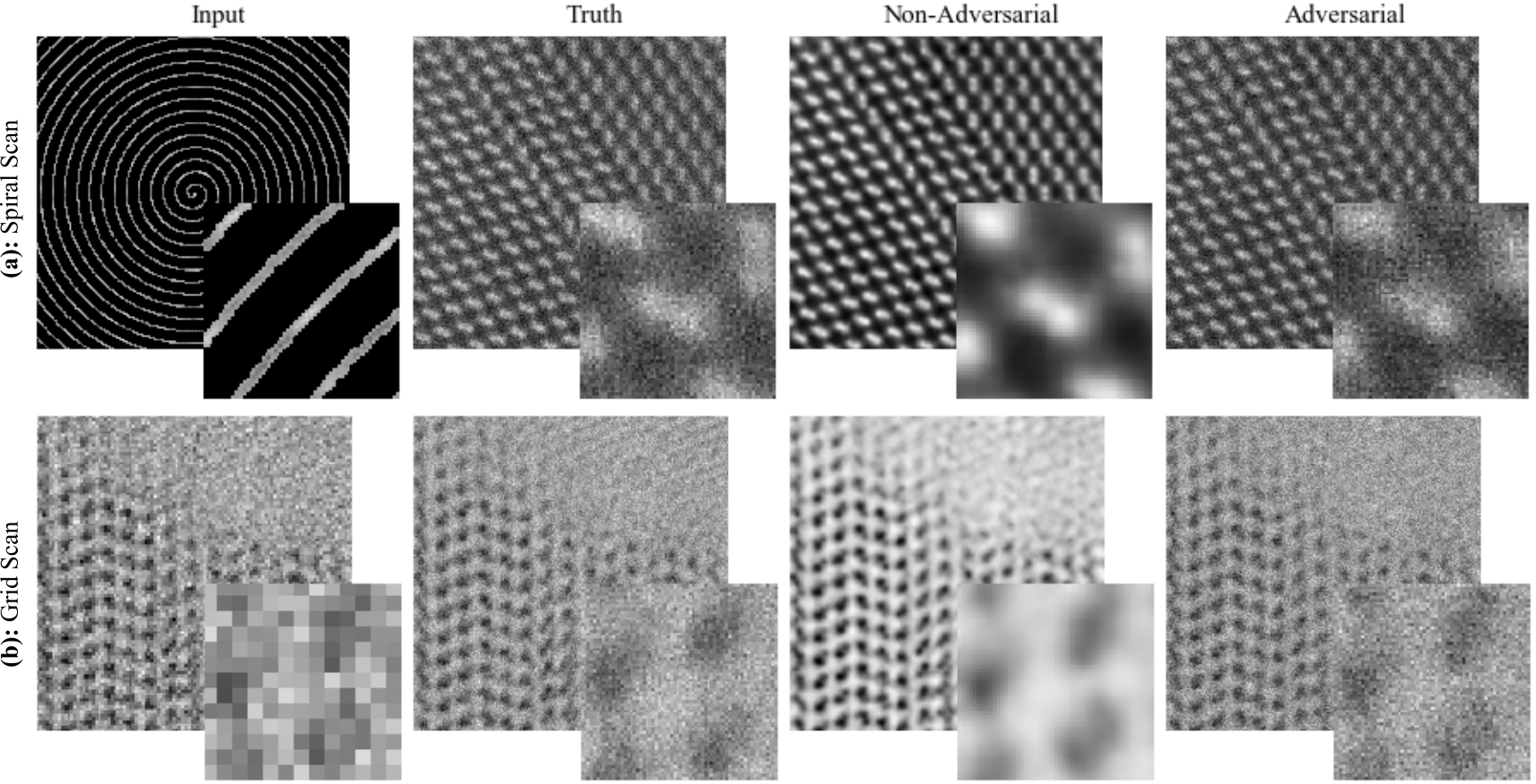}
\caption{ Example applications of DNNs to restore 512$\times$512 STEM images from sparse signals. Training as part of a generative adversarial network\cite{gui2020review, saxena2020generative, pan2019recent, wang2019generative} yields more realistic outputs than training a single DNN with mean squared errors. Enlarged 64$\times$64 regions from the top left of each crop are shown to ease comparison. a) Input is a Gaussian blurred 1/20 coverage spiral\cite{ede2020partial}. b) Input is a 1/25 coverage grid\cite{ede2019deep}. This figure is adapted from our earlier works under Creative Commons Attribution 4.0\cite{cc2020by} licenses. }
\label{fig:compressed_example}
\end{figure*}

\subsection{Compressed Sensing}\label{sec:compressed_sensing}

Compressed sensing\cite{atta2020comparison, vidyasagar2019introduction, rani2018systematic, eldar2012compressed, donoho2006compressed} is the efficient reconstruction of a signal from a subset of measurements. Applications include faster medical imaging\cite{johnson2020improving, ye2019compressed, lustig2007sparse}, image compression\cite{yuan2020image, gunasheela2019compressed}, increasing image resolution\cite{wang2020deep, yang2019deep}, lower medical radiation exposure\cite{shin2020low, cong2019deep, barkan2013adaptive}, and low-light vision\cite{almasri2020robust, chen2018learning}. In STEM, compressed sensing has enabled electron beam exposure and scan time to be decreased by 10-100$\times$ with minimal information loss\cite{ede2020partial, ede2019deep}. Thus, compressed sensing can be essential to investigations where the high current density of electron probes damages specimens\cite{egerton2019radiation, peet2019energy, zhang2018radiation, lehnert2017electron, hermannsdorfer2016effect, johnston2016dose, jenkins2000characterisation, Egerton_2004}. Even if the effects of beam damage can be corrected by postprocessing, the damage to specimens is often permanent. Examples of beam-sensitive materials include organic crystals\cite{s2019low}, metal-organic frameworks\cite{mayoral2017cs}, nanotubes\cite{gnanasekaran2018quantification}, and nanoparticle dispersions\cite{ilett2019cryo}. In electron microscopy, compressed sensing is especially effective due to high signal redundancy\cite{ede2020warwick}. For example, most electron microscopy images are sampled at 5-10$\times$ their Nyquist rates\cite{landau1967sampling} to ease visual inspection, decrease sub-Nyquist aliasing\cite{amidror2015sub}, and avoid undersampling.

Perhaps the most popular approach to compressed sensing is upsampling or infilling a uniformly spaced grid of signals\cite{fadnavis2014image, getreuer2011linear, turkowski1990filters}. Interpolation methods include Lancsoz\cite{fadnavis2014image}, nearest neighbour\cite{beretta2016nearest}, polynomial interpolation\cite{alfeld1984trivariate}, Wiener\cite{cruz2017single} and other resampling methods\cite{zulkifli2019rational, costella2017magickernel, olivier2012nearest}. However, a variety of other strategies to minimize STEM beam damage have also been proposed, including dose fractionation\cite{Jones2018} and a variety of sparse data collection methods\cite{Trampert2018}. Perhaps the most intensively investigated approach to the latter is sampling a random subset of pixels, followed by reconstruction using an inpainting algorithm\cite{Stevens2018, Trampert2018, Hwang2017, Hujsak_2016, anderson2013, Stevens2013}. Random sampling of pixels is nearly optimal for reconstruction by compressed sensing algorithms\cite{Candes2007}. However, random sampling exceeds the design parameters of standard electron beam deflection systems, and can only be performed by collecting data slowly\cite{Kovarik2016, sang2016dynamic}, or with the addition of a fast deflection or blanking system\cite{Hujsak_2016, Beche2016}.

Sparse data collection methods that are more compatible with conventional STEM electron beam deflection systems have also been investigated. For example, maintaining a linear fast scan deflection whilst using a widely-spaced slow scan axis with some small random `jitter'\cite{Kovarik2016, Stevens2018}. However, even small jumps in electron beam position can lead to a significant difference between nominal and actual beam positions in a fast scan. Such jumps can be avoided by driving functions with continuous derivatives, such as those for spiral and Lissajous scan paths\cite{ede2020partial, Li2018, sang2016dynamic, Sang2017a, Hujsak_2016}. Sang\cite{sang2016dynamic, Sang2017a} considered a variety of scans including Archimedes and Fermat spirals, and scans with constant angular or linear displacements, by driving electron beam deflectors with a field-programmable gate array\cite{gandhare2019survey} (FPGA) based system\cite{sang2016dynamic}. Spirals with constant angular velocity place the least demand on electron beam deflectors. However, dwell times, and therefore electron dose, decreases with radius. Conversely, spirals created with constant spatial speeds are prone to systematic image distortions due to lags in deflector responses. In practice, fixed doses are preferable as they simplify visual inspection and limit the dose dependence of STEM noise\cite{seki2018theoretical}.

Deep learning can leverage an understanding of physics to infill images\cite{qiao2020deep, wu2019deep, adler2017block}. Example applications include increasing scanning electron microscopy\cite{de2019resolution, fang2019deep, gao2020deep} (SEM), STEM\cite{ede2019deep, ede2020adaptive} and TEM\cite{suveer2019super} resolution, and infilling continuous sparse scans\cite{ede2020partial}. Example applications of DNNs to complete sparse spiral and grid scans are shown in figure~\ref{fig:compressed_example}. However, caution should be used when infilling large regions as ANNs may generate artefacts if a signal is unpredictable\cite{ede2020partial}. A popular alternative to deep learning for infilling large regions is exemplar-based infilling\cite{ahmed2020quality, pinjarkar2019robust, zhang2019exemplar, criminisi2004region}. However, exemplar-based infilling often leaves artefacts\cite{lu2020detection} and is usually limited to leveraging information from single images. Smaller regions are often infilled by fast marching\cite{telea2004image}, Navier-Stokes infilling\cite{bertalmio2001navier}, or interpolation\cite{alfeld1984trivariate}.

\subsection{Labelling}

Deep learning has been the basis of state-of-the-art classification\cite{he2019bag, sun2019evolving, rawat2017deep, druzhkov2016survey} since convolutional neural networks (CNNs) enabled a breakthrough in classification accuracy on ImageNet\cite{krizhevsky2012imagenet}. Most classifiers are single feedforward neural networks (FNNs) that learn to predict discrete labels. In electron microscopy, applications include classifying image region quality\cite{yokoyama2020development, sanchez2020micrographcleaner}, material structures\cite{aguiar2019decoding, vasudevan2018mapping}, and image resolution\cite{avramov2019deep}. However, siamese\cite{koch2015siamese, chopra2005learning, bromley1994signature} and dynamically parameterized\cite{cai2018memory} networks can more quickly learn to recognise images. Finally, labelling ANNs can learn to predict continuous features, such as mechanical properties\cite{li2019predicting}. Labelling ANNs are often combined with other methods. For example, ANNs can be used to automatically identify particle locations\cite{sanchez2018deep, wang2016deeppicker, tegunov2019real, george2020cassper} to ease subsequent processing.

\begin{figure*}[tbh!]
\centering
\includegraphics[width=0.8\textwidth]{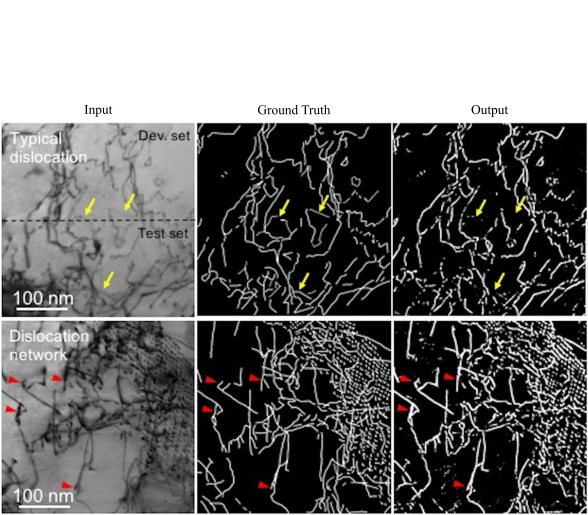}
\caption{ Example applications of a semantic segmentation DNN to STEM images of steel to classify dislocation locations. Yellow arrows mark uncommon dislocation lines with weak contrast, and red arrows indicate that fixed widths used for dislocation lines are sometimes too narrow to cover defects. This figure is adapted with permission\cite{roberts2019deep} under a Creative Commons Attribution 4.0\cite{cc2020by} license. }
\label{fig:semantic_example}
\end{figure*}

\subsection{Semantic Segmentation}\label{sec:semantic_segmentation}

Semantic segmentation is the classification of pixels into discrete categories. In electron microscopy, applications include the automatic identification of local features\cite{madsen2018deep, ziatdinov2017deep}, such as defects\cite{ziatdinov2019building, meyer2008direct}, dopants\cite{meyer2011experimental}, material phases\cite{he2014situ}, material structures\cite{nagao2015experimental, li2017direct}, dynamic surface phenomena\cite{schneider2014atomic}, and chemical phases in nanoparticles\cite{hussaini2018determination}. Early approaches to semantic segmentation used simple rules. However, such methods were not robust to a high variety of data\cite{pham2000current}. Subsequently, more adaptive algorithms based on soft-computing\cite{mesejo2015biomedical} and fuzzy algorithms\cite{zheng2015image} were developed to use geometric shapes as priors. However, these methods were limited by programmed features and struggled to handle the high variety of data.

To improve performance, DNNs have been trained to semantically segment images\cite{hao2020brief, sultana2020evolution, minaee2020image, guo2018review, chen2018encoder, chen2017rethinking, badrinarayanan2017segnet, ronneberger2015u}. Semantic segmentation DNNs have been developed for focused ion beam scanning electron microscopy\cite{yi2020adversarial, khadangi2020net, roels2019cost} (FIB-SEM), SEM\cite{yu2020high, fakhry2016residual, urakubo2019uni, roels2019cost}, STEM\cite{roberts2019defectnet, roberts2019deep}, and TEM\cite{george2020cassper, ibtehaz2020multiresunet, groschner2020methodologies, khadangi2020net, horwath2019understanding, roels2019cost, chen2017convolutional}. For example, applications of a DNN to semantic segmentation of STEM images of steel are shown in figure~\ref{fig:semantic_example}. Deep learning based semantic segmentation also has a high variety of applications outside of electron microscopy, including autonomous driving\cite{feng2020deep, yang2020lightningnet, hofmarcher2019visual, blum2019fishyscapes, zhou2019automated}, dietary monitoring\cite{pfisterer2019fully, aslan2018semantic}, magnetic resonance images\cite{ghosh2020automated, memis2020semantic, duran2020prostate, bevilacqua2019comparison, liu2018deep}, medical images\cite{taghanaki2020deep, tajbakhsh2020embracing, du2020medical} such as prenatal ultrasound\cite{yang2020hybrid, wang2019joint, venturini2019multi, yang2018towards}, and satellite image translation\cite{tasar2020standardgan, barthakur2020deep, wu2019towards, wurm2019semantic, zhou2018d}. Most DNNs for semantic segmentation are trained with images segmented by humans. However, human labelling may be too expensive, time-consuming, or inappropriate for sensitive data. Unsupervised semantic segmentation can avoid these difficulties by learning to segment images from an additional dataset of segmented images\cite{joyce2018deep} or image-level labels\cite{araslanov2020single, chen2020exploiting, jing2019coarse, oh2017exploiting}. However, unsupervised semantic segmentation networks are often less accurate than supervised networks.

\begin{figure*}[tbh!]
\centering
\includegraphics[width=\textwidth]{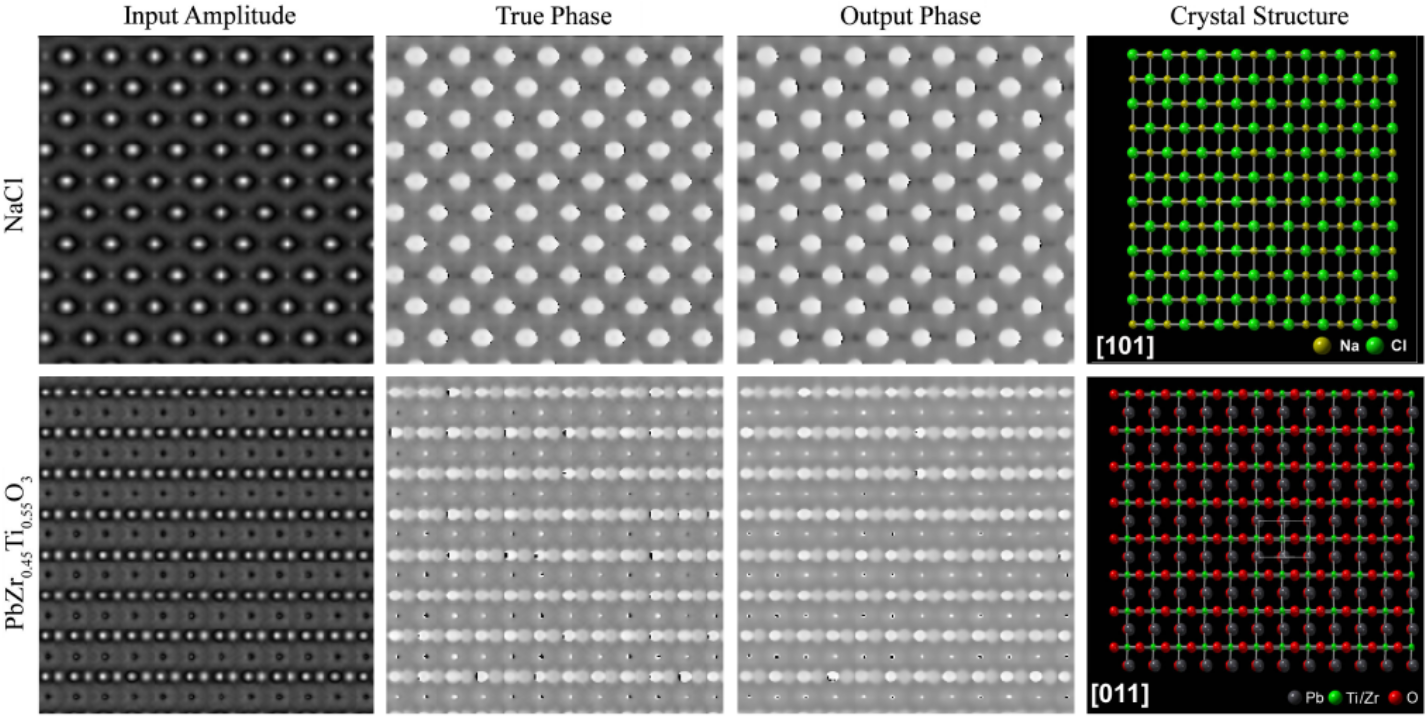}
\caption{ Example applications of a DNN to reconstruct phases of exit wavefunction from intensities of single TEM images. Phases in $[-\pi, \pi)$ rad are depicted on a linear greyscale from black to white, and Miller indices label projection directions. This figure is adapted from our earlier work\cite{ede2020exit} under a Creative Commons Attribution 4.0\cite{cc2020by} license. }
\label{fig:wavefunction_example}
\end{figure*}

\subsection{Exit Wavefunction Reconstruction}

Electrons exhibit wave-particle duality\cite{frabboni2007young, matteucci1998experiment}, so electron propagation is often described by wave optics\cite{lehmann2002tutorial}. Applications of electron wavefunctions exiting materials\cite{tonomura1987applications} include determining projected potentials and corresponding crystal structure information\cite{lentzen2000reconstruction, auslender2019measuring}, information storage, point spread function deconvolution, improving contrast, aberration correction\cite{fu1991correction}, thickness measurement\cite{mccartney1994absolute}, and electric and magnetic structure determination\cite{park2014observation, dunin2004off}. Usually, exit wavefunctions are either iteratively reconstructed from focal series\cite{lubk2016fundamentals, koch2014towards, haigh2013recording, koch2010off, van1994object} or recorded by electron holography\cite{lehmann2002tutorial, koch2010off, ozsoy2014hybridization}. However, iterative reconstruction is often too slow for live applications, and holography is sensitive to distortions and may require expensive microscope modification. 

Non-iterative methods based on DNNs have been developed to reconstruct optical exit wavefunctions from focal series\cite{zhang2018fast} or single images\cite{rivenson2018phase, wu2018extended, sinha2017lensless}. Subsequently, DNNs have been developed to reconstruct exit wavefunctions from single TEM images\cite{ede2020exit}, as shown in figure~\ref{fig:wavefunction_example}. Indeed, deep learning is increasingly being applied to accelerated quantum mechanics\cite{beach2018qucumber, dral2020quantum, liu2020deepfeynman, bharti2020machine, carleo2019netket, schutt2019unifying}. Other examples of DNNs adding new dimensions to data include semantic segmentation described in section~\ref{sec:semantic_segmentation}, and reconstructing 3D atomic distortions from 2D images\cite{laanait2019reconstruction}. Non-iterative methods that do not use ANNs to recover phase information from single images have also been developed\cite{morgan2011direct, martin2008direct}. However, they are limited to defocused images in the Fresnel regime\cite{morgan2011direct}, or to non-planar incident wavefunctions in the Fraunhofer regime\cite{martin2008direct}.


\section{Resources}

Access to scientific resources is essential to scientific enterprise\cite{schlitz2018science}. Fortunately, most resources needed to get started with machine learning are freely available. This section provides directions to various machine learning resources, including how to access deep learning frameworks, a free GPU or tensor processing unit (TPU) to accelerate tensor computations, platforms that host datasets and source code, and pretrained models. To support the ideals of open science embodied by Plan S\cite{sparc2018coalition, schlitz2018science, PlanS2020}, we focus on resources that enhance collaboration and enable open access\cite{banks2019answers}. We also discuss how electron microscopes can interface with ANNs and the importance of machine learning resources in the context of electron microscopy. However, we expect that our insights into electron microscopy can be generalized to other scientific fields.

\subsection{Hardware Acceleration}

A DNN is an ANN with multiple layers that perform a sequence of tensor operations. Tensors can either be computed on central processing units (CPUs) or hardware accelerators\cite{chen2020survey}, such as FPGAs\cite{shi2020ftdl, kaarmukilan2020fpga, wang2019overview, guo2019dl}, GPUs\cite{cano2018survey, nvidia2017v100, heterogeneous-computing}, and TPUs\cite{gordienko2020scaling, jouppi2018motivation, jouppi2017datacenter}. Most benchmarks indicate that GPUs and TPUs outperform CPUs for typical DNNs that could be used for image processing\cite{mattson2020mlperf, mlperf, wang2019benchmarking, wang2019performance, li2019cpu} in electron microscopy. However, GPU and CPU performance can be comparable when CPU computation is optimized\cite{awan2017depth}. TPUs often outperform GPUs\cite{wang2019benchmarking}, and FPGAs can outperform GPUs\cite{nurvitadhi2017can, berten2016gpu} if FPGAs have sufficient arithmetic units\cite{nangia2018resource, grover2014design}. Typical power consumption per TFLOPS\cite{dolbeau2018theoretical} decreases in order CPU, GPU, FPGA, then TPU, so hardware acceleration can help to minimize long-term costs and environmental damage\cite{strubell2019energy}.

For beginners, Google Colab\cite{nelson2020notes, bisong2019google, tutorial2019colab, carneiro2018performance} and Kaggle\cite{kaggle_docs} provide hardware accelerators in ready-to-go deep learning environments. Free compute time on these platforms is limited as they are not intended for industrial applications. Nevertheless, the free compute time is sufficient for some research\cite{kalinin2020decoding}. For more intensive applications, it may be necessary to get permanent access to hardware accelerators. If so, many online guides detail how to install\cite{green2019how, ryan2017how} and set up an Nvidia\cite{radecic2020an} or AMD\cite{varile2019train} GPU in a desktop computer for deep learning. However, most hardware comparisons for deep learning\cite{dettmers2018a} focus on Nvidia GPUs as most deep learning frameworks use Nvidia's proprietary Compute Unified Device Architecture (CUDA) Deep Neural Network (cuDNN) primitives for deep learning\cite{chetlur2014cudnn}, which are optimized for Nvidia GPUs. Alternatively, hardware accelerators may be accessible from a university or other institutional high performance computing (HPC) centre, or via a public cloud service provider\cite{listofcloud2020, marozzo2019infrastructures, joshi2019comprehensive, gupta2020deploying}.

\begin{table}[tbh!]
\centering
\footnotesize
\begin{tabular*}{\textwidth}{@{\extracolsep{\fill}}lll}
\hline
\multicolumn{1}{c}{Framework} & \multicolumn{1}{c}{License} & \multicolumn{1}{c}{Programming Interfaces} \\
\hline
Apache SINGA\cite{ooi2015singa} & Apache 2.0\cite{singalicense} & C++, Java, Python \\
BigDL\cite{dai2019bigdl} & Apache 2.0\cite{bigdllicense} & Python, Scala \\
Caffe\cite{jia2014caffe, caffemergespytorch} & BSD\cite{caffelicense} & C++, MATLAB, Python \\
Chainer\cite{tokui2015chainer} & MIT\cite{chainerlicense} & Python \\
Deeplearning4j\cite{gibson2016deeplearning4j} & Apache 2.0\cite{dl4jlicense} & Clojure, Java, Kotlin, Python, Scala \\
Dlib\cite{king2009dlib, dlib} & BSL\cite{dliblicense} & C++ \\
Flux\cite{innes2018flux} & MIT\cite{fluxlicense} & Julia \\
MATLAB Deep Learning Toolbox\cite{beale2020matlab} & Proprietary\cite{matlablicense} & MATLAB \\
Microsoft Cognitive Toolkit\cite{seide2017keynote} & MIT\cite{cntklicense} & BrainScript, C++, Python \\
Apache MXNet\cite{chen2015mxnet} & Apache 2.0\cite{mxnetlicense} & C++, Clojure, Go, JavaScript, Julia, Matlab, Perl, Python, R, Scala \\
OpenNN\cite{opennn} & GNU LGPL\cite{opennnlicense} & C++ \\
PaddlePaddle\cite{ma2019paddlepaddle} & Apache 2.0\cite{paddlepaddlelicense} & C++ \\
PyTorch\cite{paszke2019pytorch} & BSD\cite{pytorchlicense} & C++, Python \\
TensorFlow\cite{abadi2016tensorflow, abadi2016tensorflow_large} & Apache 2.0\cite{tensorflowlicense} & C++, C\#, Go, Haskell, Julia, MATLAB, Python, Java, JavaScript, R, Ruby, Rust, Scala, Swift \\
Theano\cite{team2016theano, ketkar2017introduction} & BSD\cite{theanolicense} & Python \\
Torch\cite{collobert2002torch} & BSD\cite{torchlicense} & C, Lua \\
Wolfram Mathematica\cite{mathematica2020documentation} & Proprietary\cite{wolframlicense} & Wolfram Language \\
\hline
\end{tabular*}
\caption{ Deep learning frameworks with programming interfaces. Most frameworks have open source code and many support multiple programming languages. }
\label{table:ai_framworks}
\end{table}

\subsection{Deep Learning Frameworks}

A deep learning framework\cite{li2020deep, nguyen2019machine, dai2019benchmarking, kharkovyna2019top, zacharias2018survey, hatcher2018survey, parvat2017survey, erickson2017toolkits} (DLF) is an interface, library or tool for DNN development. Features often include automatic differentiation\cite{baydin2017automatic}, heterogeneous computing, pretrained models, and efficient computing\cite{barham2019machine} with CUDA\cite{afif2020computer, cook2012cuda, nickolls2008scalable}, cuDNN\cite{chetlur2014cudnn, jorda2019performance}, OpenMP\cite{de2018ongoing, dagum1998openmp}, or similar libraries. Popular DLFs tabulated in table~\ref{table:ai_framworks} often have open source code and support multiple programming interfaces. Overall, TensorFlow\cite{abadi2016tensorflow, abadi2016tensorflow_large} is the most popular DLF\cite{the2019he}. However, PyTorch\cite{paszke2019pytorch} is the most popular DLF at top machine learning conferences\cite{the2019he, paperswithcodetrends}. Some DLFs also have extensions that ease development or extend functionality. For example, TensorFlow extensions\cite{libraries2020extensions} that ease development include Keras\cite{chollet2015keras}, Sonnet\cite{sonnet2020github}, Tensor2Tensor\cite{vaswani2018tensor2tensor} and TFLearn\cite{tang2016tf, tflearn2019}, and extensions that add functionality include Addons\cite{addons2020}, Agents\cite{TFAgents}, Dopamine\cite{castro2018dopamine}, Federated\cite{mcmahan2017federated, TFFederated, caldas2018leaf}, Probability\cite{dillon2017tensorflow}, and TRFL\cite{deepmind2018open}. In addition, DLFs are supplemented by libraries for predictive data analysis, such as scikit-learn\cite{scikit-learn}.

A limitation of the DLFs in table~\ref{table:ai_framworks} is that users must use programming interfaces. This is problematic as many electron microscopists have limited, if any, programming experience. To increase accessibility, a range of graphical user interfaces (GUIs) have been created for ANN development. For example, ANNdotNET\cite{anndotnet}, Create ML\cite{createml}, Deep Cognition\cite{deepcognition}, Deep Network Designer\cite{deepnetworkdesigner}, DIGITS\cite{digits}, ENNUI\cite{ennui}, Expresso\cite{expresso}, Neural Designer\cite{neuraldesigner}, Waikato Environment for Knowledge Analysis\cite{witten2016data, hall2009weka, holmes1994weka} (WEKA) and ZeroCostDL4Mic\cite{von2020zerocostdl4mic}. The GUIs offer less functionality and scope for customization than programming interfaces. However, GUI-based DLFs are rapidly improving. Moreover, existing GUI functionality is more than sufficient to implement popular FNNs, such as image classifiers\cite{rawat2017deep} and encoder-decoders\cite{ye2019understanding, ye2018deep, chen2018encoder, chen2017rethinking, badrinarayanan2017segnet, ronneberger2015u, sutskever2014sequence}.

\subsection{Pretrained Models}

Training ANNs is often time-consuming and computationally expensive\cite{strubell2019energy}. Fortunately, pretrained models are available from a range of open access collections\cite{list2020}, such as Model Zoo\cite{modelzoo2020}, Open Neural Network Exchange\cite{onnx2020, bai2020onnx, shah2017microsoft, boyd2017microsoft} (ONNX) Model Zoo\cite{onnxzoo2020}, TensorFlow Hub\cite{gordon2018introducing, TFHub}, and TensorFlow Model Garden\cite{tensorflow2020model}. Some researchers also provide pretrained models via project repositories\cite{ede2019improving, ede2020exit, ede2020partial, ede2020warwick, ede2019deep}. Pretrained models can be used immediately or to transfer learning\cite{liang2019survey, zhuang2019comprehensive, tan2018survey, marcelino2018transfer, weiss2016survey, yosinski2014transferable, da2020agents} to new applications. For example, by fine-tuning and augmenting the final layer of a pretrained model\cite{shermin2019enhanced}. Benefits of transfer learning can include decreasing training time by orders of magnitude, reducing training data requirements, and improving generalization\cite{yosinski2014transferable, ada2019generalization}.

Using pretrained models is complicated by ANNs being developed with a variety of DLFs in a range of programming languages. However, most DLFs support interoperability. For example, by supporting the saving of models to a common format or to formats that are interoperable with the Neural Network Exchange Format\cite{nnef} (NNEF) or ONNX formats. Many DLFs also support saving models to HDF5\cite{hdf5, h5py}, which is popular in the pycroscopy\cite{somnath2019usid, pycroscopy2020} and HyperSpy\cite{hyperspy2020, de2017electron} libraries used by electron microscopists. The main limitation of interoperability is that different DLFs may not support the same functionality. For example, Dlib\cite{king2009dlib, dlib} does not support recurrent neural networks\cite{rezk2020recurrent, du2019recurrent, yu2019review, choe2017probabilistic, choi2017awesome, lipton2015critical} (RNNs).

\subsection{Datasets}

Randomly initialized ANNs\cite{hanin2018start} must be trained, validated, and tested with large, carefully partitioned datasets to ensure that they are robust to general use\cite{raschka2018model}. Most ANN training starts from random initialization, rather than transfer learning\cite{liang2019survey, zhuang2019comprehensive, tan2018survey, marcelino2018transfer, weiss2016survey, yosinski2014transferable, da2020agents}, as:
\begin{enumerate}
\item Researchers may be investigating modifications to ANN architecture or ability to learn.
\item Pretrained models may be unavailable or too difficult to find. 
\item Models may quickly achieve sufficient performance from random initialization. For example, training an encoder-decoder based on Xception\cite{chollet2017xception} to improve electron micrograph signal-to-noise\cite{ede2019improving} can require less training than for PASCAL VOC 2012\cite{everingham2015pascal} semantic segmentation\cite{chen2018encoder}.
\item There may be a high computing budget, so transfer learning is unnecessary\cite{goyal2017accurate, laanait2019exascale}.
\end{enumerate}
There are millions of open access datasets\cite{castelvecchi2018google, noy2020discovering} and a range of platforms that host\cite{plesa2020machine, dua2020uci, kaggle2020datasets, visual2020data, vanschoren2014openml} or aggregate\cite{stanford2020the, elite2020datasets, iderhoff2020natural, dl2017datasets} machine learning datasets. Openly archiving datasets drives scientific enterprise by reducing need to repeat experiments\cite{hughes2010measurements, jcgm2008100, vaux2012replicates, urbach1981utility, musgrave1975popper}, enabling new applications through data mining\cite{senior2020improved, voss2020database}, and standardizing performance benchmarks\cite{papers2020leaderboards}. For example, popular datasets used to standardize image classification performance benchmarks include CIFAR-10\cite{krizhevsky2014cifar, krizhevsky2009learning}, MNIST\cite{lecun2010mnist} and ImageNet\cite{russakovsky2015imagenet}. A high range of both domain-specific and general platforms that host scientific data for free are listed by the Open Access Directory\cite{oad2020repos} and Nature Scientific Data\cite{nature2020recom}. For beginners, we recommend Zenodo\cite{zenodo} as it is free, open access, has an easy-to-use interface, and will host an unlimited number of datasets smaller than 50 GB for at least 20 years\cite{zenodofaqs}. 

There are a range of platforms dedicated to hosting electron microscopy datasets, including the Caltech Electron Tomography Database\cite{ortega2019etdb} (ETDB-Caltech), Electron Microscopy Data Bank\cite{EMDataResource2020, lawson2016emdatabank, esquivel2015navigating, lawson2010emdatabank, henrick2003emdep, tagari2002new} (EMDataBank), and the Electron Microscopy Public Image Archive\cite{iudin2016empiar} (EMPIAR). However, most electron microscopy datasets are small, esoteric or are not partitioned for machine learning\cite{ede2020warwick}. Nevertheless, a variety of large machine learning datasets for electron microscopy are being published in independent repositories\cite{ede2020warwick, aversa2018first, levin2016nanomaterial}, including Warwick Electron Microscopy Datasets\cite{ede2020warwick} (WEMD) that we curated. In addition, a variety of databases host information that supports electron microscopy. For example, crystal structure databases provide data in standard formats\cite{cerius2020modeling, crystalmaker2020file}, such as Crystallography Information Files\cite{bernstein2016specification, hall2016implementation, brown2002cif, hall1991crystallographic} (CIFs). Large crystal structure databases\cite{bruno2017crystallography, crystallographic2020databases, crystal2020structure} containing over $10^5$ crystal structures include the Crystallography Open Database\cite{Quiros2018, Merkys2016, Grazulis2015, Grazulis2012, Grazulis2009, Downs2003} (COD), Inorganic Crystal Structure Database\cite{zagorac2019recent, allmann2007introduction, hellenbrandt2004inorganic, belsky2002new, bergerhoff1987crystallographic} (ICSD), and National Institute of Standards and Technology (NIST) Crystal Data\cite{mighell1996nist, NIST2020standard}.

\begin{table}[tbh!]
\centering
\footnotesize
\begin{tabular*}{\textwidth}{@{\extracolsep{\fill}}lll}
\hline
\multicolumn{1}{c}{Platform} & \multicolumn{1}{c}{Website} & \multicolumn{1}{c}{For Machine Learning} \\
\hline
Amazon Mechanical Turk & \url{https://www.mturk.com} & General tasks \\
Appen & \url{https://appen.com} & Machine learning data preparation \\
Clickworker & \url{https://www.clickworker.com} & Machine learning data preparation \\
Fiverr & \url{https://www.fiverr.com} & General tasks \\
Hive & \url{https://thehive.ai} & Machine learning data preparation \\
iMerit & \url{https://imerit.net} & Machine learning data preparation \\
JobBoy & \url{https://www.jobboy.com} & General tasks \\
Minijobz & \url{https://minijobz.com} & General tasks \\
Microworkers & \url{https://www.microworkers.com} & General tasks \\
OneSpace & \url{https://freelance.onespace.com} & General tasks \\
Playment & \url{https://playment.io} & Machine learning data preparation \\
RapidWorkers & \url{https://rapidworkers.com} & General tasks \\
Scale & \url{https://scale.com} & Machine learning data preparation \\
Smart Crowd & \url{https://thesmartcrowd.lionbridge.com} & General tasks \\
Trainingset.ai & \url{https://www.trainingset.ai} & Machine learning data preparation \\
ySense & \url{https://www.ysense.com} & General tasks \\
\hline
\end{tabular*}
\caption{Microjob service platforms. The size of typical tasks varies for different platforms and some platforms specialize in preparing machine learning datasets.}
\label{table:microservices}
\end{table}

To achieve high performance, it may be necessary to curate a large dataset for ANN training\cite{sun2017revisiting}. However, large datasets like DeepMind Kinetics\cite{kay2017kinetics}, ImageNet\cite{russakovsky2015imagenet}, and YouTube 8M\cite{abu2016youtube} may take a team months to prepare. As a result, it may not be practical to divert sufficient staff and resources to curate a high-quality dataset, even if curation is partially automated\cite{rehm2020qurator, van2020deepdicomsort, pezoulas2019medical, bhat2019adex, thirumuruganathan2018data, lee2018scaling, abu2016youtube, freitas2016big}. To curate data, human capital can be temporarily and cheaply increased by using microjob services\cite{eumicrocredic}. For example, through microjob platforms tabulated in table~\ref{table:microservices}. Increasingly, platforms are emerging that specialize in data preparation for machine learning. Nevertheless, microjob services may be inappropriate for sensitive data or tasks that require substantial domain-specific knowledge.

\subsection{Source Code}

Software is part of our cultural, industrial, and scientific heritage\cite{di2017software}. Source code should therefore be archived where possible. For example, on an open source code platform such as Apache Allura\cite{apacheallura}, AWS CodeCommit\cite{codecommit}, Beanstalk\cite{beanstalk}, BitBucket\cite{bitbucket}, GitHub\cite{github}, GitLab\cite{gitlab}, Gogs\cite{gogs}, Google Cloud Source Repositories\cite{google2020cloudrepos}, Launchpad\cite{launchpad}, Phabricator\cite{phabricator}, Savannah\cite{savannah} or SourceForge\cite{sourceforge}. These platforms enhance collaboration with functionality that helps users to watch\cite{sheoran2014understanding} and contribute improvements\cite{vale2020relation, bao2019large, elazhary2019not, pinto2016more, kobayakawa2017github, lu2019studying, qiu2019signals} to source code. The choice of platform is often not immediately important for small electron microscopy projects as most platforms offer similar functionality. Nevertheless, functionality comparisons of open source platforms are available\cite{alamer2017open, compareopensource, apachecomparison}. For beginners, we recommend GitHub as it is actively developed, scalable to large projects and has an easy-to-use interface. 

\subsection{Finding Information}

Most web traffic\cite{alexa2020top, alexa2020how} goes to large-scale web search engines\cite{haider2019invisible, vincent2019measuring, jain2013role, brin1998the, frobe2020effect} such as Bing, DuckDuckGo, Google, and Yahoo. This includes searches for scholarly content\cite{kostagiolas2020impact, gul2020retrieval, shafi2019retrieval}. We recommend Google for electron microscopy queries as it appears to yield the best results for general\cite{steiner2020seek, wu2019evaluating, rahim2019evaluation}, scholarly\cite{shafi2019retrieval, gul2020retrieval} and other\cite{tazehkandi2020evaluating} queries. However, general search engines can be outperformed by dedicated search engines for specialized applications. For example, for finding academic literature\cite{gusenbauer2019google, hook2018dimensions, bates2017will}, data\cite{verheggen2020anatomy}, jobs\cite{li2020deepjob, agazzi2020study}, publication venues\cite{forrester2017new}, patents\cite{kang2020patent, kang2020prior, shalaby2019patent, khode2017literature}, people\cite{kong2019academic, makri2019global, acquisti2020experiment}, and many other resources. The use of search engines is increasingly political\cite{mustafaraj2020case, kulshrestha2019search, puschmann2019beyond} as they influence which information people see. However, most users appear to be satisfied with their performance\cite{ray20202020}.

Introductory textbooks are outdated\cite{johnson2019lectures, lin2019teaching} insofar that most information is readily available online. We find that some websites are frequent references for up-to-date and practical information:
\begin{enumerate}
    \item Stack Overflow\cite{stackoverflow, wu2019developers, zhang2019reading, zhang2019empirical, ragkhitwetsagul2019toxic, zhang2018code} is a source of working code snippets and a useful reference when debugging code.
    \item Papers With Code State-of-the-Art\cite{papers2020leaderboards} leaderboards rank the highest performing ANNs with open source code for various benchmarks. 
    \item Medium\cite{medium} and its subsidiaries publish blogs with up-to-date and practical advice about machine learning.
    \item The Machine Learning subreddit\cite{reddit2020machine} hosts discussions about machine learning. In addition, there is a Learn Machine Learning subreddit\cite{reddit2020learn} aimed at beginners.
    \item Dave Mitchell's DigitalMicrograph Scripting Website\cite{mitchell2005scripting, dmscripts2020list} hosts a collection of scripts and documentation for programming electron microscopes. 
    \item The Internet Archive\cite{internetarchive, kanhabua2016search} maintains copies of software and media, including webpages via its Wayback Machine\cite{wayback, bowyer2020wayback, grotke2011web}.
    \item Distill\cite{distill2020about} is a journal dedicated to providing clear explanations about machine learning. Monetary prizes are awarded for excellent communication and refinement of ideas.
\end{enumerate}
This list enumerates popular resources that we find useful, so it may introduce personal bias. However, alternative guides to useful resources are available\cite{lewinson2020my, chadha2020handpicked, besbes2020here}. We find that the most common issues finding information are part of an ongoing reproducibility crisis\cite{hutson2018artificial, baker2016reproducibility} where machine learning researchers do not publish their source code or data. Nevertheless, third party source code is sometimes available. Alternatively, ANNs can reconstruct source code from some research papers\cite{sethi2017dlpaper2code}.

\subsection{Scientific Publishing}

The number of articles published per year in reputable peer-reviewed\cite{publons20182018, tennant2018state, walker2015emerging, vesper2018peer, tan2019performance} scientific journals\cite{kim2019scientific, rallison2015journals} has roughly doubled every nine years since the beginning of modern science\cite{bornmann2015growth}. There are now over 25000 peer-reviewed journals\cite{rallison2015journals} with varying impact factors\cite{kaldas2020journal, orbay2020building, lei2020should}, scopes and editorial policies. Strategies to find the best journal to publish in include using online journal finders\cite{pubrica2019top}, seeking the advice of learned colleagues, and considering where similar research has been published. Increasingly, working papers are also being published in open access preprint archives\cite{hoy2020rise, fry2019praise, rodriguez2019preprints}. For example, the arXiv\cite{arxiv2020about, ginsparg2011arxiv} is a popular preprint archive for computer science, mathematics, and physics. Advantages of preprints include ensuring that research is openly available, increasing discovery and citations\cite{fraser2020relationship, wang2020impact, furnival2020open, fu2019meta, niyazov2016open}, inviting timely scientific discussion, and raising awareness to reduce unnecessary duplication of research. Many publishers have adapted to the popularity of preprints\cite{hoy2020rise} by offering open access publication options\cite{robinson2020state, siler2020pricing, green2019open, gadd2018influence} and allowing, and in some cases encouraging\cite{iop2020why}, the prior publication of preprints. Indeed, some journals are now using the arXiv to host their publications\cite{gibney2016open}. 

A variety of software can help authors prepare scientific manuscripts\cite{martinez2019tools}. However, we think the most essential software is a document preparation system. Most manuscripts are prepared with Microsoft Word\cite{word} or similar software\cite{investintech202010}. However, Latex\cite{pignalberi2019introduction, bransen2018pimp, lamport1994latex} is a popular alternative among computer scientists, mathematicians and physicists\cite{matthews2019craft}. Most electron microscopists at the University of Warwick appear to prefer Word. A 2014 comparison of Latex and Word found that Word is better at all tasks other than typesetting equations\cite{knauff2014efficiency}. However, in 2017 it become possible to use Latex to typeset equations within Word\cite{matthews2019craft}. As a result, Word appears to be more efficient than Latex for most manuscript preparation. Nevertheless, Latex may still be preferable to authors who want fine control over typesetting\cite{kovic2017why, allington2016the}. As a compromise, we use Overleaf\cite{overleaf} to edit Latex source code, then copy our code to Word as part of proofreading to identify issues with grammar and wording.

\begin{figure*}[tbh!]
\centering
\includegraphics[width=\textwidth]{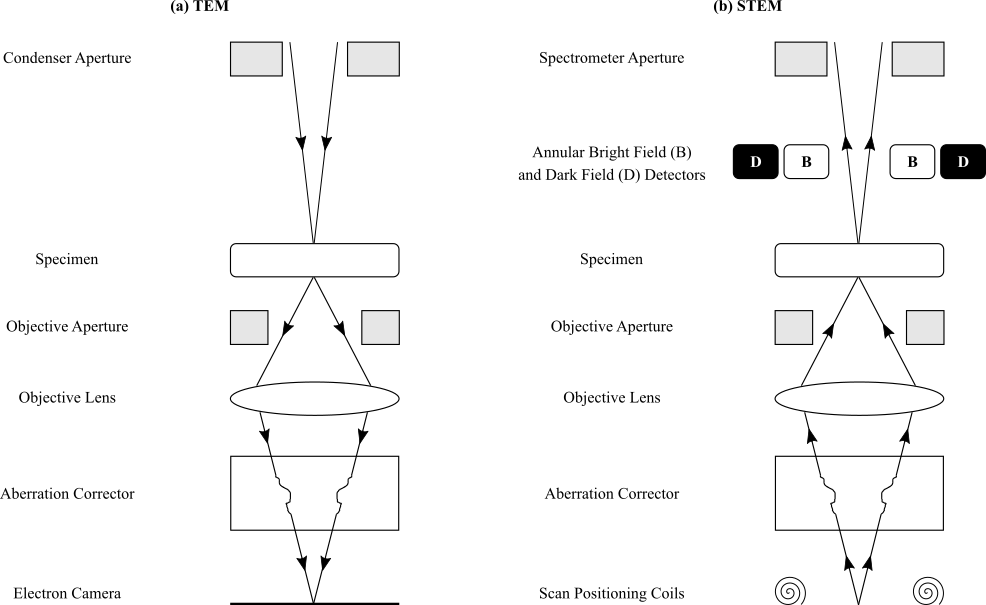}
\caption{ Reciprocity of TEM and STEM electron optics. }
\label{fig:tem_vs_stem}
\end{figure*}

\section{Electron Microscopy}

An electron microscope is an instrument that uses electrons as a source of illumination to enable the study of small objects. Electron microscopy competes with a large range of alternative techniques for material analysis\cite{venkateshaiah2020microscopic, alqaheem2020microscopy, morrison2019characterisation}, including atomic force microscopy\cite{maghsoudy2018review, rugar1990atomic, krull2020artificial} (AFM); Fourier transformed infrared (FTIR) spectroscopy\cite{dutta2017fourier, griffiths2007fourier}; nuclear magnetic resonance\cite{chien2020recent, lambert2019nuclear, mlynarik2017introduction, rabi1938new} (NMR); Raman spectroscopy\cite{smith2019modern, jones2019raman, ameh2019review, rostron2016raman, zhang2016review, epp2016x, keren2008noninvasive}; and x-ray diffraction\cite{khan2020experimental, scarborough2017dynamic} (XRD), dispersion\cite{leani2019energy}, fluorescence\cite{vanhoof20202020, shackley2018x} (XRF), and photoelectron spectroscopy\cite{greczynski2020x, baer2019practical} (XPS). Quantitative advantages of electron microscopes can include higher resolution and depth of field, and lower radiation damage than light microscopes\cite{du2020relative}. In addition, electron microscopes can record images, enabling visual interpretation of complex structures that may otherwise be intractable. This section will briefly introduce varieties of electron microscopes, simulation software, and how electron microscopes can interface with ANNs.

\subsection{Microscopes}

\begin{figure*}[tbh!]
\centering
\includegraphics[width=0.55\textwidth]{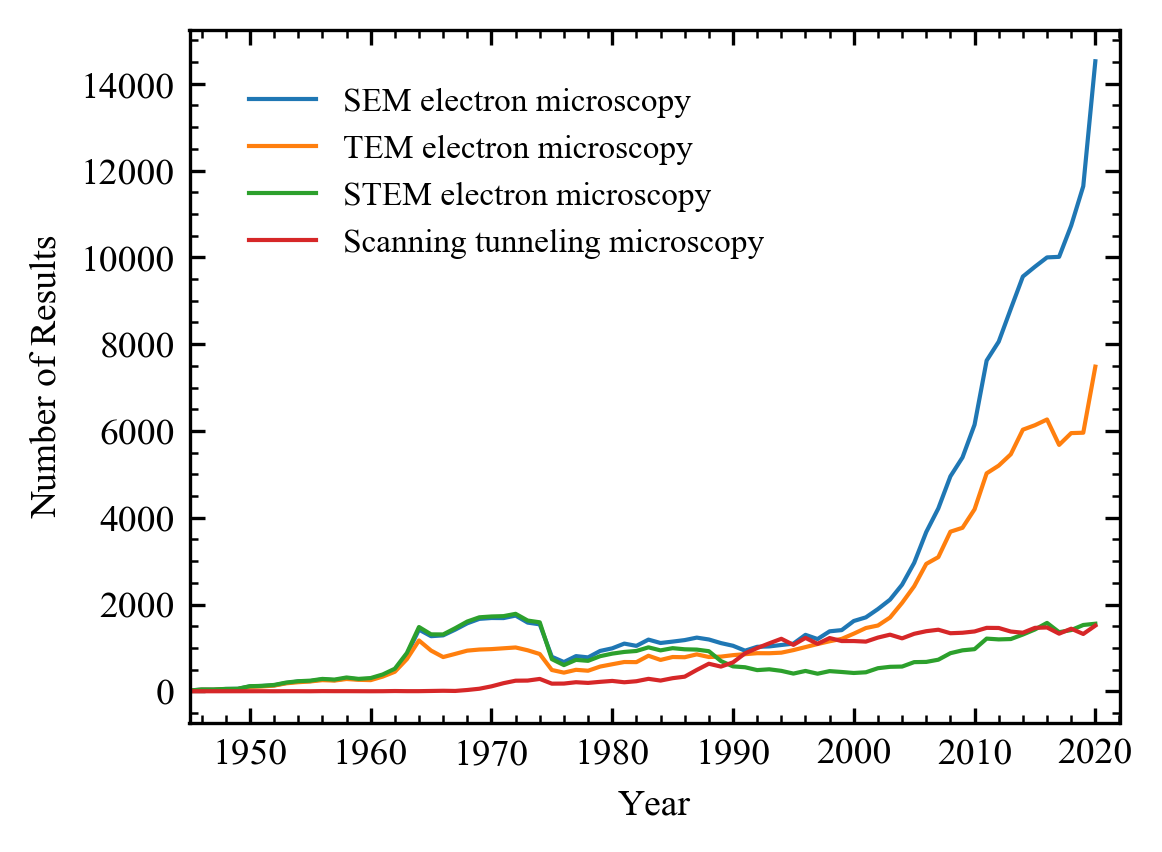}
\caption{ Numbers of results per year returned by Dimensions.ai abstract searches for SEM, TEM, STEM, STM and REM qualitate their popularities. The number of results for 2020 is extrapolated using the mean rate before 14th July 2020. }
\label{fig:em_popularity}
\end{figure*}

There are a variety of electron microscopes that use different illumination mechanisms. For example, reflection electron microscopy\cite{hsu1992technique, yagi1987reflection} (REM), scanning electron microscopy\cite{mohammed2018scanning, goldstein2017scanning} (SEM), scanning transmission electron microscopy\cite{keyse2018introduction, pennycook2011scanning} (STEM), scanning tunnelling microscopy\cite{sutter2019scanning, voigtlander2018invited} (STM), and transmission electron microscopy\cite{carter2016transmission, tang2017transmission, harris2015transmission} (TEM). To roughly gauge popularities of electron microscope varieties, we performed abstract searches with Dimenions.ai\cite{herzog2020dimensions, bode2018guide, hook2018dimensions, adams2018dimensions} for their abbreviations followed by \enquote{electron microscopy} e.g. \enquote{REM electron microscopy}. Numbers of results per year in figure~\ref{fig:em_popularity} qualitate that popularity increases in order REM, STM, STEM, TEM, then SEM. It may be tempting to attribute the popularity of SEM over TEM to the lower cost of SEM\cite{sem2020gleichmann}, which increases accessibility. However, a range of considerations influence the procurement of electron microscopes\cite{owen2018purchasing} and hourly pricing at universities\cite{openuni_emprices, newcastle_emprices, sahlgrenska_emprices, harvard_emprices, cambridge_emprices} is similar for SEM and TEM. 

In SEM, material surfaces are scanned by sequential probing with a beam of electrons, which are typically accelerated to 0.2-40 keV. The SEM detects quanta emitted from where the beam interacts with the sample. Most SEM imaging uses low-energy secondary electrons. However, reflection electron microscopy\cite{hsu1992technique, yagi1987reflection} (REM) uses elastically backscattered electrons and is often complimented by a combination of reflection high-energy electron diffraction\cite{ichimiya2004reflection, braun1999applied, xiang2016reflection} (RHEED), reflection high-energy electron loss spectroscopy\cite{mavsek2003reflection, atwater1991reflection} (RHEELS) and spin-polarized low-energy electron microscopy\cite{yu2020aberration, bauer2019leem, li2017study} (SPLEEM). Some SEMs also detect Auger electrons\cite{matsui2018auger, macdonald1971auger}. To enhance materials characterization, most SEMs also detect light. The most common light detectors are for cathodoluminescence and energy dispersive r-ray\cite{scimeca2018energy, chen2016quantitative} (EDX) spectroscopy. Nonetheless, some SEMs also detect Bremsstrahlung radiation\cite{eggert2018benefits}.

Alternatively, TEM and STEM detect electrons transmitted through specimens. In conventional TEM, a single region is exposed to a broad electron beam. In contrast, STEM uses a fine electron beam to probe a series of discrete probing locations. Typically, electrons are accelerated across a potential difference to kinetic energies, $E_k$, of 80-300 keV. Electrons also have rest energy $E_\text{e} = m_\text{e} c^2$, where $m_\text{e}$ is electron rest mass and $c$ is the speed of light. The total energy, $E_t = E_\text{e} + E_k$, of free electrons is related to their rest mass energy by a Lorentz factor, $\gamma$,
\begin{align}
    E_t &= \gamma m_\text{e} c^2 \,, \\
    \gamma &= (1 - v^2/c^2)^{1/2} \,,
\end{align}
where $v$ is the speed of electron propagation in the rest frame of an electron microscope. Electron kinetic energies in TEM and STEM are comparable to their rest energy, $E_\text{e} = 511$ keV\cite{mohr2016codata}, so relativistic phenomena\cite{romano2018introduction, french2017special} must be considered to accurately describe their dynamics. 

Electrons exhibit wave-particle duality\cite{frabboni2007young, matteucci1998experiment}. Thus, in an ideal electron microscope, the maximum possible detection angle, $\theta$, between two point sources separated by a distance, $d$, perpendicular to the electron propagation direction is diffraction-limited. The resolution limit for imaging can be quantified by Rayleigh's criterion\cite{rayleigh1879xxxi, ram2006beyond, hyperphysics2020rayleigh}
\begin{equation}
    \theta \simeq 1.22\frac{\lambda}{d},
\end{equation}
where resolution increases with decreasing wavelength, $\lambda$. Electron wavelength decreases with increasing accelerating voltage, as described by the relativistic de Broglie relation\cite{guemez2016principle, mackinnon1976broglie, hyperphysics2020debroglie},
\begin{equation}
    \lambda = hc \left( E_k^2 + 2 E_\text{e} E_k \right)^{-1/2} \,,
\end{equation}
where $h$ is Planck's constant\cite{mohr2016codata}. Electron wavelengths for typical acceleration voltages tabulated by JEOL are in picometres\cite{jeolwavelengths}. In comparison, Cu K-$\alpha$ x-rays, which are often used for XRD, have wavelengths near 0.15 nm\cite{mendenhall2017high}. In theory, electrons can therefore achieve over 100$\times$ higher resolution than x-rays. Electrons and x-rays are both ionizing; however, electrons often do less radiation damage to thin specimens than x-rays\cite{du2020relative}. Tangentially, TEM and STEM often achieve over 10 times higher resolution than SEM\cite{thermofisher2020transmission} as transmitted electrons in TEM and STEM are easier to resolve than electrons returned from material surfaces in SEM.

In practice, TEM and STEM are also limited by incoherence\cite{latychevskaia2017spatial, van2011persistent, krumeich2011properties} introduced by inelastic scattering, electron energy spread, and other mechanisms. TEM and STEM are related by an extension of Helmholtz reciprocity\cite{greffet1998field, clarke1985helmholtz} where the source plane in a TEM corresponds to the detector plane in a STEM\cite{rose2005reciprocity}, as shown in figure~\ref{fig:tem_vs_stem}. Consequently, TEM coherence is limited by electron optics between the specimen and image, whereas STEM coherence is limited by the illumination system. For conventional TEM and STEM imaging, electrons are normally incident on a specimen\cite{peters2017structure}. Advantages of STEM imaging can include higher contrast and resolution than TEM imaging, and lower radiation damage\cite{yakovlev2010advantages}. As a result, STEM is increasing being favoured over TEM for high-resolution studies. However, we caution that definitions of TEM and STEM resolution can be disparate\cite{voelkl2017stem}.

In addition to conventional imaging, TEM and STEM include a variety of operating modes for different applications. For example, TEM operating configurations include electron diffraction\cite{bendersky2001electron}; convergent beam electron diffraction\cite{hubert2019structure, beanland2013digital, tanaka1994convergent} (CBED); tomography\cite{hovden2020electron, koneti2019fast, song2019electron, chen2019complete, ercius2015electron, weyland2004electron, wang2020consensus, doerr2017cryo, oktem2015mathematics}; and bright field\cite{tichelaar2020tem, fujii2018toward, tang2017transmission, vander1998soot}, dark field\cite{tang2017transmission, vander1998soot} and annular dark field\cite{bals2004annular} imaging. Similarly, STEM operating configurations include differential phase contrast\cite{yucelen2018phase, krajnak2016pixelated, lazic2016phase, muller2019comparison}; tomography\cite{hovden2020electron, song2019electron, ercius2015electron, weyland2004electron}; and bright field\cite{zhou2016sample, okunishi2009visualization} or dark field\cite{van2016unscrambling} imaging. Further, electron cameras\cite{mcmullan2016direct, mcmullan2009detective} are often supplemented by secondary signal detectors. For example, elemental composition is often mapped by EDX spectroscopy, electron energy loss spectroscopy\cite{torruella2018clustering, pomarico2018ultrafast} (EELS) or wavelength dispersive spectroscopy\cite{koguchi2016analytical, tanaka2008x} (WDS). Similarly, electron backscatter diffraction\cite{schwartz2009electron, humphreys2001review, winkelmann2016physics} (EBSD) can detect strain\cite{wright2011review, wilkinson2009mapping, wilkinson2006high} and crystallization\cite{wisniewski20202, basu2016determination, zou2009dynamic}.

\subsection{Contrast Simulation}

The propagation of electron wavefunctions though electron microscopes can be described by wave optics\cite{rose2008optics}. Further, the most popular approach to modelling measurement contrast is multislice simulation\cite{kirkland2006image, kirkland2016computation}, where an electron wavefunction is iteratively perturbed as it travels through a model of a specimen. Multislice software for electron microscopy includes ACEM\cite{kirkland2016computation, kirkland2010advanced, computem}, clTEM\cite{dyson2014advances, clTEM_repo}, cudaEM\cite{cudaem}, Dr. Probe\cite{barthel2018dr, drprobe}, EMSoft\cite{singh2017emsoft, emsoft2020github}, JEMS\cite{jems2015stadelmann}, JMULTIS\cite{zuo2013electron}, MULTEM\cite{lobato2016accurate, lobato2016progress, lobato2015multem}, NCEMSS\cite{o1988advances, edm}, NUMIS\cite{numis}, Prismatic\cite{pryor2017streaming, ophus2017fast, prismatic}, QSTEM\cite{qstem}, SimulaTEM\cite{gomez2010simulatem}, STEM-CELL\cite{stem-cell}, Tempas\cite{tempas}, and xHREM\cite{ishizuka2002practical, ishizuka2001prospects, ishizuka1982multislice, ishizuka1980contrast, ishizuka1977new, hrem2020simulation}. We find that most multislice software is a recreation and slight modification of common functionality, possibly due to a publish-or-perish culture in academia\cite{gianola2020publish, nielsen2020predatory, genova2019problem}. Bloch-wave simulation\cite{zuo1995beam, yang2017quantitative, peng2004haadf, cheng2018bohmian, kirkland2016computation, beanland2020felix} is an alternative to multislice simulation that can reduce computation time and memory requirements for crystalline materials\cite{morimura2009bloch}.

\subsection{Automation}

Most modern electron microscopes support Gatan Microscopy Suite (GMS) Software\cite{gms_webpage}. GMS enables electron microscopes to be programmed by DigitalMicrograph Scripting, a propriety Gatan programming language akin to a simplified version of C++. A variety of DigitalMicrograph scripts, tutorials and related resources are available from Dave Mitchell's DigitalMicrograph Scripting Website\cite{mitchell2005scripting, dmscripts2020list}, FELMI/ZFE's Script Database\cite{felmi_webpage} and Gatan's Script library\cite{gatanscriptslibrary}. Some electron microscopists also provide DigitalMicrograph scripting resources on their webpages\cite{potapov2020temdm, koch2016electron, schaffer2015how}. However, DigitalMicrograph scripts are slow insofar that they are interpreted at runtime, and there is limited native functionality for parallel and distributed computing. As a result, extensions to DigitalMicrograph scripting are often developed in other programming languages that offer more functionality.

Historically, most extensions were developed in C++\cite{mitchell2014guide}. This was problematic as there is limited documentation, the standard approach used outdated C++ software development kits such as Visual Studio 2008, and programming expertise required to create functions that interface with DigitalMicrograph scripts limited accessibility. To increase accessibility, recent versions of GMS now support python\cite{miller2019real}. This is convenient as it enables ANNs developed with python to readily interface with electron microscopes. For ANNs developed with C++, users have the option to either create C++ bindings for DigitalMicrograph script or for python. Integrating ANNs developed in other programming languages is more complicated as DigitalMicrograph provides almost no support. However, that complexity can be avoided by exchanging files from DigitalMicrograph script to external libraries via a random access memory (RAM) disk\cite{hoffman2019ram} or secondary storage\cite{coughlin2020digital}. 

Increasing accessibility, there are collections of GMS plugins with GUIs for automation and analysis\cite{hrem2020a, potapov2020temdm, koch2016electron, schaffer2015how}. In addition, various individual plugins are available\cite{rene2019tcp, schorb2019software, peters2018dm, wolf2014weighted, wolf2013tomography}. Some plugins are open source, so they can be adapted to interface with ANNs. However, many high-quality plugins are proprietary and closed source, limiting their use to automation of data collection and processing. Plugins can also be supplemented by a variety of libraries and interfaces for electron microscopy signal processing. For example, popular general-purpose software includes ImageJ\cite{schindelin2015imagej}, pycroscopy\cite{somnath2019usid, pycroscopy2020} and HyperSpy\cite{hyperspy2020, de2017electron}. In addition, there are directories for tens of general-purpose and specific electron microscopy programs\cite{em2020software, software2020tools, queensland2020centre}.

\section{Components}

Most modern ANNs are configured from a variety of DLF components. To take advantage of hardware accelerators\cite{chen2020survey}, most ANNs are implemented as sequences of parallelizable layers of tensor operations\cite{ben2019demystifying}. Layers are often parallelized across data and may be parallelized across other dimensions\cite{dryden2019channel}. This section introduces popular nonlinear activation functions, normalization layers, convolutional layers, and skip connections. To add insight, we provide comparative discussion and address some common causes of confusion. 

\subsection{Nonlinear Activation}

In general, DNNs need multiple layers to be universal approximators\cite{kidger2019universal, lin2018resnet, hanin2017approximating, lu2017expressive, pinkus1999approximation, leshno1993multilayer, hornik1991approximation, hornik1989multilayer, cybenko1989approximation}. Nonlinear activation functions\cite{nwankpa2018activation, hayou2019impact} are therefore essential to DNNs as successive linear layers can be contracted to a single layer. Activation functions separate artificial neurons, similar to biological neurons\cite{roos2019deep}. To learn efficiently, most DNNs are tens or hundreds of layers deep\cite{lin2017does, eldan2016power, telgarsky2016benefits, ba2014deep}. High depth increases representational capacity\cite{lin2017does}, which can help training by gradient descent as DNNs evolve as linear models\cite{lee2019wide} and nonlinearities can create suboptimal local minima where data cannot be fit by linear models\cite{yun2018small}. There are infinitely many possible activation functions. However, most activation functions have low polynomial order, similar to physical Hamiltonians\cite{lin2017does}.

Most ANNs developed for electron microscopy are for image processing, where the most popular nonlinearities are rectifier linear units\cite{nair2010rectified, glorot2011deep} (ReLUs). The ReLU activation, $f(x)$, of an input, $x$, and its gradient, $\partial_x f(x)$, are\\
\noindent\begin{subequations}
  \begin{tabularx}{\textwidth}{Xp{2cm}X}
  \begin{equation}
      f(x) = \max(0, x)
  \end{equation}
  & &
  \begin{equation}
    \frac{\partial f(x)}{\partial x} = \begin{cases}
    0, & \text{if } x \le 0 \\
    1, & \text{if } x > 0
\end{cases}
  \end{equation}
  \end{tabularx}
\end{subequations}\\
Popular variants of ReLUs include Leaky ReLU\cite{maas2013rectifier},\\
\noindent\begin{subequations}
  \begin{tabularx}{\textwidth}{Xp{2cm}X}
  \begin{equation}
      f(x) = \max(\alpha x, x)
  \end{equation}
  & &
  \begin{equation}
    \frac{\partial f(x)}{\partial x} = \begin{cases}
    \alpha, & \text{if } x \le 0 \\
    1, & \text{if } x > 0
\end{cases}
  \end{equation}
  \end{tabularx}
\end{subequations}\\
where $\alpha$ is a hyperparameter, parametric ReLU\cite{he2015delving} (PreLU) where $\alpha$ is a learned parameter, dynamic ReLU where $\alpha$ is a learned function of inputs\cite{chen2020dynamic}, and randomized leaky ReLU\cite{xu2015empirical} (RReLU) where $\alpha$ is chosen randomly. Typically, learned PreLU $\alpha$ are higher the nearer a layer is to ANN inputs\cite{he2015delving}. Motivated by limited comparisons that do not show a clear performance difference between ReLU and leaky ReLU\cite{pedamonti2018comparison}, some blogs\cite{chris2019leaky} argue against using leaky ReLU due to its higher computational requirements and complexity. However, an in-depth comparison found that leaky ReLU variants consistently slightly outperform ReLU\cite{xu2015empirical}. In addition, the non-zero gradient of leaky ReLU for $x \le 0$ prevents saturating, or \enquote{dying}, ReLU\cite{arnekvist2020effect, lu2019dying, douglas2018relu}, where the zero gradient of ReLUs stops learning. 

There are a variety of other piecewise linear ReLU variants that can improve performance. For example, ReLU$h$ activations are limited to a threshold\cite{krizhevsky2010convolutional}, $h$, so that\\
\noindent\begin{subequations}
  \begin{tabularx}{\textwidth}{Xp{2cm}X}
  \begin{equation}
      f(x) = \min(\max(0, x), h)
  \end{equation}
  & &
  \begin{equation}
    \frac{\partial f(x)}{\partial x} = \begin{cases}
    0, & \text{if } x \le 0 \\
    1, & \text{if } 0 < x \le h \\
    0, & \text{if } x > h
\end{cases}
  \end{equation}
  \end{tabularx}
\end{subequations}\\
Thresholds near $h=6$ are often effective, so popular choice is ReLU6. Another popular activation is concatenated ReLU\cite{shang2016understanding} (CReLU), which is the concatenation of $\text{ReLU}(x)$ and $\text{ReLU}(-x)$. Other ReLU variants include adaptive convolutional\cite{gao2020adaptive}, bipolar\cite{eidnes2017shifting}, elastic\cite{jiang2018deep}, and Lipschitz\cite{basirat2020relu} ReLUs. However, most ReLU variants are uncommon as they are more complicated than ReLU and offer small, inconsistent, or unclear performance gains. Moreover, it follows from the universal approximator theorems\cite{kidger2019universal, lin2018resnet, hanin2017approximating, lu2017expressive, pinkus1999approximation, leshno1993multilayer, hornik1991approximation, hornik1989multilayer, cybenko1989approximation} that disparity between ReLU and its variants approaches zero as network depth increases.

In shallow networks, curved activation functions with non-zero Hessians often accelerate convergence and improve performance. A popular activation is the exponential linear unit\cite{clevert2015fast} (ELU),\\
\noindent\begin{subequations}
  \begin{tabularx}{\textwidth}{Xp{2cm}X}
  \begin{equation}
      f(x) = \begin{cases}
    \alpha(\exp(x)-1), & \text{if } x \le 0 \\
    x, & \text{if } x \ge 0
\end{cases}
  \end{equation}
  & &
  \begin{equation}
    \frac{\partial f(x)}{\partial x} = \begin{cases}
    \alpha\exp(x), & \text{if } x \le 0 \\
    1, & \text{if } x \ge 0
\end{cases}
  \end{equation}
  \end{tabularx}
\end{subequations}\\
where $\alpha$ is a learned parameter. Further, a scaled ELU\cite{klambauer2017self} (SELU),\\
\noindent\begin{subequations}
  \begin{tabularx}{\textwidth}{Xp{2cm}X}
  \begin{equation}
      f(x) = \begin{cases}
    \lambda \alpha(\exp(x)-1), & \text{if } x \le 0 \\
    \lambda x, & \text{if } x \ge 0
\end{cases}
  \end{equation}
  & &
  \begin{equation}
    \frac{\partial f(x)}{\partial x} = \begin{cases}
    \lambda \alpha\exp(x), & \text{if } x \le 0 \\
    \lambda, & \text{if } x \ge 0
\end{cases}
  \end{equation}
  \end{tabularx}
\end{subequations}\\
with fixed $\alpha=1.67326$ and scale factor $\lambda=1.0507$ can be used to create self-normalizing neural networks (SNNs). A SNN cannot be derived from ReLUs or most other activation functions. Activation functions with curvature are especially common in ANNs with only a couple of layers. For example, activation functions in radial basis function (RBF) networks\cite{hryniowski2019deeplabnet, dash2016radial, orr1996introduction, jang1993functional}, which are efficient universal approximators, are often Gaussians, multiquadratics, inverse multiquadratics, or square-based RBFs\cite{wuraola2018computationally}. Similarly, support vector machines\cite{cervantes2020comprehensive, scholkopf2018learning, tavara2019parallel} (SVMs) often use RBFs, or sigmoids,\\
\noindent\begin{subequations}
  \begin{tabularx}{\textwidth}{Xp{2cm}X}
  \begin{equation}
      f(x) = \frac{1}{1+\exp(-x)} \label{eqn:logistic_sigmoid}
  \end{equation}
  & &
  \begin{equation}
    \frac{\partial f(x)}{\partial x} = f(x)\left(1 - f(x)\right) \label{eqn:logistic_sigmoid_grad}
  \end{equation}
  \end{tabularx}
\end{subequations}\\
Sigmoids can also be applied to limit the support of outputs. Unscaled, or \enquote{logistic}, sigmoids are often denoted $\sigma(x)$ and are related to $\tanh$ by $\tanh(x) = 2\sigma(2x) - 1$. To avoid expensive $\exp(-x)$ in the computation of tanh, we recommend K-tanH\cite{kundu2019k}, LeCun tanh\cite{lecun2012efficient}, or piecewise linear approximation\cite{abdelouahab2017tanh, gulcehre2016noisy}. 

The activation functions introduced so far are scalar functions than can be efficiently computed in parallel for each input element. However, functions of vectors, $\textbf{x} = \{ x_1, x_2, ...\}$, are also popular. For example, softmax activation\cite{dunne1997pairing},\\
\noindent\begin{subequations}
  \begin{tabularx}{\textwidth}{Xp{2cm}X}
  \begin{equation}
      f(\textbf{x}) = \frac{\exp(\textbf{x})}{\text{sum}(\exp(\textbf{x}))} 
  \end{equation}
  & &
  \begin{equation}
    \frac{f(\textbf{x})}{\partial x_j} = \sum_i f(\textbf{x})_i(\delta_{ij} - f(\textbf{x})_j)
  \end{equation}
  \end{tabularx}
\end{subequations}\\
is often applied before computing cross-entropy losses for classification networks. Similarly, L$n$ vector normalization,\\
\noindent\begin{subequations}
  \begin{tabularx}{\textwidth}{Xp{2cm}X}
  \begin{equation}
      f(\textbf{x}) = \frac{\textbf{x}}{||\textbf{x}||_n}
  \end{equation}
  & &
  \begin{equation}
    \frac{f(\textbf{x})}{\partial x_j} = \frac{1}{||\textbf{x}||_n} \left( 1 - \frac{x_j^n}{||\textbf{x}||_n^n} \right)
  \end{equation}
  \end{tabularx}
\end{subequations}\\
with $n = 2$ is often applied to vectors to ensure that they lie on a unit sphere\cite{ede2020exit}. Finally, max pooling\cite{dumoulin2016guide, graham2014fractional},\\
\noindent\begin{subequations}
  \begin{tabularx}{\textwidth}{Xp{2cm}X}
  \begin{equation}
      f(\textbf{x}) = \max( \textbf{x} )
  \end{equation}
  & &
  \begin{equation}
    \frac{f(\textbf{x})}{\partial x_j} = \begin{cases}
    1, & \text{if } j = \text{argmax}(\textbf{x}) \\
    0, & \text{if } j \neq \text{argmax}(\textbf{x})
\end{cases}
  \end{equation}
  \end{tabularx}
\end{subequations}\\
is another popular multivariate activation function that is often used for downsampling. However, max pooling has fallen out of favour as it is often outperformed by strided convolutional layers\cite{springenberg2014striving}. Other vector activation functions include squashing nonlinearities for dynamic routing by agreement in capsule networks\cite{sabour2017dynamic} and cosine similarity\cite{luo2018cosine}. 

There are many other activation functions that are not detailed here for brevity. Further, finding new activation functions is an active area of research\cite{nader2020searching, ramachandran2018searching}. Notable variants include choosing activation functions from a set before training\cite{bingham2020discovering, ertuugrul2018novel} and learning activation functions\cite{bingham2020discovering, lau2018review, chung2016deep, agostinelli2014learning, wu1997beyond}. Activation functions can also encode probability distributions\cite{lee2019probact, kingma2014auto, springenberg2013improving} or include noise\cite{gulcehre2016noisy}. Finally, there are a variety of other deterministic activation functions\cite{bawa2019linearized, ramachandran2018searching}. In electron microscopy, most ANNs enable new or enhance existing applications. Subsequently, we recommend using computationally efficient and established activation functions unless there is a compelling reason to use a specialized activation function.

\subsection{Normalization}\label{sec:normalization}

Normalization\cite{kurita2018overview, ren2016normalizing, liao2016streaming} standardizes signals, which can accelerate convergence by gradient descent and improve performance. Batch normalization\cite{santurkar2018does, ioffe2015batch, bjorck2018understanding, yang2019mean, ioffe2019batch, lian2019revisit} is the most popular normalization layer in image processing DNNs trained with minibatches of $N$ examples. Technically, a \enquote{batch} is an entire training dataset and a \enquote{minibatch} is a subset; however, the \enquote{mini} is often omitted where meaning is clear from context. During training, batch normalization applies a transform,
\begin{align}
    \mu_B &= \frac{1}{N} \sum\limits_{i=1}^N x_i\,, \label{eqn:bn_mean} \\
    \sigma_B^2 &= \frac{1}{N} \sum\limits_{i=1}^N (x_i - \mu_B)^2\,, \\
    \hat{\textbf{x}} &= \frac{\textbf{x} - \mu_B}{(\sigma_B^2 + \epsilon)^{1/2}}\,, \label{eqn:bn_apply_moments}\\
    \text{BatchNorm}(\textbf{x}) &= \gamma \hat{\textbf{x}} + \beta\,,
\end{align}
where $\textbf{x} = \{x_1, ..., x_N\}$ is a batch of layer inputs, $\gamma$ and $\beta$ are a learnable scale and shift, and $\epsilon$ is a small constant added for numerical stability. During inference, batch normalization applies a transform,
\begin{align}
    \text{BatchNorm}(\textbf{x}) = \frac{\gamma}{(\text{Var}[x] + \epsilon)^{1/2}} \textbf{x} + \left( \beta - \frac{\gamma \text{E}[x]}{(\text{Var}[x] + \epsilon)^{1/2}} \right)\,, \label{eqn:bn_inference}
\end{align}
where \text{E}[x] and \text{Var}[x] are expected batch means and variances. For convenience, \text{E}[x] and \text{Var}[x] are often estimated with exponential moving averages that are tracked during training. However, \text{E}[x] and \text{Var}[x] can also be estimated by propagating examples through an ANN after training.

Increasing batch size stabilizes learning by averaging destabilizing loss spikes over batches\cite{ede2020adaptive}. Batched learning also enables more efficient utilization of modern hardware accelerators. For example, larger batch sizes improve utilization of GPU memory bandwidth and throughput\cite{gao2018low, jouppi2017datacenter, fang2019serving}. Using large batches can also be more efficient than many small batches when distributing training across multiple CPU clusters or GPUs due to communication overheads. However, the performance benefits of large batch sizes can come at the cost of lower test accuracy as training with large batches tends to converge to sharper minima\cite{das2016distributed, keskar2016large}. As a result, it is often best not to use batch sizes higher than $N \approx 32$ for image classification\cite{masters2018revisiting}. However, learning rate scaling\cite{goyal2017accurate} and layer-wise adaptive learning rates\cite{you2017scaling} can increase accuracy of training with fixed larger batch sizes. Batch size can also be increased throughout training without compromising accuracy\cite{devarakonda2017adabatch} to exploit effective learning rates being inversely proportional to batch size\cite{devarakonda2017adabatch, goyal2017accurate}. Alternatively, accuracy can be improved by creating larger batches from replicated instances of training inputs with different data augmentations\cite{hoffer2019augment}.

There are a few caveats to batch normalization. Originally, batch normalization was applied before activation\cite{ioffe2015batch}. However, applying batch normalization after activation often slightly improves performance\cite{hasani2019empirical, dmytro2016batch}. In addition, training can be sensitive to the often-forgotten $\epsilon$ hyperparameter\cite{nado2020evaluating} in equation~\ref{eqn:bn_apply_moments}. Typically, performance decreases as $\epsilon$ is increased above $\epsilon \approx 0.001$; however, there is a sharp increase in performance around $\epsilon=0.01$ on ImageNet. Finally, it is often assumed that batches are representative of the training dataset. This is often approximated by shuffling training data to sample independent and identically distributed (i.i.d.) samples. However, performance can often be improved by prioritizing sampling\cite{zha2019experience, schaul2015prioritized}. We observe that batch normalization is usually effective if batch moments, $\mu_B$ and $\sigma_B$, have similar values for every batch.

Batch normalization is less effective when training batch sizes are small, or do not consist of independent samples. To improve performance, standard moments in equation~\ref{eqn:bn_apply_moments} can be renormalized\cite{ioffe2017batch} to expected means, $\mu$, and standard deviations, $\sigma$,
\begin{align}
    \hat{\textbf{x}} &\leftarrow r \hat{\textbf{x}} + d \,, \\
    r &= \text{clip}_{[1/r_\text{max}, r_\text{max}]} \left( \frac{\sigma_B}{\sigma} \right) \,,\\
    d &= \text{clip}_{[-d_\text{max}, d_\text{max}]} \left( \frac{\mu_B - \mu}{\sigma} \right) \,,
\end{align}
where gradients are not backpropagated with respect to (w.r.t.) the renormalization parameters, $r$ and $d$. Moments, $\mu$ and $\sigma$ are tracked by exponential moving averages and clipping to $r_\text{max}$ and $d_\text{max}$ improves learning stability. Usually, clipping values are increased from starting values of $r_\text{max}=1$ and $d_\text{max}=0$, which correspond to batch normalization, as training progresses. Another approach is virtual batch normalization\cite{salimans2016improved} (VBN), which estimates $\mu$ and $\sigma$ from a reference batch of samples and does not require clipping. However, VBN is computationally expensive as it requires computing a second batch of statistics at every training iteration. Finally, online\cite{chiley2019online} and streaming\cite{liao2016streaming} normalization enable training with small batch sizes by replace $\mu_B$ and $\sigma_B$ in equation~\ref{eqn:bn_apply_moments} with their exponential moving averages. 

There are alternatives to the $L_2$ batch normalization of equations~\ref{eqn:bn_mean}-\ref{eqn:bn_inference} that standardize to different Euclidean norms. For example, $L_1$ batch normalization\cite{hoffer2018norm} computes 
\begin{align}
    s_1 &= \frac{1}{N} \sum\limits_{i=1}^N |x_i - \mu_B|\,, \\
    \hat{\textbf{x}} &= \frac{\textbf{x} - \mu_B}{ C_{L_1} s_1}\,,
\end{align}
where $C_{L_1} = (\pi / 2)^{1/2}$. Although the $C_{L_1}$ factor could be learned by ANN parameters, its inclusion accelerates convergence of the original implementation of $L_1$ batch normalization\cite{hoffer2018norm}. Another alternative is $L_\infty$ batch normalization\cite{hoffer2018norm}, which computes
\begin{align}
    s_\infty &= \text{mean}( \text{top}_k( |\textbf{x} - \mu_B| ) )\,, \\
    \hat{\textbf{x}} &= \frac{\textbf{x} - \mu_B}{ C_{L_\infty} s_\infty }\,,
\end{align}
where $C_{L_\infty}$ is a scale factor, and $\text{top}_k (\textbf{x})$ returns the $k$ highest elements of $\textbf{x}$. Hoffer \textit{et al} suggest $k=10$\cite{hoffer2018norm}. Some $L_1$ batch normalization proponents claim that $L_1$ batch normalization outperforms\cite{santurkar2018does} or achieves similar performance\cite{hoffer2018norm} to $L_2$ batch normalization. However, we found that $L_1$ batch normalization often lowers performance in our experiments. Similarly, $L_\infty$ batch normalization often lowers performance\cite{hoffer2018norm}. Overall, $L_1$ and $L_\infty$ batch normalization do not appear to offer a substantial advantage over $L_2$ batch normalization.

\begin{figure*}[tbh!]
\centering
\includegraphics[width=0.95\textwidth]{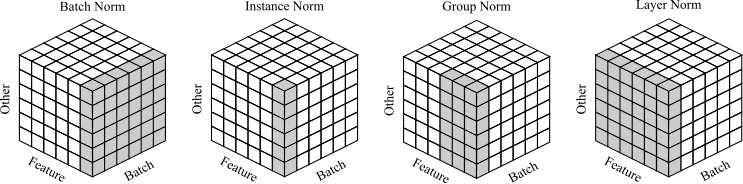}
\caption{ Visual comparison of various normalization methods highlighting regions that they normalize. Regions can be normalized across batch, feature and other dimensions, such as height and width. }
\label{fig:norm_comparison}
\end{figure*}

A variety of layers normalize samples independently, including layer, instance, and group normalization. They are compared with batch normalization in figure~\ref{fig:norm_comparison}. Layer normalization\cite{ba2016layer, xu2019understanding} is a transposition of batch normalization that is computed across feature channels for each training example, instead of across batches. Batch normalization is ineffective in RNNs; however, layer normalization of input activations often improves accuracy\cite{ba2016layer}. Instance normalization\cite{ulyanov2016instance} is an extreme version of layer normalization that standardizes each feature channel for each training example. Instance normalization was developed for style transfer\cite{jing2019neural, gatys2016image, gatys2015neural, zhu2017unpaired, li2017demystifying} and makes ANNs insensitive to input image contrast. Group normalization\cite{wu2018group} is intermediate to instance and layer normalization insofar that it standardizes groups of channels for each training example.

The advantages of a set of multiple different normalization layers, $\Omega$, can be combined by switchable normalization\cite{luo2018normalization, luo2018differentiable}, which standardizes to
\begin{align}
    \hat{\textbf{x}} &= \frac{\textbf{x} - \sum\limits_{z \in \Omega} \lambda_z^\mu \mu_z}{ \sum\limits_{z \in \Omega} \lambda_z^\sigma \sigma_z }\,,
\end{align}
where $\mu_z$ and $\sigma_z$ are means and standard deviations computed by normalization layer $z$, and their respective importance ratios, $\lambda_z^\mu$ and $\lambda_z^\sigma$, are trainable parameters that are softmax activated to sum to unity. Combining batch and instance normalization statistics outperforms batch normalization for a range of computer vision tasks\cite{nam2018batch}. However, most layers strongly weighted either batch or instance normalization, with most preferring batch normalization. Interestingly, combining batch, instance and layer normalization statistics\cite{luo2018normalization, luo2018differentiable} results in instance normalization being preferred in earlier layers, whereas layer normalization was preferred in the later layers, and batch normalization was preferred in the middle layers. Smaller batch sizes lead to a preference towards layer normalization and instance normalization. Limitingly, using multiple normalization layers increases computation. To limit expense, we therefore recommend either defaulting to batch normalization, or progressively using single instance, batch, or layer normalization layers. 

A significant limitation of batch normalization is that it is not effective in RNNs. This is a limited issue as most electron microscopists are developing CNNs for image processing. However, we anticipate that RNNs may become more popular in electron microscopy following the increasing popularity of reinforcement learning\cite{hao2019we}. In addition to general-purpose alternatives to batch normalization that are effective in RNNs, such as layer normalization, there are a variety of dedicated normalization schemes. For example, recurrent batch normalization\cite{cooijmans2016recurrent, liao2016bridging} uses distinct normalization layers for each time step. Alternatively, batch normalized RNNs\cite{laurent2016batch} only have normalization layers between their input and hidden states. Finally, online\cite{chiley2019online} and streaming\cite{liao2016streaming} normalization are general-purpose solutions that improve the performance of batch normalization in RNNs by applying batch normalization based on a stream of past batch statistics.

Normalization can also standardize trainable weights, \textbf{w}. For example, weight normalization\cite{salimans2016weight},
\begin{equation}
    \text{WeightNorm}(\textbf{w}) = \frac{g}{||\textbf{w}||_2} \textbf{w} \,,
\end{equation}
decouples the L2 norm, $g$, of a variable from its direction. Similarly, weight standardization\cite{qiao2019weight} subtracts means from variables and divides them by their standard deviations,
\begin{equation}
    \text{WeightStd}(\textbf{w}) =  \frac{\textbf{w} - \text{mean}( \textbf{w})}{\text{std}(\textbf{w})} \,,
\end{equation}
similar to batch normalization. Weight normalization often outperforms batch normalization at small batch sizes. However, batch normalization consistently outperforms weight normalization at larger batch sizes used in practice\cite{gitman2017comparison}. Combining weight normalization with running mean-only batch normalization can accelerate convergence\cite{salimans2016weight}. However, similar final accuracy can be achieved without mean-only batch normalization at the cost of slower convergence, or with the use of zero-mean preserving activation functions\cite{eidnes2017shifting, hoffer2018norm}. To achieve similar performance to batch normalization, norm-bounded weight normalization\cite{hoffer2018norm} can be applied to DNNs with scale-invariant activation functions, such as ReLU. Norm-bounded weight normalization fixes $g$ at initialization to avoid learning instability\cite{gitman2017comparison, hoffer2018norm}, and scales outputs with the final DNN layer.

Limitedly, weight normalization encourages the use of a small number of features to inform activations\cite{miyato2018spectral}. To encourage higher feature utilization, spectral normalization\cite{miyato2018spectral},
\begin{equation}
    \text{SpectralNorm}(\textbf{w}) = \frac{\textbf{w}}{\sigma(\textbf{w})}\,,
\end{equation} 
divides tensors by their spectral norms, $\sigma(\textbf{w})$. Further, spectral normalization limits Lipschitz constants\cite{wood1996estimation}, which often improves generative adversarial network\cite{gui2020review, saxena2020generative, pan2019recent, wang2019generative} (GAN) training by bounding backpropagated discriminator gradients\cite{miyato2018spectral}. The spectral norm of $\textbf{v}$ is the maximum value of a diagonal matrix, $\boldsymbol\Sigma$, in the singular value decomposition\cite{afham2019singular, afham2020singular, wall2003singular, klema1980singular} (SVG),
\begin{equation}
    \textbf{v} = \textbf{U} \boldsymbol\Sigma \textbf{V}^*\,,
\end{equation}
where $\textbf{U}$ and $\textbf{V}$ are orthogonal matrices of orthonormal eigenvectors for $\textbf{v}\textbf{v}^T$ and $\textbf{v}^T\textbf{v}$, respectively. To minimize computation, $\boldsymbol\sigma (\textbf{w})$ is often approximated by the power iteration method\cite{yoshida2017spectral, golub2000eigenvalue},
\begin{align}
    \hat{\textbf{v}} &\leftarrow \frac{\textbf{w}^\text{T} \hat{\textbf{u}}}{||\textbf{w}^\text{T}\hat{\textbf{u}}||_2}\,, \label{eqn:power_iter_v}\\
    \hat{\textbf{u}} &\leftarrow \frac{\textbf{w}\hat{\textbf{v}}}{||\textbf{w}\hat{\textbf{v}}||_2}\,, \label{eqn:power_iter_u}\\
    \sigma(\textbf{w}) &\simeq \hat{\textbf{u}}^T \textbf{w} \hat{\textbf{v}}\,,
\end{align}
where one iteration of equations~\ref{eqn:power_iter_v}-\ref{eqn:power_iter_u} per training iteration is usually sufficient. 

Parameter normalization can complement or be combined with signal normalization. For example, scale normalization\cite{nguyen2019transformers},
\begin{equation}
    \text{ScaleNorm}(\textbf{x}) = \frac{g}{||\textbf{x}||_2} \textbf{x} \,,
\end{equation}
learns scales, $g$, for activations, and is often combined with weight normalization\cite{nguyen2017improving, salimans2016weight} in transformer networks. Similarly, cosine normalization\cite{luo2018cosine},
\begin{equation}
    \text{CosineNorm}(\textbf{x}) = \frac{\textbf{w}}{||\textbf{w}||_2} \cdot \frac{\textbf{x}}{||\textbf{x}||_2} \,,
\end{equation}
computes products of L2 normalized parameters and signals. Both scale and cosine normalization can outperform batch normalization.

\begin{figure*}[tbh!]
\centering
\includegraphics[width=\textwidth]{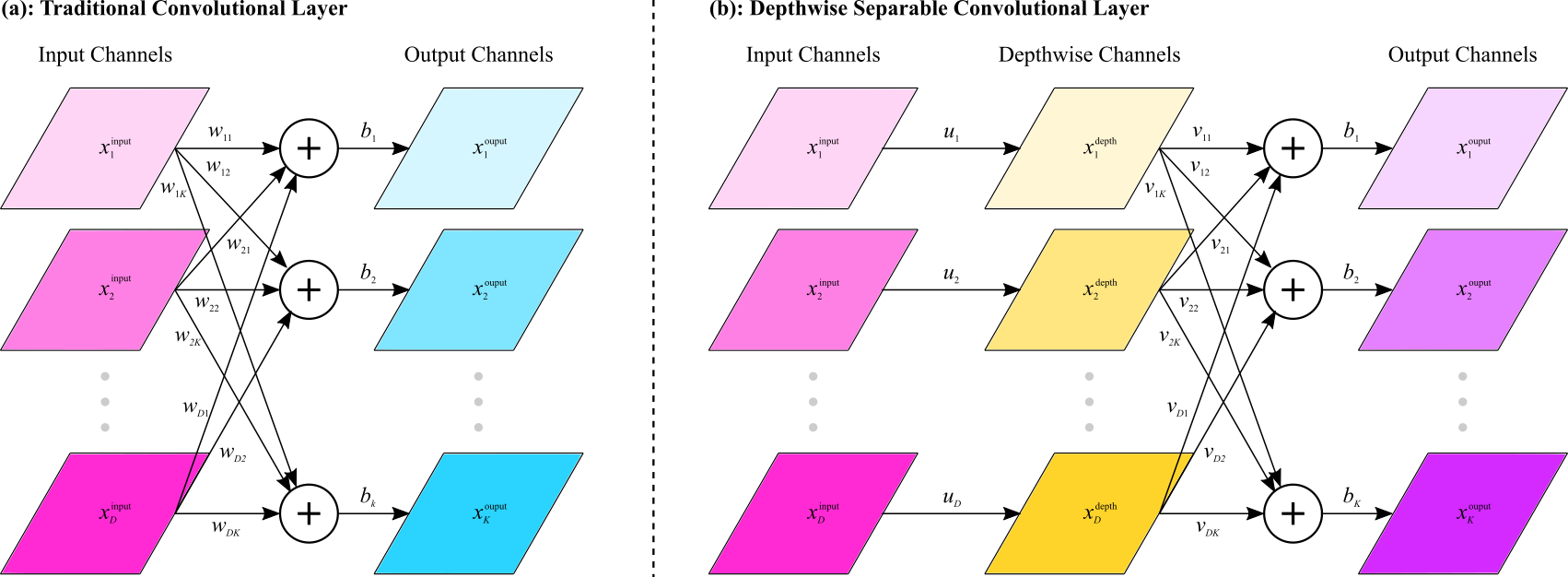}
\caption{ Visualization of convolutional layers. a) Traditional convolutional layer where output channels are sums of biases and convolutions of weights with input channels. b) Depthwise separable convolutional layer where depthwise convolutions compute one convolution with weights for each input channel. Output channels are sums of biases and pointwise convolutions weights with depthwise channels. }
\label{fig:convolutions}
\end{figure*}

\subsection{Convolutional Layers}

A convolutional neural network\cite{Stewart2019simple, wu2017introduction, mccann2017convolutional, o2015introduction} (CNN) is trained to weight convolutional kernels to exploit local correlations, such as spatial correlations in electron micrographs\cite{ede2020warwick}. Historically, the development of CNNs was inspired by primate visual cortices\cite{hubel1968receptive}, where partially overlapping neurons are only stimulated by visual stimuli within their receptive fields. Based on this idea, Fukushima published his Neocognitron\cite{fukushima1980self, fukushima1982neocognitron, fukushima1988neocognitron, fukushima2003neocognitron} in 1980. Convolutional formulations were then published by Atlas \textit{et al} in 1988 for a single-layer CNN\cite{atlas1988artificial}, and LeCun \textit{et al} in 1998 for a multi-layer CNN\cite{lecun1998gradient, lecun1999object}. Subsequently, GPUs were applied to accelerate convolutions in 2010\cite{cirecsan2010deep}, leading to a breakthrough in classification performance on ImageNet with AlexNet in 2012\cite{krizhevsky2012imagenet}. Indeed, the deep learning era is often partitioned into before and after AlexNet\cite{alom2018history}. Deep CNNs are now ubiquitous. For example, there are review papers on applications of CNNs to action recognition in videos\cite{yao2019review}, cytometry\cite{gupta2019deep}, image and video compression\cite{ma2019image, liu2020deep}, image background subtraction\cite{bouwmans2019deep}, image classification\cite{rawat2017deep}, image style transfer\cite{jing2019neural}, medical image analysis\cite{anwar2018medical, soffer2019convolutional, yamashita2018convolutional, bernal2019deep, du2020medical, fu2020deep, badar2020application, taghanaki2020deep, tajbakhsh2020embracing, litjens2017survey, liu2018applications}, object detection\cite{zhao2019object, wang2019salient}, semantic image segmentation\cite{guo2018review, du2020medical, tajbakhsh2020embracing, taghanaki2020deep}, and text classification\cite{minaee2020deep}.

In general, the convolution of two functions, $f$ and $g$, is
\begin{equation}
    (f*g)(x) \coloneqq \int\limits_{s \in \Omega} f(s)g(x-s) \diff s \,, \label{eqn:general_conv}
\end{equation}
and their cross-correlation is
\begin{equation}
    (f \circ g)(x) \coloneqq \int\limits_{s \in \Omega} f(s)g(x+s) \diff s \,, \label{eqn:general_cross}
\end{equation}
where integrals have unlimited support, $\Omega$. In a CNN, convolutional layers sum convolutions of feature channels with trainable kernels, as shown in figure~\ref{fig:convolutions}. Thus, $f$ and $g$ are discrete functions and the integrals in equations~\ref{eqn:general_conv}-\ref{eqn:general_cross} can be replaced with limited summations. Since cross-correlation is equivalent to convolution if the kernel is flipped in every dimension, and CNN kernels are usually trainable, convolution and cross-correlation is often interchangeable in deep learning. For example, a TensorFlow function named \enquote{tf.nn.convolution} computes cross-correlations\cite{tensorflow2020convolutional}. Nevertheless, the difference between convolution and cross-correlation can be source of subtle errors if convolutional layers from a DLF are used in an image processing pipeline with static asymmetric kernels.

\begin{figure}[tbh!]
\centering
\footnotesize
\includegraphics[width=\textwidth]{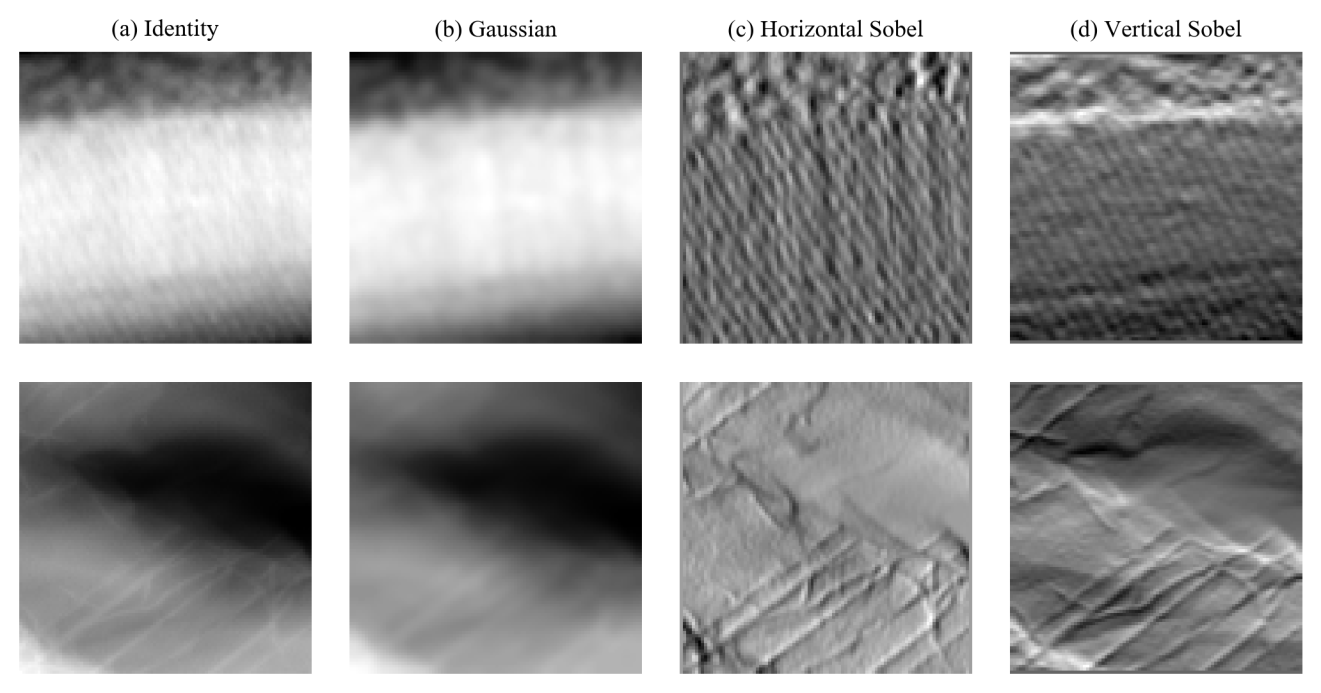}
\caption{ Two 96$\times$96 electron micrographs a) unchanged, and filtered by b) a 5$\times$5 symmetric Gaussian kernel with a 2.5 px standard deviation, c) a 3$\times$3 horizontal Sobel kernel, and d) a 3$\times$3 vertical Sobel kernel. Intensities in a) and b) are in [0, 1], whereas intensities in c) and d) are in [-1, 1]. }
\label{table:kernel_examples}
\end{figure}

Kernels designed by humans\cite{mcandrew2015computational} are often convolved in image processing pipelines. For example, convolutions of electron micrographs with Gaussian and Sobel kernels are shown in figure~\ref{table:kernel_examples}. Gaussian kernels compute local averages, blurring images and suppressing high-frequency noise. For example, a 5$\times$5 symmetric Gaussian kernel with a 2.5 px standard deviation is\\
\begin{equation}
\begin{bmatrix}
0.1689 \\
0.2148 \\
0.2326 \\
0.2148 \\
0.1689 
\end{bmatrix} \begin{bmatrix}
0.1689 & 0.2148 & 0.2326 & 0.2148 & 0.1689 
\end{bmatrix} = \begin{bmatrix}
0.0285 & 0.0363 & 0.0393 & 0.0363 & 0.0285 \\
0.0363 & 0.0461 & 0.0500 & 0.0461 & 0.0363 \\
0.0393 & 0.0500 & 0.0541 & 0.0500 & 0.0393 \\
0.0363 & 0.0461 & 0.0500 & 0.0461 & 0.0363 \\
0.0285 & 0.0363 & 0.0393 & 0.0363 & 0.0285 
\end{bmatrix}\,. \label{eqn:gaussian_kernel}
\end{equation}
Alternatives to Gaussian kernels for image smoothing\cite{opencv2013smoothing} include mean, median and bilateral filters. Sobel kernels compute horizontal and vertical spatial gradients that can be used for edge detection\cite{vairalkar2012edge}. For example, 3$\times$3 Sobel kernels are\\
\noindent\begin{subequations}
  \begin{tabularx}{\textwidth}{Xp{2cm}X}
  \begin{equation}
      \begin{bmatrix}
1 \\
2 \\
1  
\end{bmatrix} \begin{bmatrix}
1 & 0 & -1 
\end{bmatrix} = 
\begin{bmatrix}
1 & 0 & -1 \\
2 & 0 & -2 \\
1 & 0 & -1 
\end{bmatrix} \label{eqn:sobel_kernel1}
  \end{equation}
  & &
  \begin{equation}
    \begin{bmatrix}
1 \\
0 \\
-1  
\end{bmatrix} \begin{bmatrix}
1 & 2 & 1 
\end{bmatrix} = \begin{bmatrix}
1 & 2 & 1 \\
0 & 0 & 0 \\
-1 & -2 & -1 
\end{bmatrix} \label{eqn:sobel_kernel2}
  \end{equation}
  \end{tabularx}
\end{subequations}\\
Alternatives to Sobel kernels offer similar utility, and include extended Sobel\cite{bogdan2019custom}, Scharr\cite{jahne1999principles, scharr2000optimale}, Kayyali\cite{kawalec2014edge}, Roberts cross\cite{roberts1963machine} and Prewitt\cite{prewitt1970object} kernels. Two-dimensional Gaussian and Sobel kernels are examples of linearly separable, or \enquote{flattenable}, kernels, which can be split into two one-dimensional kernels, as shown in equations~\ref{eqn:gaussian_kernel}-\ref{eqn:sobel_kernel2}. Kernel separation can decrease computation in convolutional layers by convolving separated kernels in series, and CNNs that only use separable convolutions are effective\cite{jin2014flattened, chen2020xsepconv, jaderberg2014speeding}. However, serial convolutions decrease parallelization and separable kernels have fewer degrees of freedom, decreasing representational capacity. Thus, separated kernels are usually at least 5$\times$5, and separated 3$\times$3 kernels are unusual. Even-sized kernels, such as 2$\times$2 and 4$\times$4, are rare as symmetric padding is needed to avoid information erosion caused by spatial shifts of feature maps\cite{wu2019convolution}. 

A traditional 2D convolutional layer maps inputs, $x^\text{input}$, with height $H$, width, $W$, and depth, $D$, to
\begin{equation}
    x_{kij}^\text{output} = b_k + \sum\limits_{d=1}^{D} \sum\limits_{m=1}^{M} \sum\limits_{n=1}^{N} w_{dkmn} x_{d(i+m-1)(j+n-1)}^\text{input}\,, i \in [1, H - M + 1]\,, j \in [1, W - N + 1] \,, \label{eqn:conv_layer}
\end{equation}
where $K$ output channels are indexed by $k \in [1, K]$, is the sum of a bias, $b$, and convolutions of each input channel with $M \times N$ kernels with weights, $w$. For clarity, a traditional convolutional layer is visualized in figure~\ref{fig:convolutions}a. Convolutional layers for 1D, 3D and higher-dimensional kernels\cite{kossaifi2019efficient} have a similar form to 2D kernels, where kernels are convolved across each dimension. Most inputs to convolutional layers are padded\cite{chris2020using, liu2018partial} to avoid reducing spatial resolutions by kernel sizes, which could remove all resolution in deep networks. Padding is computationally inexpensive and eases implementations of ANNs that would otherwise combine layers with different sizes, such as FractalNet\cite{larsson2016fractalnet}, Inception\cite{szegedy2017inception, szegedy2016rethinking, szegedy2015going}, NASNet\cite{zoph2018learning}, recursive CNNs\cite{kim2016deeply, tai2017image}, and ResNet\cite{he2016deep}. Pre-padding inputs results in higher performance than post-padding outputs\cite{dwarampudi2019effects}. Following AlexNet\cite{krizhevsky2012imagenet}, most convolutional layers are padded with zeros for simplicity. Reflection and replication padding achieve similar results to zero padding\cite{liu2018partial}. However, padding based on partial convolutions\cite{liu2018image} consistently outperforms other methods\cite{liu2018partial}. 

Convolutional layers are similar to fully connected layers used in multilayer perceptrons\cite{peng2017multilayer, pratama2019automatic} (MLPs). For comparison with equation~\ref{eqn:conv_layer}, a fully connected, or \enquote{dense}, layer in a MLP computes
\begin{equation}
    x_{k}^\text{output} = b_k + \sum\limits_{d=1}^D w_{dk} x_d^\text{input} \,, \label{eqn:dense_layer}
\end{equation}
where every input element is connected to every output element. Convolutional layers reduce computation by making local connections within receptive fields of convolutional kernels, and by convolving kernels rather than using different weights at each input position. Intermediately, fully connected layers can be regularized to learn local connections\cite{neyshabur2020towards}. Fully connected layers are sometimes used at the middle of encoder-decoders\cite{guo20193d}. However, such fully connected layers can often be replaced by multiscale atrous, or \enquote{holey}, convolutions\cite{dumoulin2016guide} in an atrous spatial pyramid pooling\cite{chen2017rethinking, chen2018encoder} (ASPP) module to decrease computation without a significant decrease in performance. Alternatively, weights in fully connected layers can be decomposed into multiple smaller tensors to decrease computation without significantly decreasing performance\cite{oseledets2011tensor, novikov2015tensorizing}. 

Convolutional layers can perform a variety of convolutional arithmetic\cite{dumoulin2016guide}. For example, strided convolutions\cite{kong2017take} usually skip computation of outputs that are not at multiples of an integer spatial stride. Most strided convolutional layers are applied throughout CNNs to sequentially decrease spatial extent, and thereby decrease computational requirements. In addition, strided convolutions are often applied at the start of CNNs\cite{szegedy2017inception, chollet2017xception, szegedy2016rethinking, szegedy2015going} where most input features can be resolved at a lower resolution than the input. For simplicity and computational efficiency, stride is typically constant within a convolutional layer; however, increasing stride away from the centre of layers can improve performance\cite{zaniolo2020use}. To increase spatial resolution, convolutional layers often use reciprocals of integer strides\cite{shi2016deconvolution}. Alternatively, spatial resolution can be increased by combining interpolative upsampling with an unstrided convolutional layer\cite{aitken2017checkerboard, odena2016deconvolution}, which can help to minimize output artefacts. 

Convolutional layers couple the computation of spatial and cross-channel convolutions. However, partial decoupling of spatial and cross-channel convolutions by distributing inputs across multiple convolutional layers and combining outputs can improve performance. Partial decoupling of convolutions is prevalent in many seminal DNN architectures, including FractalNet\cite{larsson2016fractalnet}, Inception\cite{szegedy2017inception, szegedy2016rethinking, szegedy2015going}, NASNet\cite{zoph2018learning}. Taking decoupling to an extreme, depthwise separable convolutions\cite{chollet2017xception, howard2017mobilenets, guo2018network} shown in figure~\ref{fig:convolutions}b compute depthwise convolutions, 
\begin{align}
    x_{dij}^\text{depth} &= \sum\limits_{m=1}^{M} \sum\limits_{n=1}^{N} u_{dmn} x_{d(i+m-1)(j+n-1)}^\text{input} \,, i \in [1, H - M + 1]\,, j \in [1, W - N + 1] \,,
\end{align}
then compute pointwise 1$\times$1 convolutions for $D$ intermediate channels,
\begin{align}
    x_{kij}^\text{output} &= b_k + \sum\limits_{d=1}^{D}  v_{dk}^\text{point} x_{dij}^\text{depth}\,,
\end{align}
where $K$ output channels are indexed by $k \in [1, K]$. Depthwise convolution kernels have weights, $u$, and the depthwise layer is often followed by extra batch normalization before pointwise convolution to improve performance and accelerate convergence\cite{howard2017mobilenets}. Increasing numbers of channels with pointwise convolutions can increase accuracy\cite{howard2017mobilenets}, at the cost of increased computation. Pointwise convolutions are a special case of traditional convolutional layers in equation~\ref{eqn:conv_layer} and have convolution kernel weights, $v$, and add biases, $b$. Naively, depthwise separable convolutions require fewer weight multiplications than traditional convolutions\cite{geeksforgeeks2020depthwise, liu2020depthwise}. However, extra batch normalization and serialization of one convolutional layer into depthwise and pointwise convolutional layers mean that depthwise separable convolutions and traditional convolutions have similar computing times\cite{chollet2017xception, liu2020depthwise}.

Most DNNs developed for computer vision use fixed-size inputs. Although fixed input sizes are often regarded as an artificial constraint, it is similar to animalian vision where there is an effectively constant number of retinal rods and cones\cite{gunther2019eye, lamb2016rods, cohen1972rods}. Typically, the most practical approach to handle arbitrary image shapes is to train a DNN with crops so that it can be tiled across images. In some cases, a combination of cropping, padding and interpolative resizing can also be used. To fully utilize unmodified variable size inputs, a simple is approach to train convolutional layers on variable size inputs. A pooling layer, such as global average pooling, can then be applied to fix output size before fully connected or other layers that might require fixed-size inputs. More involved approaches include spatial pyramid pooling\cite{he2015spatial} or scale RNNs\cite{zhang2018image}. However, typical electron micrographs are much larger than 299$\times$299, which often makes it unfeasible for electron microscopists with a few GPUs to train high-performance DNNs on full-size images. For comparison, Xception was trained on 299$\times$299 images with 60 K80 GPUs for over one month.

The Fourier transform\cite{tanaka2017introduction}, $\hat{f}(k_1,...,k_N)$, at an $N$-dimensional Fourier space vector, $\{k_1, ..., k_N\}$, is related to a function, $f(x_1,...,x_N)$, of an $N$-dimensional signal domain vector, $\{x_1, ..., x_N\}$, by
\begin{align}
\hat{f}(k_1,...,k_N) &= \left(\frac{|b|}{(2 \pi)^{1-a}}\right)^{N/2} \int\limits_{-\infty}^{\infty} ... \int\limits_{-\infty}^{\infty} f(x_1,...,x_N) \exp(+ibk_1x_i + ... + ibk_Nx_N) \diff x_1 ... \diff x_N \,, \\
f(x_1,...,x_N) &= \left(\frac{|b|}{(2 \pi)^{1+a}}\right)^{N/2} \int\limits_{-\infty}^{\infty} ... \int\limits_{-\infty}^{\infty} \hat{f}(k_1,...,k_N) \exp(- ibk_1x_i - ... - ibk_Nx_N) \diff k_1 ... \diff k_N \,,
\end{align}
where $\pi=3.141...$, and $i=(-1)^{1/2}$ is the imaginary number. Two parameters, $a$ and $b$, can parameterize popular conventions that relate the Fourier and inverse Fourier transforms. Mathematica documentation nominates conventions\cite{mathematica_ft_conventions} for general applications $(a, b)$, pure mathematics $(1, -1)$, classical physics $(-1, 1)$, modern physics $(0, 1)$, systems engineering $(1, -1)$, and signal processing $(0, 2\pi)$. We observe that most electron microscopists follow the modern physics convention of $a=0$ and $b=1$; however, the choice of convention is arbitrary and does not matter if it is consistent within a project. For discrete functions, Fourier integrals are replaced with summations that are limited to the support of a function.

Discrete Fourier transforms of uniformly spaced inputs are often computed with a fast Fourier transform (FFT) algorithm, which can be parallelized for CPUs\cite{frigo2005design} or GPUs\cite{moreland2003fft, stokfiszewski2017fast, chen2010large, gu2010empirically}. Typically, the speedup of FFTs on GPUs over CPUs is higher for larger signals\cite{puchala2015effectiveness, ogata2008efficient}. Most popular FFTs are based on the Cooley-Turkey algorithm\cite{cooley1965algorithm, duhamel1990fast}, which recursively divides FFTs into smaller FFTs. We observe that some electron microscopists consider FFTs to be limited to radix-2 signals that can be recursively halved; however, FFTs can use any combination of factors for the sizes of recursively smaller FFTs. For example, clFFT\cite{clFFT} FFT algorithms support signal sizes that are any sum of powers of 2, 3, 5, 7, 11 and 13. 

Convolution theorems can decrease computation by enabling convolution in the Fourier domain\cite{highlander2016very}. To ease notation, we denote the Fourier transform of a signal, $\textbf{I}$, by $\text{FT}(\textbf{I})$, and the inverse Fourier transform by $\text{FT}^{-1}(\textbf{I})$. Thus, the convolution theorems for two signals, $\textbf{I}_1$ and $\textbf{I}_2$, are\cite{weisstein2020convolution}
\begin{align}
\text{FT}(\textbf{I}_1 * \textbf{I}_2) &= \text{FT}(\textbf{I}_1) \cdot \text{FT}(\textbf{I}_2) \,, \\
\text{FT}(\textbf{I}_1 \cdot \textbf{I}_2) &= \text{FT}(\textbf{I}_1) * \text{FT}(\textbf{I}_2) \,,
\end{align}
where the signals can be feature channels and convolutional kernels. Fourier domain convolutions, $\textbf{I}_1 * \textbf{I}_2 = \text{FT}^{-1}\left( \text{FT}(\textbf{I}_1) \cdot \text{FT}(\textbf{I}_2) \right)$, are increasingly efficient, relative to signal domain convolutions, as kernel and image sizes increase\cite{highlander2016very}. Indeed, Fourier domain convolutions are exploited to enable faster training with large kernels in Fourier CNNs\cite{pratt2017fcnn, highlander2016very}. However, Fourier CNNs are rare as most researchers use small 3$\times$3 kernels, following University of Oxford Visual Geometry Group (VGG) CNNs\cite{simonyan2014very}. 

\begin{figure*}[tbh!]
\centering
\includegraphics[width=\textwidth]{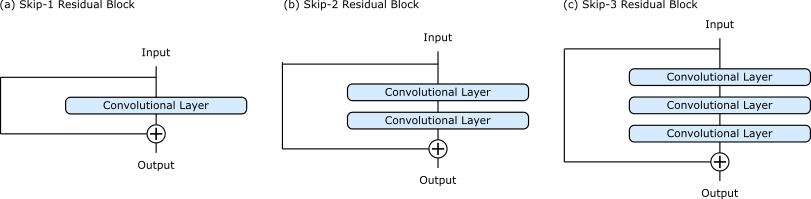}
\caption{ Residual blocks where a) one, b) two, and c) three convolutional layers are skipped. Typically, convolutional layers are followed by batch normalization then activation.  }
\label{fig:residual_block_diagram}
\end{figure*}

\subsection{Skip Connections}

Residual connections\cite{he2016deep} add a signal after skipping ANN layers, similar to cortical skip connections\cite{thomson2010neocortical, fitzpatrick1996functional}. Residuals improve DNN performance by preserving gradient norms during backpropagation\cite{zaeemzadeh2020norm, hanin2018start} and avoiding bad local minima\cite{kawaguchi2019depth} by smoothing DNN loss landscapes\cite{li2018visualizing}. In practice, residuals enable DNNs to behave like an ensemble of shallow networks\cite{veit2016residual} that learn to iteratively estimate outputs\cite{greff2016highway}. Mathematically, a residual layer learns parameters, $\textbf{w}_l$, of a perturbative function, $f_l(\textbf{x}_l, \textbf{w}_l)$, that maps a signal, $\textbf{x}_l$, at depth $l$ to depth $l+1$,
\begin{equation}
    \textbf{x}_{l+1} = \textbf{x}_l + f_l(\textbf{x}_l, \textbf{w}_l) \,.
\end{equation}
Residuals were developed for CNNs\cite{he2016deep}, and examples of residual connections that skip one, two and three convolutional layers are shown in figure~\ref{fig:residual_block_diagram}. Nonetheless, residuals are also used in MLPs\cite{martinez2017simple} and RNNs\cite{yue2018residual, kim2017residual, wu2016google}. Representational capacity of perturbative functions increases as the number of skipped layers increases. As result, most residuals skip two or three layers. Skipping one layer rarely improves performance due to its low representational capacity\cite{he2016deep}.

There are a range of residual connection variants that can improve performance. For example, highway networks\cite{srivastava2015training, srivastava2015highway} apply a gating function to skip connections, and dense networks\cite{huang2019convolutional, huang2017densely, tong2017image} use a high number of residual connections from multiple layers. Another example is applying a 1$\times$1 convolutional layer to $x_l$ before addition\cite{he2016deep, chollet2017xception} where $f_l(x_l, w_l)$ spatially resizes or changes numbers of feature channels. However, resizing with norm-preserving convolutional layers\cite{zaeemzadeh2020norm} before residual blocks can often improve performance. Finally, long additive\cite{jiang2017end} residuals that connect DNN inputs to outputs are often applied to DNNs that learn perturbative functions. 

A limitation of preserving signal information with residuals\cite{yang2017mean, xiao2018dynamical} is that residuals make DNNs learn perturbative functions, which can limit accuracy of DNNs that learn non-perturbative functions if they do not have many layers. Feature channel concatenation is an alternative approach that is not perturbative, and that supports combination of layers with different numbers of feature channels. In encoder-decoders, a typical example is concatenating features computed near the start with layers near the end to help resolve output features\cite{ibtehaz2020multiresunet, chen2018encoder, chen2017rethinking, ronneberger2015u}. Concatenation can also combine embeddings of different\cite{wu2019concatenate, terwilliger2017vertex} or variants of\cite{rivenson2018phase} input features from multiple DNNs. Finally, peephole connections in RNNs can improve performance by using concatenation to combine cell state information with other cell inputs\cite{gers2002learning, gers2001lstm}.

\section{Architecture}

There is a high variety of ANN architectures\cite{sengupta2020review, shrestha2019review, dargan2019survey, alom2019state} that are trained to minimize losses for a range of applications. Many of the most popular ANNs are also the simplest, and information about them is readily available. For example, encoder-decoder\cite{ye2019understanding, ye2018deep, chen2018encoder, chen2017rethinking, badrinarayanan2017segnet, ronneberger2015u, sutskever2014sequence} or classifier\cite{rawat2017deep} ANNs usually consist of single feedforward sequences of layers that map inputs to outputs. This section introduces more advanced ANNs used in electron microscopy, including actor-critics, GANs, RNNs, and variational autoencoders (VAEs). These ANNs share weights between layers or consist of multiple subnetworks. Other notable architectures include recursive CNNs\cite{kim2016deeply, tai2017image}, Network-in-Networks\cite{lin2013network} (NiNs), and transformers\cite{vaswani2017attention, alammar2018illustrated}. Although they will not be detailed in this review, their references may be good starting points for research.

\begin{figure*}[tbh!]
\centering
\includegraphics[width=0.5\textwidth]{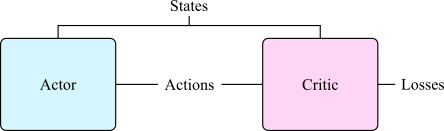}
\caption{ Actor-critic architecture. An actor outputs actions based on input states. A critic then evaluates action-state pairs to predict losses. }
\label{fig:actor-critic_diagram}
\end{figure*}

\subsection{Actor-Critic}\label{sec:actor-critic}

Most ANNs are trained by gradient descent using backpropagated gradients of a differentiable loss function cf. section~\ref{sec:gradient_descent}. However, some losses are not differentiable. Examples include losses of actors directing their vision\cite{mnih2014recurrent, ba2014multiple}, and playing competitive\cite{alphastarblog} or score-based\cite{lillicrap2015continuous, heess2015memory} computer games. To overcome this limitation, a critic\cite{konda2000actor} can be trained to predict differentiable losses from action and state information, as shown in figure~\ref{fig:actor-critic_diagram}. If the critic does not depend on states, it is a surrogate loss function\cite{grabocka2019learning, neftci2019surrogate}. Surrogates are often fully trained before actor optimization, whereas critics that depend on actor-state pairs are often trained alongside actors to minimize the impact of catastrophic forgetting\cite{liang2018generative} by adapting to changing actor policies and experiences. Alternatively, critics can be trained with features output by intermediate layers of actors to generate synthetic gradients for backpropagation\cite{jaderberg2017decoupled}.

\begin{figure*}[tbh!]
\centering
\includegraphics[width=0.62\textwidth]{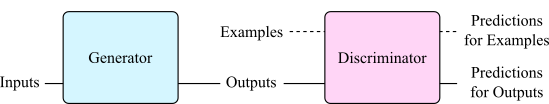}
\caption{ Generative adversarial network architecture. A generator learns to produce outputs that look realistic to a discriminator, which learns to predict whether examples are real or generated. }
\label{fig:gan_diagram}
\end{figure*}

\subsection{Generative Adversarial Network}

Generative adversarial networks\cite{gui2020review, saxena2020generative, pan2019recent, wang2019generative} (GANs) consist of generator and discriminator subnetworks that play an adversarial game, as shown in figure~\ref{fig:gan_diagram}. Generators learn to generate outputs that look realistic to discriminators, whereas discriminators learn to predict whether examples are real or generated. Most GANs are developed to generate visual media with realistic characteristics. For example, partial STEM images infilled with a GAN are less blurry than images infilled with a non-adversarial generator trained to minimize MSEs\cite{ede2020partial} cf. figure~\ref{fig:compressed_example}. Alternatively, computationally inexpensive loss functions designed by humans, such as structural similarity index measures\cite{wang2004image} (SSIMs) and Sobel losses\cite{ede2020warwick}, can improve generated output realism. However, it follows from the universal approximator theorems\cite{kidger2019universal, lin2018resnet, hanin2017approximating, lu2017expressive, pinkus1999approximation, leshno1993multilayer, hornik1991approximation, hornik1989multilayer, cybenko1989approximation} that training with ANN discriminators can often yield more realistic outputs. 

There are many popular GAN loss functions and regularization mechanisms\cite{pan2020loss, dong2019towards, mescheder2018training, kurach2019large, roth2017stabilizing}. Traditionally, GANs were trained to minimize logarithmic discriminator, $D$, and generator, $G$, losses\cite{goodfellow2014generative},
\begin{align}
    L_D &= -\log D(\textbf{x}) - \log (1-D(G(\textbf{z}))) \,, \\
    L_G &= \log (1-D(G(\textbf{z}))) \,,
\end{align}
where $\textbf{z}$ are generator inputs, $G(\textbf{z})$ are generated outputs, and $\textbf{x}$ are example outputs. Discriminators predict labels, $D(\textbf{x})$ and $D(G(\textbf{z}))$, where target labels are 0 and 1 for generated and real examples, respectively. Limitedly, logarithmic losses are numerically unstable for $D(\textbf{x}) \rightarrow 0$ or $D(G(\textbf{z})) \rightarrow 1$, as the denominator, $f(x)$, in $\partial_x \log f(x) = \partial_x f(x) / f(x)$ vanishes. In addition, discriminators must be limited to $D(\textbf{x}) > 0$ and $D(G(\textbf{z})) < 1$, so that logarithms are not complex. To avoid these issues, we recommend training discriminators with squared difference losses\cite{mao2018effectiveness, mao2017least},
\begin{align}
    L_D &= (D(\textbf{x})-1)^2 + D(G(\textbf{z}))^2 \,, \\
    L_G &= (D(G(\textbf{z}))-1)^2 \,.
\end{align}
However, there are a variety of other alternatives to logarithmic loss functions that are also effective\cite{pan2020loss, dong2019towards}.

A variety of methods have been developed to improve GAN training\cite{wiatrak2019stabilizing, salimans2016improved}. The most common issues are catastrophic forgetting\cite{liang2018generative} of previous learning, and mode collapse\cite{bang2018mggan} where generators only output examples for a subset of a target domain. Mode collapse often follows discriminators becoming Lipschitz discontinuous. Wasserstein GANs\cite{arjovsky2017wasserstein} avoid mode collapse by clipping trainable variables, albeit often at the cost of 5-10 discriminator training iterations per generator training iteration. Alternatively, Lipschitz continuity can be imposed by adding a gradient penalty\cite{gulrajani2017improved} to GAN losses, such as differences of L2 norms of discriminator gradients from unity, 
\begin{align}
    \tilde{x} &= G(\textbf{z}) \,, \\
    \hat{\textbf{x}} &= \epsilon \textbf{x} + (1-\epsilon) \tilde{\textbf{x}} \,, \\
    L_D &= D(\tilde{\textbf{x}}) - D(\textbf{x}) + \lambda (|| \partial_{\hat{\textbf{x}}} D(\hat{\textbf{x}})||_2 - 1)^2 \,, \\
    L_G &= -D(G(\textbf{z})) \,,
\end{align}
where $\epsilon \in [0,1]$ is a uniform random variate, $\lambda$ weights the gradient penalty, and $\tilde{\textbf{x}}$ is an attempt to generate $x$. However, using a gradient penalty introduces additional gradient backpropagation that increases discriminator training time. There are also a variety of computationally inexpensive tricks that can improve training, such as adding noise to labels\cite{salimans2016improved, szegedy2016rethinking, hazan2017adversarial} or balancing discriminator and generator learning rates\cite{ede2020exit}. These tricks can help to avoid discontinuities in discriminator output distributions that can lead to mode collapse; however, we observe that these tricks do not reliably stabilize GAN training.

Instead, we observe that spectral normalization\cite{miyato2018spectral} reliably stabilizes GAN discriminator training in our electron microscopy research\cite{ede2020partial, ede2020exit, ede2019deep}. Spectral normalization controls Lipschitz constants of discriminators by fixing the spectral norms of their weights, as introduced in section~\ref{sec:normalization}. Advantages of spectral normalization include implementations based on the power iteration method\cite{yoshida2017spectral, golub2000eigenvalue} being computationally inexpensive, not adding a regularizing loss function that could detrimentally compete\cite{chen2017gradnorm, lee2020multitask} with discrimination losses, and being effective with one discriminator training iterations per generator training iteration\cite{miyato2018spectral, zhang2019self}. Spectral normalization is popular in GANs for high-resolution image synthesis, where it is also applied in generators to stabilize training\cite{brock2018large}.

There are a variety of GAN architectures\cite{hindupur2018gan}. For high-resolution image synthesis, computation can be decreased by training multiple discriminators to examine image patches at different scales\cite{wang2018high, ede2020partial}. For domain translation characterized by textural differences, a cyclic GAN\cite{zhu2017unpaired, bashkirova2018unsupervised} consisting of two GANs can map from one domain to the other and vice versa. Alternatively, two GANs can share intermediate layers to translate inputs via a shared embedding domain\cite{liu2017unsupervised}. Cyclic GANs can also be combined with a siamese network\cite{koch2015siamese, chopra2005learning, bromley1994signature} for domain translation beyond textural differences\cite{amodio2019travelgan}. Finally, discriminators can introduce auxiliary losses to train DNNs to generalize to examples from unseen domains\cite{tzeng2017adversarial, ganin2015unsupervised, tzeng2015simultaneous}.


\begin{figure*}[tbh!]
\centering
\includegraphics[width=\textwidth]{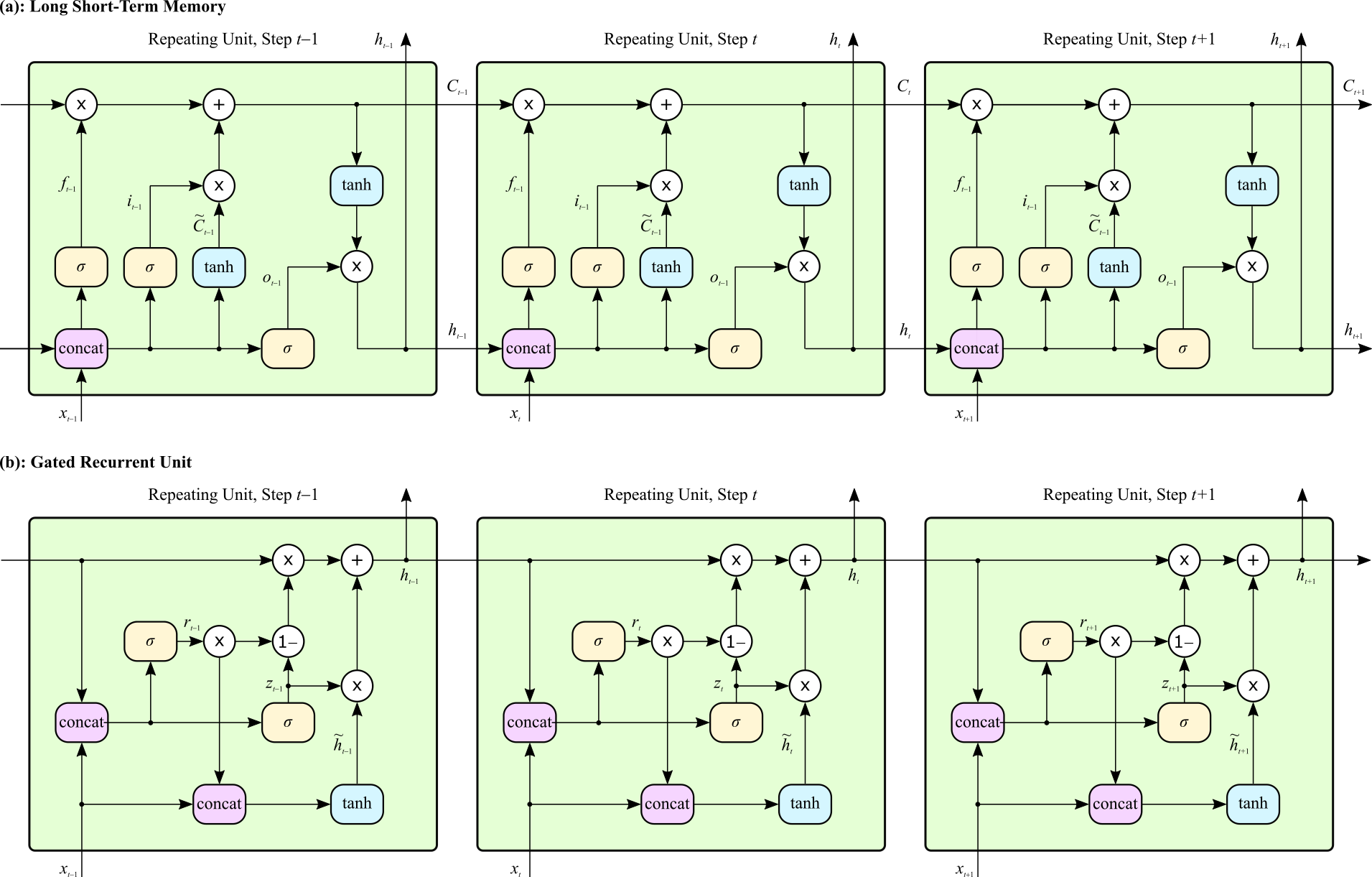}
\caption{ Architectures of recurrent neural networks with a) long short-term memory (LSTM) cells, and b) gated recurrent units (GRUs). }
\label{fig:lstm_and_gru}
\end{figure*}

\subsection{Recurrent Neural Network}

Recurrent neural networks\cite{rezk2020recurrent, du2019recurrent, yu2019review, choe2017probabilistic, choi2017awesome, lipton2015critical} reuse an ANN cell to process each step of a sequence. Most RNNs learn to model long-term dependencies by gradient backpropagation through time\cite{werbos1990backpropagation} (BPTT). The ability of RNNs to utilize past experiences enables them to model partially observed and variable length Markov decision processes\cite{saldi2019asymptotic, jaakkola1995reinforcement} (MDPs). Applications of RNNs include directing vision\cite{mnih2014recurrent, ba2014multiple}, image captioning\cite{xu2015show, vinyals2015show}, language translation\cite{basmatkar2019survey}, medicine\cite{antczak2018deep}, natural language processing\cite{wu2020deep, otter2020survey}, playing computer games\cite{alphastarblog}, text classification\cite{minaee2020deep}, and traffic forecasting\cite{iyer2020forecasting}. Many RNNs are combined with CNNs to embed visual media\cite{ba2014multiple} or words\cite{mandal2019deep, rhanoui2019cnn}, or to process RNN outputs\cite{zhang2018combination, qu2018question}. RNNs can also be combined with MLPs\cite{mnih2014recurrent}, or text embeddings\cite{sieg2019from} such as BERT\cite{devlin2018bert, sieg2019from}, continuous bag-of-words\cite{mikolov2013distributed, mnih2013learning, grave2018learning} (CBOW), doc2vec\cite{le2014distributed, lau2016empirical}, GloVe\cite{pennington2014glove}, and word2vec\cite{mikolov2013efficient, mikolov2013distributed}.

The most popular RNNs consist of long short-term memory\cite{sherstinsky2020fundamentals, olah2015understanding, gers2000learning, hochreiter1997long} (LSTM) cells or gated recurrent units\cite{olah2015understanding, cho2014learning, dey2017gate, heck2017simplified} (GRUs). LSTMs and GRUs are popular as they solve the vanishing gradient problem\cite{pascanu2013difficulty, hanin2018neural, hanin2018start} and have consistently high performance\cite{britz2017massive, jozefowicz2015empirical, chung2014empirical, gruber2020gru, weiss2018practical, bayer2009evolving}. Their architectures are shown in figure~\ref{fig:lstm_and_gru}. At step $t$, an LSTM outputs a hidden state, $h_t$, and cell state, $C_t$, given by
\begin{align}
    \textbf{f}_t &= \sigma (\textbf{w}_f \cdot [\textbf{h}_{t-1}, \textbf{x}_t] + \textbf{b}_f) \,, \\
    \textbf{i}_t &= \sigma (\textbf{w}_i \cdot [\textbf{h}_{t-1}, \textbf{x}_t] + \textbf{b}_i ) \,, \\
    \tilde{\textbf{c}}_t &= \tanh (\textbf{w}_C \cdot [\textbf{h}_{t-1}, \textbf{x}_t] + \textbf{b}_C) \,, \\
    \textbf{C}_t & = \textbf{f}_t \textbf{C}_{t-1} + \textbf{i}_t \tilde{\textbf{C}_t} \,, \\
    \textbf{o}_t &= \sigma (\textbf{w}_o \cdot [\textbf{h}_{t-1}, \textbf{x}_t] + \textbf{b}_o) \,, \\
    \textbf{h}_t &= \textbf{o}_t \tanh(\textbf{C}_t) \,,
\end{align}
where $\textbf{C}_{t-1}$ is the previous cell state, $\textbf{h}_{t-1}$ is the previous hidden state, $\textbf{x}_t$ is the step input, and $\sigma$ is a logistic sigmoid function of equation~\ref{eqn:logistic_sigmoid}, $[\textbf{x}, \textbf{y}]$ is the concatenation of $\textbf{x}$ and $\textbf{y}$ channels, and $(\textbf{w}_f, \textbf{b}_f)$, $(\textbf{w}_i, \textbf{b}_i)$, $(\textbf{w}_C, \textbf{b}_C)$ and $(\textbf{w}_o, \textbf{b}_o)$ are pairs of weights and biases. A GRU performs fewer computations than an LSTM and does not have separate cell and hidden states,
\begin{align}
    \textbf{z}_t &= \sigma (\textbf{w}_z \cdot [\textbf{h}_{t-1}, \textbf{x}_t] + \textbf{b}_z) \,, \\
    \textbf{r}_t &= \sigma (\textbf{w}_r \cdot [\textbf{h}_{t-1}, \textbf{x}_t] + \textbf{b}_r) \,, \\
    \tilde{\textbf{h}}_t &= \tanh (\textbf{w}_h \cdot [\textbf{r}_t \textbf{h}_{t-1}, \textbf{x}_t] + \textbf{b}_h) \,, \\
    \textbf{h}_t &= (1 - \textbf{z}_t) \textbf{h}_{t-1} + \textbf{z}_t \tilde{\textbf{h}}_t \,,
\end{align}
where $(\textbf{w}_z, \textbf{b}_z)$, $(\textbf{w}_r, \textbf{b}_r)$, and $(\textbf{w}_h, \textbf{b}_h)$ are pairs of weights and biases. Minimal gated units (MGUs) can further reduce computation\cite{zhou2016minimal}. A large-scale analysis of RNN architectures for language translation found that LSTMs consistently outperform GRUs\cite{britz2017massive}. GRUs struggle with simple languages that are learnable by LSTMs as the combined hidden and cell states of GRUs make it more difficult for GRUs to perform unbounded counting\cite{weiss2018practical}. However, further investigations found that GRUs can outperform LSTMs on tasks other than language translation\cite{jozefowicz2015empirical}, and that GRUs can outperform LSTMs on some datasets\cite{greff2016lstm, chung2014empirical, gruber2020gru}. Overall, LSTM performance is usually comparable to that of GRUs.

There are a variety of alternatives to LSTM and GRUs. Examples include continuous time RNNs\cite{mozer2017discrete, funahashi1993approximation, quinn2001evolving, beer1997dynamics, harvey1994seeing} (CTRNNs), Elman\cite{elman1990finding} and Jordan\cite{jordan1997serial} networks, independently RNNs\cite{li2018independently} (IndRNNs), Hopfield networks\cite{sathasivam2008logic}, recurrent MLPs\cite{tutschku1995recurrent} (RMLPs). However, none of the variants offer consistent performance benefits over LSTMs for general sequence modelling. Similarly, augmenting LSTMs with additional connections, such as peepholes\cite{gers2002learning, gers2001lstm} and projection layers\cite{jia2017long}, does not consistently improve performance. For electron microscopy, we recommend defaulting to LSTMs as we observe that their performance is more consistently high than performance of other RNNs. However, LSTM and GRU performance is often comparable, so GRUs are also a good choice to reduce computation.

There are a variety of architectures based on RNNs. Popular examples include deep RNNs\cite{pascanu2014construct} that stack RNN cells to increase representational ability, bidirectional RNNs\cite{schuster1997bidirectional, bahdanau2015neural, graves2005framewise, thireou2007bidirectional} that process sequences both forwards and in reverse to improve input utilization, and using separate encoder and decoder subnetworks\cite{cho2014learning, cho2014properties} to embed inputs and generate outputs. Hierarchical RNNs\cite{zhang2018learning, chung2016hierarchical, sordoni2015hierarchical, paine2005hierarchical, schmidhuber1992learning} are more complex models that stack RNNs to efficiently exploit hierarchical sequence information, and include multiple timescale RNNs\cite{yamashita2008emergence, shibata2013hierarchical} (MTRNNs) that operate at multiple sequence length scales. Finally, RNNs can be augmented with additional functionality to enable new capabilities. For example, attention\cite{chaudhari2019attentive, luong2015effective, bahdanau2014neural, xu2015show} mechanisms can enable more efficient input utilization. Further, creating a neural Turing machine (NTM) by augmenting a RNN with dynamic external memory\cite{graves2016hybrid, graves2014neural} can make it easier for an agent to solve dynamic graphs.

\begin{figure*}[tbh!]
\centering
\includegraphics[width=\textwidth]{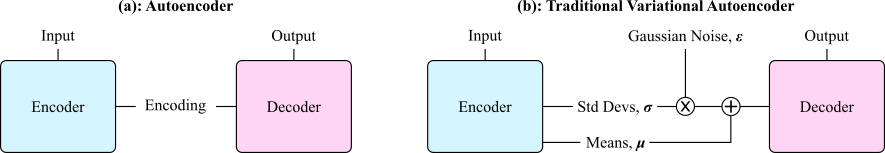}
\caption{ Architectures of autoencoders where an encoder maps an input to a latent space and a decoder learns to reconstruct the input from the latent space. a) An autoencoder encodes an input in a deterministic latent space, whereas a b) traditional variational autoencoder encodes an input as means, $\mu$, and standard deviations, $\sigma$, of Gaussian multivariates, $\mu + \sigma \cdot \epsilon$, where $\epsilon$ is a standard normal multivariate. }
\label{fig:ae_and_vae_diagram}
\end{figure*}

\subsection{Autoencoders}

Autoencoders\cite{tschannen2018recent, hinton2006reducing, kramer1991nonlinear} (AEs) learn to efficiently encode inputs, $\textbf{I}$, without supervision. An AE consists of a encoder, $E$, and decoder, $D$, as shown in figure~\ref{fig:ae_and_vae_diagram}a. Most encoders and decoders are jointly trained\cite{zhou2014joint} to restore inputs from encodings, $E(\textbf{I})$, to minimize a MSE loss,
\begin{equation}
    L_\text{AE} = \text{MSE}( D(E(\textbf{I})), \textbf{I} ) \,,
\end{equation}
by gradient descent. In practice, DNN encoders and decoders yield better compression\cite{hinton2006reducing} than linear techniques, such as principal component analysis\cite{jolliffe2016principal} (PCA), or shallow ANNs. Indeed, deep AEs can outperform JPEG image compression\cite{theis2017lossy}. Denoising autoencoders\cite{vincent2010stacked, vincent2008extracting, gondara2016medical, cho2013simple, cho2013boltzmann} (DAEs) are a popular AE variant that can learn to remove artefacts by artificially corrupting inputs inside encoders. Alternatively, contractive autoencoders\cite{rifai2011contractive, rifai2011higher} (CAEs) can decrease sensitivity to input values by adding a loss to minimize gradients w.r.t. inputs. Most DNNs that improve electron micrograph signal-to-noise are DAEs.

In general, semantics of AE outputs are pathological functions of encodings. To generate outputs with well-behaved semantics, traditional VAEs\cite{kingma2014auto, kingma2019introduction, doersch2016tutorial} learn to encode means, $\boldsymbol\mu$, and standard deviations, $\boldsymbol\sigma$, of Gaussian multivariates. Meanwhile, decoders learn to reconstruct inputs from sampled multivariates, $\boldsymbol\mu + \boldsymbol\sigma \cdot \boldsymbol\epsilon$, where $\boldsymbol\epsilon$ is a standard normal multivariate. Traditional VAE architecture is shown in figure~\ref{fig:ae_and_vae_diagram}b. Usually, VAE encodings are regularized by adding Kullback-Leibler (KL) divergence of encodings from standard multinormals to an AE loss function,
\begin{equation}\label{eqn:traditional_vae}
    L_\text{VAE} = \text{MSE}( D(\boldsymbol\mu + \boldsymbol\sigma \cdot \boldsymbol\epsilon ), \textbf{I} ) + \frac{\lambda_\text{KL}}{2Bu} \sum\limits_{i=1}^B \sum\limits_{j=1}^u \mu_{ij}^2 + \sigma_{ij}^2 - \log(\sigma_{ij}^2) - 1 \,,
\end{equation}
where $\lambda_\text{KL}$ weights the contribution of the KL divergence loss for a batch size of $B$, and a latent space with $u$ degrees of freedom. However, variants of Gaussian regularization can improve clustering\cite{ede2020warwick}, and sparse autoencoders\cite{makhzani2013k, nair20093d, arpit2016regularized, zeng2018facial} (SAEs) that regularize encoding sparsity can encode more meaningful features. To generate realistic outputs, a VAE can be combined with a GAN to create a VAE-GAN\cite{yin2020survey, yu2019vaegan, larsen2016autoencoding}. Adding a loss to minimize differences between gradients of generated and target outputs is computationally inexpensive alternative that can generate realistic outputs for some applications\cite{ede2020warwick}.

A popular application of VAEs is data clustering. For example, VAEs can encode hash tables\cite{zhuang2020new, jin2019deep, khobahi2019model, wang2018deep, pattersonsemantic} for search engines, and we use VAEs as the basis of our electron micrograph search engines\cite{ede2020warwick}. Encoding clusters visualized by tSNE can be labelled to classify data\cite{ede2020warwick}, and encoding deviations from clusters can be used for anomaly detection\cite{fan2020video, yao2019unsupervised, xu2018unsupervised, an2015variational, gauerhof2020reverse}. In addition, learning encodings with well-behaved semantics enables encodings to be used for semantic manipulation\cite{klys2018learning, gauerhof2020reverse}. Finally, VAEs can be used as generative models to create synthetic populations\cite{borysov2019generate, salim2018synthetic}, develop new chemicals\cite{gomez2018automatic, zhavoronkov2019deep, griffiths2020constrained, lim2018molecular}, and synthesize underrepresented data to reduce imbalanced learning\cite{wan2017variational}.

\section{Optimization}

Training, testing, deployment and maintenance of machine learning systems is often time-consuming and expensive\cite{zhang2020machine, amershi2019software, breck2017ml, sculley2015hidden}. The first step is usually preparing training data and setting up data pipelines for ANN training and evaluation. Typically, ANN parameters are randomly initialized for optimization by gradient descent, possibly as part of an automatic machine learning algorithm. Reinforcement learning is a special optimization case where the loss is a discounted future reward. During training, ANN components are often regularized to stabilize training, accelerate convergence, or improve performance. Finally, trained models can be streamlined for efficient deployment. This section introduces each step. We find that electron microscopists can be apprehensive about robustness and interpretability of ANNs, so we also provide subsections on model evaluation and interpretation.

\begin{figure*}[tbh!]
\centering
\includegraphics[width=0.85\textwidth]{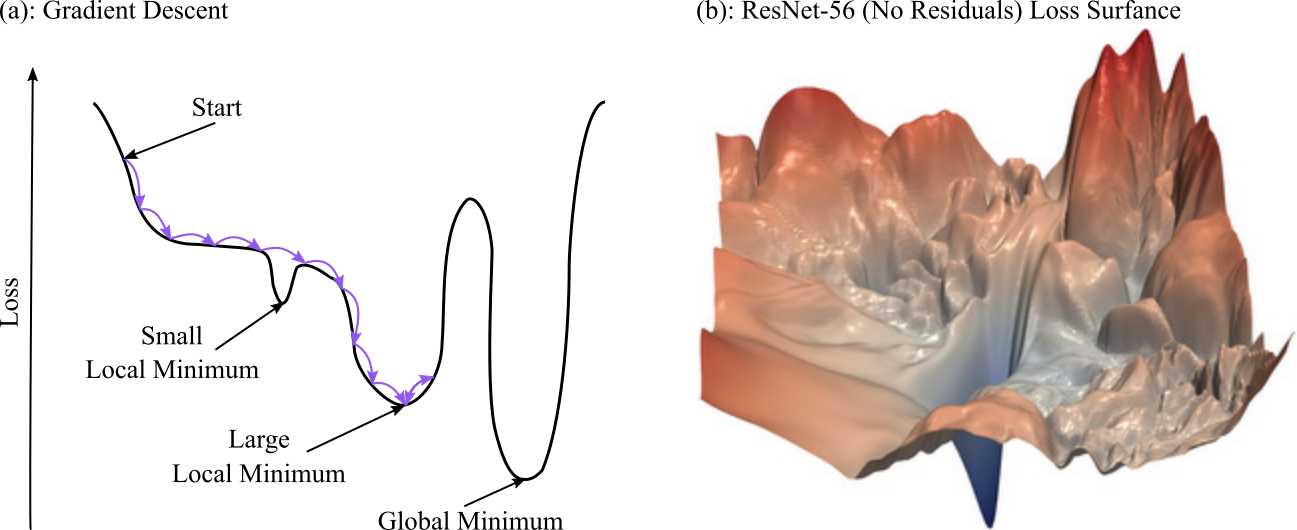}
\caption{ Gradient descent. a) Arrows depict steps across one dimension of a loss landscape as a model is optimized by gradient descent. In this example, the optimizer traverses a small local minimum; however, it then gets trapped in a larger sub-optimal local minimum, rather than reaching the global minimum. b) Experimental DNN loss surface for two random directions in parameter space showing many local minima\cite{li2018visualizing}. The image in part b) is reproduced with permission under an MIT license\cite{li2017mitlicense}. }
\label{fig:gradient_descent_example}
\end{figure*}

\begin{algorithm}
\caption{ Optimization by gradient descent. }
\begin{algorithmic}
\STATE Initialize a model, $f(\textbf{x})$, with trainable parameters, $\boldsymbol\theta_1$.
\FOR{training step $t = 1, T$}
    \STATE Forwards propagate a randomly sampled batch of inputs, $\textbf{x}$, through the model to compute outputs, $\textbf{y} = f(\textbf{x})$.
    \STATE Compute loss, $L_t$, for outputs.
    \STATE Use the differentiation chain rule\cite{rodriguez2010semiotic} to backpropagate gradients of the loss to trainable parameters, $\boldsymbol\theta_{t-1}$.
    \STATE Apply an optimizer to the gradients to update $\boldsymbol\theta_{t-1}$ to $\boldsymbol\theta_t$.
\ENDFOR
\end{algorithmic}
\label{alg:gradient_descent}
\end{algorithm}

\begin{figure}[tbh!]
\renewcommand\figurename{Algorithms}
\renewcommand\thefigure{1}
\begin{minipage}[t]{.45\textwidth}
Vanilla SGD\cite{kiefer1952stochastic, robbins1951stochastic} $[\eta]$
\begin{align}
\theta_t = \theta_{t-1} - \eta \partial_\theta L_t    
\end{align}
Momentum\cite{polyak1964some} $[\eta, \gamma]$
\begin{align}
v_t &= \gamma v_{t-1} + \eta \partial_\theta L_t \\
\theta_t &= \theta_{t-1} - v_t  
\end{align}
Nesterov momentum\cite{sutskever2013importance, su2014differential, tensorflow2018nesterov} $[\eta, \gamma]$
\begin{align}
\phi &= \theta_{t-1} + \eta \gamma v_{t-1} \\
v_t &= \gamma v_{t-1} + \partial_\theta L_t \\
\theta_t &= \phi - \eta v_t  (1 + \gamma)
\end{align}
Quasi-hyperbolic momentum\cite{ma2018quasi} $[\eta, \beta, \nu]$
\begin{align}
g_t &= \beta g_{t-1} + (1 - \beta) \partial_\theta L_t \\
\theta_t &= \theta_{t-1} - \eta ( v g_t + (1-v)\partial_\theta L_t )
\end{align}
AggMo\cite{lucas2018aggregated} $[\eta, \beta^{(1)}, ..., \beta^{(K)}]$
\begin{align}
v_t^{(i)} &= \beta^{(i)} v_{t-1}^{(i)} - (\partial_\theta L_t) \\
\theta_t & = \theta_{t-1} + \frac{\eta}{K} \sum\limits_{i=1}^K v_t^{(i)}
\end{align}
\end{minipage}%
\hspace{0.1\textwidth}\begin{minipage}[t]{.45\textwidth}
RMSProp\cite{hinton2012neural} $[\eta, \beta, \epsilon]$
\begin{align}
v_t &= \beta v_{t-1} + (1-\beta) (\partial_\theta L_t)^2 \\
\theta_t &= \theta_{t-1} - \frac{\eta}{(v_t + \epsilon)^{1/2}} \partial_\theta L_t
\end{align}
ADAM\cite{kingma2014adam} $[\eta, \beta_1, \beta_2, \epsilon]$
\begin{align}
m_t &= \beta_1 m_{t-1} + (1-\beta_1) \partial_\theta L_t \\
v_t &= \beta_2 v_{t-1} + (1-\beta_2) (\partial_\theta L_t)^2 \\
\hat{m}_t & = \frac{m_t}{1-\beta_1^t} \\
\hat{v}_t & = \frac{v_t}{1-\beta_2^t} \\
\theta_t &= \theta_{t-1} - \frac{\eta}{\hat{v}_t^{1/2} + \epsilon} \hat{m}_t
\end{align}
AdaMax\cite{kingma2014adam} $[\eta, \beta_1, \beta_2]$
\begin{align}
m_t &= \beta_1 m_{t-1} + (1-\beta_1) \partial_\theta L_t \\
u_t &= \max(\beta_2 u_{t-1}, |\partial_\theta L_t|) \\
\hat{m}_t & = \frac{m_t}{1-\beta_1^t} \\
\theta_t &= \theta_{t-1} - \frac{\eta}{u_t} \hat{m}_t
\end{align}
\end{minipage}
\caption{ Update rules of various gradient descent optimizers for a trainable parameter, $\theta_t$, at iteration $t$, gradients of losses w.r.t. the parameter, $\partial_\theta L_t$, and learning rate, $\eta$. Hyperparameters are listed in square brackets. }
\label{eqns:optimizers}
\end{figure}

\subsection{Gradient Descent}\label{sec:gradient_descent}

Most ANNs are iteratively trained by gradient descent\cite{sun2019survey, bottou2018optimization, ruder2016overview, baydin2017automatic, curry1944method, lemarechal2012cauchy}, as described by algorithm~\ref{alg:gradient_descent} and shown in figure~\ref{fig:gradient_descent_example}. To minimize computation, results at intermediate stages of forward propagation, where inputs are mapped to outputs, are often stored in memory. Storing the forwards pass in memory enables backpropagation memoization by sequentially computing gradients w.r.t. trainable parameters. To reduce memory costs for large ANNs, a subset of intermediate forwards pass results can be saved as starting points to recompute other stages during backpropagation\cite{chen2016training, cybertron2019saving}. Alternatively, forward pass computations can be split across multiple devices\cite{jin2018spatially}. Optimization by gradient descent plausibly models learning in some biological systems\cite{whittington2019theories}. However, gradient descent is not generally an accurate model of biological learning\cite{green2008exercising, bassett2011dynamic, o2004reward}.

There are many popular gradient descent optimizers for deep learning\cite{sun2019survey, bottou2018optimization, ruder2016overview}. Update rules for eight popular optimizers are summarized in figure~\ref{eqns:optimizers}. Other optimizers include AdaBound\cite{luo2019adaptive}, AMSBound\cite{luo2019adaptive}, AMSGrad\cite{reddi2019convergence}, Lookahead\cite{zhang2019lookahead}, NADAM\cite{dozat2016incorporating}, Nostalgic Adam\cite{huang2018nostalgic}, Power Gradient Descent\cite{baiesi2019power}, Rectified ADAM\cite{liu2019variance} (RADAM), and trainable optimizers\cite{bello2017neural, andrychowicz2016learning, li2016learning, hochreiter2001learning, duan2016rl}. Gradient descent is effective in the high-dimensional optimization spaces of overparameterized ANNs\cite{zou2018stochastic} as the probability of getting trapped in a sub-optimal local minima decreases as the number of dimensions increases. The simplest optimizer is \enquote{vanilla} stochastic gradient descent (SGD), where a trainable parameter perturbation, $\Delta \theta_t = \theta_t - \theta_{t-1}$, is the product of a learning rate, $\eta$, and derivative of a loss, $L_t$, w.r.t. the trainable parameter, $\partial_\theta L_t$. However, vanilla SGD convergence is often limited by unstable parameter oscillations as it a low-order local optimization method\cite{watt2020two}. Further, vanilla SGD has no mechanism to adapt to varying gradient sizes, which vary effective learning rates as $\Delta \theta \propto \partial_\theta L_t$. 

To accelerate convergence, many optimizers introduce a momentum term that weights an average of gradients with past gradients\cite{goh2017momentum, sutskever2013importance, qian1999momentum}. Momentum-based optimizers in figure~\ref{eqns:optimizers} are momentum, Nesterov momentum\cite{sutskever2013importance, su2014differential}, quasi-hyperbolic momentum\cite{ma2018quasi}, AggMo\cite{lucas2018aggregated}, ADAM\cite{kingma2014adam}, and AdaMax\cite{kingma2014adam}. To standardize effective learning rates for every layer, adaptive optimizers normalize updates based on an average of past gradient sizes. Adaptive optimizers in figure~\ref{eqns:optimizers} are RMSProp\cite{hinton2012neural}, ADAM\cite{kingma2014adam}, and AdaMax\cite{kingma2014adam}, which usually result in faster convergence and higher accuracy than other optimizers\cite{schmidt2020descending, choi2019empirical}. However, adaptive optimizers can be outperformed by vanilla SGD due to overfitting\cite{wilson2017marginal}, so some researchers adapt adaptive learning rates to their variance\cite{liu2019variance} or transition from adaptive optimization to vanilla SGD as training progresses\cite{luo2019adaptive}. For electron microscopy we recommend adaptive optimization with Nadam\cite{dozat2016incorporating}, which combines ADAM with Nesterov momentum, as it is well-established and a comparative analysis of select gradient descent optimizers found that it often achieves higher performance than other popular optimizers\cite{dogo2018comparative}. Limitingly, most adaptive optimizers slowly adapt to changing gradient sizes e.g. a default value for ADAM $\beta_2$ is 0.999\cite{kingma2014adam}. To prevent learning being destabilized by spikes in gradient sizes, adaptive optimizers can be combined with adaptive learning rate\cite{ede2020adaptive, luo2019adaptive} or gradient\cite{seetharaman2020autoclip, pascanu2013difficulty, gorbunov2020stochastic} clipping.

For non-adaptive optimizers, effective learning rates are likely to vary due to varying magnitudes of gradients w.r.t. trainable parameters. Similarly, learning by biological neurons varies as stimuli usually activate a subset of neurons\cite{yoshida2020natural}. However, all neuron outputs are usually computed for ANNs. Thus, not effectively using all weights to inform decisions is computational inefficient. Further, inefficient weight updates can limit representation capacity, slow convergence, and decrease training stability. A typical example is effective learning rates varying between layers. Following the chain rule, gradients backpropagated to the $i$th layer of a DNN from its start are
\begin{equation}
    \frac{\partial L_t}{\partial \textbf{x}_i}  = \left( \prod\limits_{l=i}^{L-1} \frac{\partial \textbf{x}_{l+1}}{\partial \textbf{x}_l}\right) \frac{\partial L_t}{\partial \textbf{x}_L} \,,
\end{equation}
for a DNN with $L$ layers. Vanishing gradients\cite{hanin2018neural, hanin2018start, pascanu2013difficulty} occur when many layers have $\partial x_{l+1} / \partial x_l \ll 1$. For example, DNNs with logistic sigmoid activations often exhibit vanishing gradients as their maximum gradient is $1/4$ cf. equation~\ref{eqn:logistic_sigmoid_grad}. Similarly, exploding gradients\cite{hanin2018neural, hanin2018start, pascanu2013difficulty} occur when many layers have $\partial x_{l+1} / \partial x_l \gg 1$. Adaptive optimizers alleviate vanishing and exploding gradients by dividing gradients by their expected sizes. Nevertheless, it is essential to combine adaptive optimizers with appropriate initialization and architecture to avoid numerical instability. 

Optimizers have a myriad of hyperparameters to be initialized and varied throughout training to optimize performance\cite{probst2018tunability} cf. figure~\ref{eqns:optimizers}. For example, stepwise exponentially decayed learning rates are often theoretically optimal\cite{ge2019step}. There are also various heuristics that are often effective, such as using a DEMON decay schedule for an ADAM first moment of the momentum decay rate\cite{chen2019decaying}, 
\begin{equation}
    \beta_1 = \frac{1-t/T}{(1-\beta_\text{init}) + \beta_\text{init}(1- t/T)} \beta_\text{init} \,,
\end{equation}
where $\beta_\text{init}$ is the initial value of $\beta_1$, $t$ is the iteration number, and $T$ is the final iteration number. Developers often optimize ANN hyperparameters by experimenting with a range of heuristic values. Hyperparameter optimization algorithms\cite{yang2020hyperparameter, chandra2019gradient, akiba2019optuna, lakhmiri2019hypernomad, ilievski2017efficient, lorenzo2017particle} can automate optimizer hyperparameter selection. However, automatic hyperparameter optimizers may not yield sufficient performance improvements relative to well-established heuristics to justify their use, especially in initial stages of development.

Alternatives to gradient descent\cite{wilamowski2010neural} are rarely used for parameter optimization as they are not known to consistently improve upon gradient descent. For example, simulated annealing\cite{blum2020learning, ingber1993simulated} has been applied to CNN training\cite{ayumi2016optimization, rere2015simulated}, and can be augmented with momentum to accelerate convergence in deep learning\cite{borysenko2020coolmomentum}. Simulated annealing can also augment gradient descent to improve performance\cite{fischetti2019embedded}. Other approaches include evolutionary\cite{sloss20202019, al2019survey} and genetic\cite{shapiro1999genetic, doerr2017fast} algorithms, which can be a competitive alternative to deep reinforcement learning where convergence is slow\cite{such2017deep}. Indeed, recent genetic algorithms have outperformed a popular deep reinforcement learning algorithm\cite{sehgal2019deep}. Another direction is to augment genetic algorithms with ANNs to accelerate convergence\cite{hu2020genetic, jennings2019genetic, nigam2019augmenting, potapov2017genetic}. Other alternatives to backpropagation include direct search\cite{powell1998direct}, the Moore-Penrose Pseudo Inverse\cite{ranganathan2018new}; particle swarm optimization\cite{junior2019particle, qolomany2017parameters, kennedy1995particle, kennedy1997particle} (PSO); and echo-state networks\cite{xu2020review, jaeger2007echo, gallicchio2017deep} (ESNs) and extreme learning machines\cite{alaba2019towards, ghosh2018survey, albadra2017extreme, tang2015extreme, huang2011extreme, huang2006extreme, huang2004extreme} (ELMs), where some randomly initialized weights are never updated.

\subsection{Reinforcement Learning}

Reinforcement learning\cite{li2017deep, mondal2020survey, haney2020deep, nguyen2020deep, botvinick2019reinforcement, recht2019tour, arulkumaran2017brief} (RL) is where a machine learning system, or \enquote{actor}, is trained to perform a sequence of actions. Applications include autonomous driving\cite{kiran2020deep, nageshrao2019autonomous, talpaert2019exploring}, communications network control\cite{luong2019applications, di2019reinforcement}, energy and environmental management\cite{han2019review, mason2019review}, playing games\cite{alphastarblog, firoiu2017beating, lample2017playing, silver2016mastering, mnih2015human, lillicrap2015continuous, mnih2013playing, tesauro2002programming}, and robotic manipulation\cite{nguyen2019review, bhagat2019deep}. To optimize a MDP\cite{saldi2019asymptotic, jaakkola1995reinforcement}, a discounted future reward, $Q_t$, at step $t$ in a MDP with $T$ steps is usually calculated from step rewards, $r_t$, with Bellman's equation,
\begin{equation}
    Q_t = \sum\limits_{t'=t}^T \gamma^{t'-t} r_{t'},
\end{equation}
where $\gamma \in [0,1)$ discounts future step rewards. To be clear, multiplying $Q_t$ by $-1$ yields a loss that can be minimized using the methods in section~\ref{sec:gradient_descent}. 

In practice, many MDPs are partially observed or have non-differentiable losses that may make it difficult to learn a good policy from individual observations. However, RNNs can often learn a model of their environments from sequences of observations\cite{heess2015memory}. Alternatively, FNNs can be trained with groups of observations that contain more information than individual observations\cite{mnih2015human, lillicrap2015continuous}. If losses are not differentiable, a critic can learn to predict differentiable losses for actor training cf. section~\ref{sec:actor-critic}. Alternatively, actions can be sampled from a differentiable probability distribution\cite{zhao2011analysis, mnih2014recurrent} as training losses given by products of losses and sampling probabilities are differentiable. There are also a variety of alternatives to gradient descent introduced at the end of section~\ref{sec:gradient_descent} that do not require differentiable loss functions.

There are a variety of exploration strategies for RL\cite{weng2020exploration, plappert2018parameter}. Adding Ornstein-Uhlenbeck\cite{uhlenbeck1930theory} (OU) noise to actions is effective for continuous control tasks optimized by deep deterministic policy gradients\cite{lillicrap2015continuous} (DDPG) or recurrent deterministic policy gradients\cite{heess2015memory} (RDPG) RL algorithms. Adding Gaussian noise achieves similar performance for optimization by TD3\cite{fujimoto2018addressing} or D4PG\cite{barth2018distributed} RL algorithms. However, a comparison of OU and Gaussian noise across a variety of tasks\cite{noriothesis} found that OU noise usually achieves similar performance to or outperforms Gaussian noise. Similarly, exploration can be induced by adding noise to ANN parameters\cite{fortunato2019noisy, hazan2019provably}. Other approaches to exploration include rewarding actors for increasing action entropy\cite{haarnoja2017reinforcement, ahmed2019understanding, hazan2019provably} and intrinsic motivation\cite{aubret2019survey, linke2019adapting, pathak2017curiosity}, where ANNs are incentified to explore actions that they are unsure about.

RL algorithms are often partitioned into online learning\cite{hoi2018online, wei2017online}, where training data is used as it is acquired; and offline learning\cite{levine2020offline, varile2020train}, where a static training dataset has already been acquired. However, many algorithms operate in an intermediate regime, where data collected with an online policy is stored in an experience replay\cite{fedus2020revisiting, nair2020accelerating, lin1992self} buffer for offline learning. Training data is often sampled at random from a replay. However, prioritizing the replay of data with high losses\cite{schaul2015prioritized} or data that results in high policy improvements\cite{zha2019experience} often improves actor performance. A default replay buffer size of around $10^6$ examples is often used; however, training is sensitive to replay buffer size\cite{zhang2017deeper}. If the replay is too small, changes in actor policy may destabilize training; whereas if the replay is too large, convergence may be slowed by delays before the actor learns from policy changes.

\subsection{Automatic Machine Learning}

There are a variety of automatic machine learning\cite{he2019automl, malekhosseini2019modeling, jaafra2019reinforcement, elsken2018neural, waring2020automated} (AutoML) algorithms that can create and optimize ANN architectures and learning policies for a dataset of input and target output pairs. Most AutoML algorithms are based on RL or evolutionary algorithms. Examples of AutoML algorithms include AdaNet\cite{weill2019adanet, weill2018introducing}, Auto-DeepLab\cite{liu2019auto}, AutoGAN\cite{gong2019autogan}, Auto-Keras\cite{jin2019auto}, auto-sklearn\cite{feurer2015efficient}, DARTS+\cite{liang2019darts+}, EvoCNN\cite{sun2019evolving}, H2O\cite{ledell2020h2o}, Ludwig\cite{molino2019ludwig}, MENNDL\cite{young2015optimizing, patton2018167}, NASBOT\cite{kandasamy2018neural}, XNAS\cite{nayman2019xnas}, and others\cite{jiang2019accuracy, liu2018progressive, zhang2018graph, baker2017accelerating, zoph2016neural}. AutoML is becoming increasingly popular as it can achieve higher performance than human developers\cite{hanussek2020can, zoph2018learning} and enables human developer time to be traded for potentially cheaper computer time. Nevertheless, AutoML is currently limited to established ANN architectures and learning policies. Consequently, we recommend that researchers either focus on novel ANN architectures and learning policies or developing ANNs for novel applications.

\subsection{Initialization}\label{sec:initialization}

How ANN trainable parameters are initialized\cite{hanin2018start, godoy2018hyper-parameters} is related to model capacity\cite{nagarajan2019generalization}. Further, initializing parameters with values that are too small or large can cause slow learning or divergence\cite{hanin2018start}. Careful initialization can also prevent training by gradient descent being destabilized by vanishing or exploding gradients\cite{hanin2018neural, hanin2018start, pascanu2013difficulty}, or high variance of length scales across layers\cite{hanin2018start}. Finally, careful initialization can enable momentum to accelerate convergence and improve performance\cite{sutskever2013importance}. Most trainable parameters are multiplicative weights or additive biases. Initializing parameters with constant values can result in every parameter in a layer receiving the same updates by gradient descent, reducing model capacity. Thus, weights are often randomly initialized. Followingly, biases are often initialized with constant values due to symmetry breaking by the weights.

Consider the projection of $n_\text{in}$ inputs, $\textbf{x}^\text{input} = \{ x_1^\text{input}, ..., x_{n_\text{in}}^\text{input} \}$, to $n_\text{out}$ outputs, $\textbf{x}^\text{output} = \{ x_1^\text{output}, ..., x_{n_\text{out}}^\text{output} \}$, by an $n_\text{in} \times n_\text{out}$ weight matrix, $\textbf{w}$. The expected variance of an output element is\cite{godoy2018hyper-parameters}
\begin{align}
    \text{Var}(\textbf{x}^\text{output}) = n_\text{in} \text{E}(\textbf{x}^\text{input})^2 \text{Var}(\textbf{w}) + n_\text{in} \text{E}(\textbf{w})^2 \text{Var}(\textbf{x}^\text{input}) + n_\text{in} \text{Var}(\textbf{w})\text{Var}(\textbf{x}^\text{input}) \,,
\end{align}
where $\text{E}(\textbf{x})$ and $\text{Var}(\textbf{x})$ denote the expected mean and variance of elements of $\textbf{x}$, respectively. For similar length scales across layers, $\text{Var}(\textbf{x}^\text{output})$ should be constant. Initially, similar variances can be achieved by normalizing ANN inputs to have zero mean, so that $\text{E}(\textbf{x}^\text{input}) = 0$, and initializing weights so that $\text{E}(\textbf{w}) = 0$ and $\text{Var}(\textbf{w}) = 1/ n_\text{in}$. However, parameters can shift during training, destabilizing learning. To compensate for parameter shift, popular normalization layers like batch normalization often impose $\text{E}(\textbf{x}^\text{input}) = 0$ and $\text{Var}(\textbf{x}^\text{input}) = 1$, relaxing need for $\text{E}(\textbf{x}^\text{input}) = 0$ or $\text{E}(\textbf{w}) = 0$. Nevertheless, training will still be sensitive to the length scale of trainable parameters.

There are a variety of popular weight initializers that adapt weights to ANN architecture. One of the oldest methods is LeCun initialization\cite{lecun2012efficient, klambauer2017self}, where weights are initialized with variance,
\begin{equation}
    \text{Var}(\textbf{w}) = \frac{1}{n_\text{in}} \,,
\end{equation}
which is argued to produce outputs with similar length scales in the previous paragraph. However, a similar argument can be made for initializing with $\text{Var}(\textbf{w}) = 1/n_\text{out}$ to produce similar gradients at each layer during the backwards pass\cite{godoy2018hyper-parameters}. As a compromise, Xavier initialization\cite{glorot2010understanding} computes an average,
\begin{equation}
    \text{Var}(\textbf{w}) = \frac{2}{n_\text{in} + n_\text{out}} \,.
\end{equation}
However, adjusting weights for $n_\text{out}$ is not necessary for adaptive optimizers like ADAM, which divide gradients by their length scales, unless gradients will vanish or explode. Finally, He initialization\cite{he2015delving} doubles the variance of weights to
\begin{equation}
    \text{Var}(\textbf{\textbf{w}}) = \frac{2}{n_\text{in}} \,,
\end{equation}
and is often used in ReLU networks to compensate for activation functions halving variances of their outputs\cite{he2015delving, kumar2017weight, godoy2018hyper-parameters}. Most trainable parameters are initialized from either a zero-centred Gaussian or uniform distribution. For convenience, the limits of such a uniform distribution are $\pm (3\text{Var}(\textbf{w}))^{1/2}$. Uniform initialization can outperform Gaussian initialization in DNNs due to Gaussian outliers harming learning\cite{godoy2018hyper-parameters}. However, issues can be avoided by truncating Gaussian initialization, often to two standard deviations, and rescaling to its original variance.

Some initializers are mainly used for RNNs. For example, orthogonal initialization\cite{saxe2013exact} often improves RNN training\cite{henaff2016recurrent} by reducing susceptibility to vanishing and exploding gradients. Similarly, identity initialization\cite{le2015simple, mikolov2014learning} can help RNNs to learn long-term dependencies. In most ANNs, biases are initialized with zeros. However, the forget gates of LSTMs are often initialized with ones to decrease forgetting at the start of training\cite{jozefowicz2015empirical}. Finally, the start states of most RNNs are initialized with zeros or other constants. However, random multivariate or trainable variable start states can improve performance\cite{pitis2016nonzero}.

There are a variety of alternatives to initialization from random multivariates. Weight normalized\cite{salimans2016weight} ANNs are a popular example of data-dependent initialization, where randomly initialized weight magnitudes and biases are chosen to counteract variances and means of an initial batch of data. Similarly, layer-sequential unit-variance (LSUV) initialization\cite{mishkin2015all} consists of orthogonal initialization followed by adjusting the magnitudes of weights to counteract variances of an initial batch of data. Other approaches standardize the norms of backpropagated gradients. For example, random walk initialization\cite{sussillo2014random} (RWI) finds scales for weights to prevent vanishing or exploding gradients in deep FNNs, albeit with varied success\cite{mishkin2015all}. Alternatively, MetaInit\cite{dauphin2019metainit} scales the magnitudes of randomly initialized weights to minimize changes in backpropagated gradients per iteration of gradient descent.

\subsection{Regularization}

There are a variety of regularization mechanisms\cite{kukavcka2017regularization, kang2019regularization, liu2019regularization, vettam2019regularized} that modify learning algorithms to improve ANN performance. One of the most popular is L$X$ regularization, which decays weights by adding a loss,
\begin{equation}
    L_X = \lambda_X \sum\limits_i \frac{|\theta_i|^X}{X} \,,
\end{equation}
weighted by $\lambda_X$ to each trainable variable, $\theta_i$. L2 regularization\cite{golatkar2019time, tanay2018new, van2017l2} is preferred\cite{van2012lost} for most DNN optimization as subtraction of its gradient, $\partial_{\theta_i} L_2 = \lambda_2 \theta_i$, is equivalent to computationally-efficient multiplicative weight decay. Nevertheless, L1 regularization is better at inducing model sparsity\cite{gribonval2012compressible} than L2 regularization, and L1 regularization achieves higher performance in some applications\cite{ng2004feature}. Higher performance can also be achieved by adding both L1 and L2 regularization in elastic nets\cite{zou2005regularization}. L$X$ regularization is most effective at the start of training and becomes less important near convergence\cite{golatkar2019time}. Finally, L1 and L2 regularization are closely related to lasso\cite{tibshirani1996regression} and ridge\cite{hoerl1970ridge} regularization, respectively, whereby trainable parameters are adjusted to limit $L_1$ and $L_2$ losses.

Gradient clipping\cite{zhang2019gradient, gorbunov2020stochastic, chen2020understanding, menon2019can} accelerates learning by limiting large gradients, and is most commonly applied to RNNs. A simple approach is to clip gradient magnitudes to a threshold hyperparameter. However, it is more common to scale gradients, $\textbf{g}_i$, at layer $i$ if their norm is above a threshold, $u$, so that\cite{pascanu2013difficulty, chen2020understanding}
\begin{equation}
    \textbf{g}_i \leftarrow \begin{cases}
    \textbf{g}_i, & \text{if } ||\textbf{g}_i||_n \le u \\
    \frac{u}{||\textbf{g}_i||_n}\textbf{g}_i, & \text{if } ||\textbf{g}_i||_n > u
\end{cases}
\end{equation}
where $n=2$ is often chosen to minimize computation. Similarly, gradients can be clipped if they are above a global norm,
\begin{align}
g_\text{norm} = \left(\sum\limits_{i=1}^L ||\textbf{g}_i||_n^n \,, \right)^{1/n}
\end{align}
computed with gradients at $L$ layers. Scaling gradient norms is often preferable to clipping to a threshold as scaling is akin to adapting layer learning rates and does not affect the directions of gradients. Thresholds for gradient clipping are often set based on average norms of backpropagated gradients during preliminary training\cite{bengio2013advances}. However, thresholds can also be set automatically and adaptively\cite{seetharaman2020autoclip, gorbunov2020stochastic}. In addition, adaptive gradient clipping algorithms can skip training iterations if gradient norms are anomalously high\cite{chen2018best}, which often indicates an imminent gradient explosion.

Dropout\cite{srivastava2014dropout, labach2019survey, li2016improved, mianjy2018implicit, warde2013empirical} often reduces overfitting by only using a fraction, $p_i$, of layer $i$ outputs during training, and multiplying all outputs by $p_i$ for inference. However, dropout often increases training time, can be sensitive to $p_i$, and sometimes lowers performance\cite{garbin2020dropout}. Improvements to dropout at the structural level, such as applying it to convolutional channels, paths, and layers, rather than random output elements, can improve performance\cite{cai2019effective}. For example, DropBlock\cite{ghiasi2018dropblock} improves performance by dropping contiguous regions of feature maps to prevent dropout being trivially circumvented by using spatially correlated neighbouring outputs. Similarly, PatchUp\cite{faramarzi2020patchup} swaps or mixes contiguous regions with regions for another sample. Dropout is often outperformed by Shakeout\cite{kang2017shakeout, kang2016shakeout}, a modification of dropout that randomly enhances or reverses contributions of outputs to the next layer.

Noise often enhances ANN training by decreasing susceptibility to spurious local minima\cite{zhou2019towards}. Adding noise to trainable parameters can improve generalization\cite{graves2013speech, graves2011practical}, or exploration for RL\cite{fortunato2019noisy}. Parameter noise is usually additive as it does not change an objective function being learned, whereas multiplicative noise can change the objective\cite{sum2019limitation}. In addition, noise can be added to inputs\cite{holmstrom1992using, vincent2010stacked}, hidden layers\cite{you2019adversarial, roth2017stabilizing}, generated outputs\cite{jenni2019stabilizing} or target outputs\cite{sun2019limited, salimans2016improved}. However, adding noise to signals does not always improve performance\cite{greff2016lstm}. Finally, modifying usual gradient noise\cite{simsekli2019tail} by adding noise to gradients can improve performance\cite{neelakantan2015adding}. Typically, additive noise is annealed throughout training, so that that final training is with a noiseless model that will be used for inference. 

There are a variety of regularization mechanisms that exploit extra training data. A simple approach is to create extra training examples by data augmentation\cite{shorten2019survey, raileanu2020automatic, antczak2019regularization}. Extra training data can also be curated, or simulated for training by domain adaption\cite{tzeng2017adversarial, ganin2015unsupervised, tzeng2015simultaneous}. Alternatively, semi-supervised learning\cite{ouali2020overview, zhu2020semi, aitchison2020statistical, bagherzadeh2019review, rasmus2015semi, lee2013pseudo} can generate target outputs for a dataset of unpaired inputs to augment training with a dataset of paired inputs and target outputs. Finally, multitask learning\cite{sun2019multiview, ruder2017overview, thung2018brief, zhang2017survey, caruana1997multitask} can improve performance by introducing additional loss functions. For instance, by adding an auxiliary classifier to predict image labels from features generated by intermediate DNN layers\cite{odena2016conditional, shu2017ac, gong2019twin, han2020unbiased}. Losses are often manually balanced; however, their gradients can also be balanced automatically and adaptively\cite{chen2017gradnorm, lee2020multitask}.

\subsection{Data Pipeline}

A data pipeline prepares data to be input to an ANN. Efficient pipelines often parallelize data preparation across multiple CPU cores\cite{tensorflow2020better}. Small datasets can be stored in RAM to decrease data access times, whereas large dataset elements are often loaded from files. Loaded data can then be preprocessed and augmented\cite{li2020feature, raileanu2020automatic, shorten2019survey, bhanja2018impact, van2016learning}. For electron micrographs, preprocessing often includes replacing non-finite elements, such as NaN and inf, with finite values; linearly transforming intensities to a common range, such as $[-1, 1]$ or zero mean and unit variance; and performing a random combination of flips and 90$\degree$ rotations to augment data by a factor of eight\cite{ede2019improving, ede2020exit, ede2020partial, ede2020warwick, ede2019deep}. Preprocessed examples can then be combined into batches. Typically, multiple batches that are ready to be input are prefetched and stored in RAM to avoid delays due to fluctuating CPU performance. 

To efficiently utilize data, training datasets are often reiterated over for multiple training epochs. Usually, training datasets are reiterated over about $10^2$ times. Increasing epochs can maximize utilization of potentially expensive training data; however, increasing epochs can lower performance due to overfitting\cite{li2020gradient, flynn2020bounding} or be too computationally expensive\cite{chollet2017xception}. Naively, batches of data can be randomly sampled with replacement during training by gradient descent. However, convergence can be accelerated by reinitializing a training dataset at the start of each training epoch and randomly sampling data without replacement\cite{nagaraj2019sgd, gurbuzbalaban2019random, haochen2019random, shamir2016without, bottou2009curiously}. Most modern DLFs, such as TensorFlow, provide efficient and easy-to-use functions to control data sampling\cite{tensorflow2020dataset}.

\subsection{Model Evaluation}

There are a variety of methods for ANN performance evaluation\cite{raschka2018model}. However, most ANNs are evaluated by 1-fold validation, where a dataset is partitioned into training, validation, and test sets. After ANN optimization with a training set, ability to generalize is measured with a validation set. Multiple validations may be performed for training with early stopping\cite{li2020gradient, flynn2020bounding} or ANN learning policy and architecture selection, so final performance is often measured with a test set to avoid overfitting to the validation set. Most researchers favour using single training, validation, and test sets to simplify standardization of performance benchmarks\cite{ede2020warwick}. However, multiple-fold validation\cite{raschka2018model} or multiple validation sets\cite{harrington2018multiple} can improve performance characterization. Alternatively, models can be bootstrap aggregated\cite{breiman1996bagging} (bagged) from multiple models trained on different subsets of training data. Bagging is usually applied to random forests\cite{breiman2001random, goel2017random, probst2019hyperparameters} or other lightweight models, and enables model uncertainly to be gauged from the variance of model outputs.

For small datasets, model performance is often sensitive to split of data between training and validation sets\cite{xu2018splitting}. Increasing training set size usually increases model accuracy, whereas increasing validation set size decreases performance uncertainty. Indeed, a scaling law can be used to estimate an optimal tradeoff\cite{guyon1997scaling} between training and validation set sizes. However, most experimenters follow a Pareto\cite{newman2005power} splitting heuristic. For example, we often use a 75:15:10 training-validation-test split\cite{ede2020warwick}. Heuristic splitting is justified for ANN training with large datasets insofar that sensitivity to splitting ratios decreases with increasing dataset size\cite{sun2017revisiting}.

\subsection{Deployment}

If an ANN is deployed\cite{opeyemi2019deployment1, opeyemi2019deployment2, wu2019machine} on multiple different devices, such as various electron microscopes, a separate model can be trained for each device\cite{strubell2019energy}, Alternatively, a single model can be trained and specialized for different devices to decrease training requirements\cite{cai2019once}. In addition, ANNs can remotely service requests from cloud containers\cite{suresh2020optimization, kumar2019effective, dubois2016model}. Integration of multiple ANNs can be complicated by different servers for different DLFs supporting different backends; however, unified interfaces are available. For example, GraphPipe\cite{oracle2020graphpipe} provides simple, efficient reference model servers for Tensorflow, Caffe2, and ONNX; a minimalist machine learning transport specification based on FlatBuffers\cite{flatbuffers2020documentation}; and efficient client implementations in Go, Python, and Java. In 2020, most ANNs developed researchers were not deployed. However, we anticipate that deployment will become a more prominent consideration as the role of deep learning in electron microscopy matures.

Most ANNs are optimized for inference by minimizing parameters and operations from training time, like MobileNets\cite{howard2017mobilenets}. However, less essential operations can also be pruned after training\cite{blalock2020state, pasandi2020modeling}. Another approach is quantization, where ANN bit depths are decreased, often to efficient integer instructions, to increase inference throughput\cite{wu2020integer, nayak2019bit}. Quantization often decreases performance; however, the amount of quantization can be adapted to ANN components to optimize performance-throughput tradeoffs\cite{zhou2017adaptive}. Alternatively, training can be modified to minimize the impact of quantization on performance\cite{yang2019quantization, zhuang2019effective, li2017training}. Another approach is to specialize bit manipulation for deep learning. For example, signed brain floating point (bfloat16) often improves accuracy on TPUs by using an 8 bit mantissa and 7 bit exponent, rather than a usual 5 bit mantissa and 10 bit exponent\cite{wang2019bfloat16}. Finally, ANNs can be adaptively selected from a set of ANNs based on available resources to balance tradeoff of performance and inference time\cite{marco2020optimizing}, similar to image optimization for web applications\cite{jackson2020how, leventic2016compression}.

\begin{figure*}[tbh!]
\renewcommand\thefigure{16} 
\centering
\includegraphics[width=\textwidth]{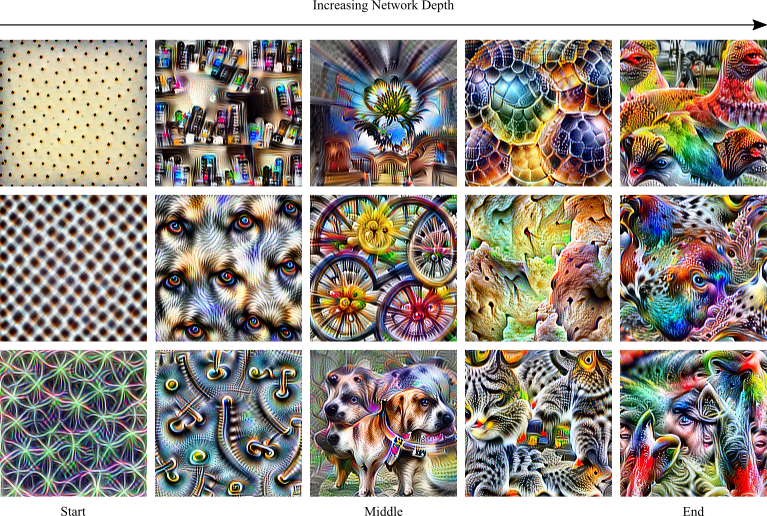}
\caption{ Inputs that maximally activate channels in GoogLeNet\cite{szegedy2015going} after training on ImageNet\cite{krizhevsky2012imagenet}. Neurons in layers near the start have small receptive fields and discern local features. Middle layers discern semantics recognisable by humans, such as dogs and wheels. Finally, layers at the end of the DNN, near its logits, discern combinations of semantics that are useful for labelling. This figure is adapted with permission\cite{olah2017feature} under a Creative Commons Attribution 4.0\cite{cc2020by} license. }
\label{fig:visualization_example}
\end{figure*}

\subsection{Interpretation}

We find that some electron microscopists are apprehensive about working with ANNs due to a lack of interpretability, irrespective of rigorous ANN validation. We try to address uncertainty by providing loss visualizations in some of our electron microscopy papers\cite{ede2020partial, ede2019improving, ede2019deep}. However, there are a variety of popular approaches to explainable artificial intelligence\cite{xie2020explainable, vilone2020explainable, arrieta2020explainable, puiutta2020explainable, gunning2019darpa, samek2019towards, hase2020evaluating} (XAI). One of the most popular approaches to XAI is saliency\cite{ullah2020brief, borji2019salient, cong2018review, borji2015salient}, where gradients of outputs w.r.t. inputs correlate with their importance. Saliency is often computed by gradient backpropagation\cite{rebuffi2020there, wang2019learning, kim2019saliency}. For example, with Grad-CAM\cite{selvaraju2017grad} or its variants\cite{morbidelli2020augmented, omeiza2019smooth, chattopadhay2018grad, patro2019u}. Alternatively, saliency can be predicted by ANNs\cite{wang2019salient, borji2019saliency, wang2019revisiting} or a variety of methods inspired by Grad-CAM\cite{chen2020adapting, ramaswamy2020ablation, wang2020score}. Applications of saliency include selecting useful features from a model\cite{cancela2020scalable}, and locating regions in inputs corresponding to ANN outputs\cite{cheng2014global}.

There are a variety of other approaches to XAI. For example, feature visualization via optimization\cite{olah2017feature, nguyen2019understanding, xiao2019gradient, erhan2009visualizing, mordvintsev2015inceptionism} can find inputs that maximally activate parts of an ANN, as shown in figure~\ref{fig:visualization_example}. Another approach is to cluster features, e.g. by tSNE\cite{maaten2008visualizing, wattenberg2016use} with the Barnes-Hut algorithm\cite{van2013barnes, barnes1986hierarchical}, and examine corresponding clustering of inputs or outputs\cite{ede2020warwick}. Finally, developers can view raw features and gradients during forward and backward passes of gradient descent, respectively. For example, CNN explainer\cite{wang2020cnn, wang2020cnn101} is an interactive visualization tool designed for non-experts to learn and experiment with CNNs. Similarly, GAN Lab\cite{kahng2018gan} is an interactive visualization tool for non-experts to learn and experiment with GANs.

\section{Discussion}

We introduced a variety of electron microscopy applications in section~\ref{sec:introduction} that have been enabled or enhanced by deep learning. Nevertheless, the greatest benefit of deep learning in electron microscopy may be general-purpose tools that enable researchers to be more effective. Search engines based on deep learning are almost essential to navigate an ever-increasing number of scientific publications\cite{bornmann2015growth}. Further, machine learning can enhance communication by filtering spam and phishing attacks\cite{gangavarapu2020applicability, dada2019machine, bhuiyan2018survey}, and by summarizing\cite{zhang2020mining, dangovski2019rotational, scholarcy2020ai-powered} and classifying\cite{romanov2019application, gonccalves2019deep, hughes2017medical, minaee2020deep} scientific documents. In addition, machine learning can be applied to education to automate and standardize scoring\cite{liu2019automated, dong2017attention,  taghipour2016neural, alikaniotis2016automatic}, detect plagiarism\cite{foltynek2019academic, meuschke2019improving, ullah2018software}, and identify at-risk students\cite{lakkaraju2015machine}.

Creative applications of deep learning\cite{foster2019generative, zhan2019deep} include making new art by style transfer\cite{jing2019neural, gatys2016image, gatys2015neural, zhu2017unpaired, li2017demystifying}, composing music\cite{dhariwal2020jukebox, briot2020deep, briot2020deepbook}, and storytelling\cite{brown2020language, radford2019better}. Similar DNNs can assist programmers\cite{chen2020deep, allamanis2018survey}. For example, by predictive source code completion\cite{tabnine2019autocompletion, svyatkovskiy2020intellicode, hammad2020deepclone, schuster2020you, svyatkovskoy2020fast, hellendoorn2019code}, and by generating source code to map inputs to target outputs\cite{balog2017deepcoder} or from labels describing desired source code\cite{murali2018neural}. Text generating DNNs can also help write scientific papers. For example, by drafting scientific passages\cite{demir2019neural} or drafting part of a paper from a list of references\cite{scinote2020manuscript}. Papers generated by early prototypes for automatic scientific paper generators, such as SciGen\cite{stribling2005scigen}, are realistic insofar that they have been accepted by scientific venues.

An emerging application of deep learning is mining scientific resources to make new scientific discoveries\cite{raghu2020survey}. Artificial agents are able to effectively distil latent scientific knowledge as they can parallelize examination of huge amounts of data, whereas information access by humans\cite{kepner2019new, adekitan2019data, xu2019prediction} is limited by human cognition\cite{granger2020toward}. High bandwidth bi-directional brain-machine interfaces are being developed to overcome limitations of human cognition\cite{musk2019integrated}; however, they are in the early stages of development and we expect that they will depend on substantial advances in machine learning to enhance control of cognition. Eventually, we expect that ANNs will be used as scientific oracles, where researchers who do not rely on their services will no longer be able to compete. For example, an ANN trained on a large corpus of scientific literature predicted multiple advances in materials science before they were reported\cite{tshitoyan2019unsupervised}. ANNs are already used for financial asset management\cite{ruf2020neural, huang2019automated} and recruiting\cite{raghavan2020mitigating, mahmoud2019performance, raub2018bots, newman2017reengineering}, so we anticipate that artificial scientific oracle consultation will become an important part of scientific grant\cite{price2019grants, zhuang2019effect} reviews. 

A limitation of deep learning is that it can introduce new issues. For example, DNNs are often susceptible to adversarial attacks\cite{zhang2020adversarial, ma2020understanding, yuan2019adversarial, akhtar2018threat, goodfellow2014explaining}, where small perturbations to inputs cause large errors. Nevertheless, training can be modified to improve robustness to adversarial attacks\cite{wen2019towards, lecuyer2019certified, li2018optimal, xie2020adversarial, deniz2020robustness}. Another potential issue is architecture-specific systematic errors. For example, CNNs often exhibit structured systematic error variation\cite{kinoshita2020fixed, aitken2017checkerboard, odena2016deconvolution, ede2020partial, ede2019improving, ede2019deep}, including higher errors nearer output edges\cite{ede2020partial, ede2019improving, ede2019deep}. However, structured systematic error variation can be minimized by GANs incentifying the generation of realistic outputs\cite{ede2020partial}. Finally, ANNs can be difficult to use as they often require downloading code with undocumented dependencies, downloading a pretrained model, and may require hardware accelerators. These issues can be avoided by serving ANNs from cloud containers. However, it may not be practical for academics to acquire funding to cover cloud service costs.

Perhaps the most important aspect of deep learning in electron microscopy is that it presents new challenges that can lead to advances in machine learning. Simple benchmarks like CIFAR-10\cite{krizhevsky2014cifar, krizhevsky2009learning} and MNIST\cite{lecun2010mnist} have been solved. Subsequently, more difficult benchmarks like Fashion-MNIST\cite{xiao2017fashion} have been introduced. However, they only partially address issues with solved datasets as they do not present fundamentally new challenges. In contrast, we believe that new problems often invite new solutions. For example, we developed adaptive learning rate clipping\cite{ede2020adaptive} (ALRC) to stabilize training of DNNs for partial scanning transmission electron microscopy\cite{ede2020partial}. The challenge was that we wanted to train a large model for high-resolution images; however, training was unstable if we used small batches needed to fit it in GPU memory. Similar challenges abound and can lead to advances in both machine learning and electron microscopy.

\section*{Data Availability}

No new data were created or analysed in this study.

\section*{Acknowledgements}

Thanks go to Jeremy Sloan and Martin Lotz for internally reviewing this article. In addition, part of the text in section~\ref{sec:compressed_sensing} is adapted from our earlier work with permission\cite{ede2020partial} under a Creative Commons Attribution 4.0\cite{cc2020by} license. Finally, the author acknowledges funding from EPSRC grant EP/N035437/1 and EPSRC Studentship 1917382.

\section*{Competing Interests}

The author declares no competing interests.

\bibliography{bibliography}

\end{document}